\documentclass[longauth]{aa} % for the long lists of affiliations 
\usepackage{graphicx}
\usepackage{lscape}

\usepackage{txfonts}
\usepackage{mathabx}
\usepackage[normalem]{ulem}
\usepackage{color}
\usepackage[dvipsnames]{xcolor}
\usepackage{soul}

\def \MJ{M$_{\mathrm{Jup}}$}
\def \MN{M$_{\mathrm{Nep}}$}
\def \ME{M$_{\Earth}$}
\def \RJ{R$_{\mathrm{Jup}}$}
\def \RN{R$_{\mathrm{Nep}}$}
\def \RE{R$_{\Earth}$}
\def \RS{R$_{\odot}$}
\def \msol{M$\mathrm{_\odot}$}

\def \kms{km\,s$^{-1}$}
\def \ms{m\,s$^{-1}$}
\def \1s{$1\,\sigma$}

\def \t0{T$_0$}

\def \toi17{TOI~1736}
\def \toipl17{TOI~1736~b}

\begin{document} 

   \title{TOI-1736 and TOI-2141: two systems including sub-Neptunes around solar analogs revealed by TESS and SOPHIE\thanks{Based on observations collected with the SOPHIE spectrograph on the 1.93 m telescope at the Observatoire de Haute-Provence (CNRS), France.}}
   %\subtitle{}
   \titlerunning{Detection and characterization of TOI-1736 and TOI-2141}

   \author{
    E. Martioli\inst{\ref{lna},\ref{iap}} % lead investigator \orcid{0000-0002-5084-168X} 
    \and G. H\'ebrard \inst{\ref{iap},\ref{ohp}} % lead co-investigator
    \and L. de\,Almeida \inst{\ref{lna}} % fundamental stellar parameters, activity, rotation
    \and N. Heidari \inst{\ref{iap}}  % SOPHIE data reduction and analysis 
    \and D. Lorenzo-Oliveira \inst{\ref{lna}} % differential analysis: stellar parameters, abundances, age
    \and F. Kiefer \inst{\ref{lesia},\ref{iap}} % GAIA astrometry for the upper limit on the mass of TOI-1736 c   
% alphabetical order
    \and J. M. Almenara \inst{\ref{ipag}} % sophie team
    \and A. Bieryla \inst{\ref{cfa}} % TRES spectra and analysis                    
    \and I. Boisse \inst{\ref{lam}} % sophie team
    \and X. Bonfils \inst{\ref{ipag}} % sophie team
    \and C. Brice\~{n}o \inst{\ref{ctio}} % SOAR high contrast imaging for TOI-2141           
    \and K. A. Collins \inst{\ref{cfa}} % LCOGT 1\,m NEB Search
    \and P. Cortés-Zuleta \inst{\ref{lam}} % sophie team
    \and S. Dalal \inst{\ref{exeter},\ref{iap}} % sophie team
    \and M. Deleuil \inst{\ref{lam}} % sophie team
    \and X. Delfosse \inst{\ref{ipag}} % sophie team
    \and O. Demangeon \inst{\ref{caup}} % sophie team
    \and J. D. Eastman \inst{\ref{cfa}} % TESS Science team
    \and T. Forveille \inst{\ref{ipag}} % sophie team
    \and E. Furlan \inst{\ref{caltech}} % high-contrast imaging of TOI-1736
    \and S. B. Howell \inst{\ref{nasaames}} % high-contrast imaging of TOI-1736  
    \and S. Hoyer \inst{\ref{lam}} % sophie team
    \and J. M. Jenkins \inst{\ref{nasaames}} % SPOC TESS team  
    \and D. W. Latham  \inst{\ref{cfa}} % TESS architects
    \and N. Law \inst{\ref{chapelhill}} % SOAR high contrast imaging for TOI-2141 
    \and A. W. Mann \inst{\ref{chapelhill}} % SOAR high contrast imaging for TOI-2141            
    \and C. Moutou \inst{\ref{irap}} % sophie team
    \and N. C. Santos \inst{\ref{caup},\ref{up}} % fundamental stellar parameters
    \and S. G. Sousa \inst{\ref{caup}} % fundamental stellar parameters
    \and K. G. Stassun \inst{\ref{vanderbilt}} % SED analysis
    \and C. Stockdale \inst{\ref{hazelwood}} % LCOGT 1\,m NEB Search
    \and G. Torres \inst{\ref{cfa}} % TESS Science team 
    \and J. D. Twicken \inst{\ref{seti},\ref{nasaames}} % TESS SPOC team  
    \and J. N. Winn \inst{\ref{princeton}} % TESS Science team     
    \and C. Ziegler \inst{\ref{sfasu}} % SOAR high contrast imaging for TOI-2141       
}
    \institute{
    Laborat\'{o}rio Nacional de Astrof\'{i}sica, Rua Estados Unidos 154, 37504-364, Itajub\'{a} - MG, Brazil, \email{emartioli@lna.br} \label{lna}
    \and Institut d'Astrophysique de Paris, CNRS, UMR 7095, Sorbonne Universit\'{e}, 98 bis bd Arago, 75014 Paris, France \label{iap}
    \and Observatoire de Haute Provence, St Michel l'Observatoire, France \label{ohp}
    \and Universit\'{e} Grenoble Alpes, CNRS, IPAG, 414 rue de la Piscine, 38400 St-Martin d'Hères, France \label{ipag}
    %\and Laboratoire d'astrophysique de Marseille, Univ. de Provence, UMR6110 CNRS, 38 r. F. Joliot Curie, 13388 Marseille cedex 13, France \label{lam}
    \and Aix Marseille Univ, CNRS, CNES, LAM, 38 rue Frédéric Joliot-Curie, 13388 Marseille, France  \label{lam} 
    \and Instituto de Astrof\'isica e Ci\^encias do Espa\c{c}o, Universidade do Porto, CAUP, Rua das Estrelas, 4150-762 Porto, Portugal \label{caup}
    \and Departamento de F\'isica e Astronomia, Faculdade de Ci\^encias, Universidade do Porto, Rua do Campo Alegre, 4169-007 Porto, Portugal \label{up}
    \and LESIA, Observatoire de Paris, Universit\'e PSL, CNRS, Sorbonne Universit\'e, Universit\'e Paris Cit\'e, 5 place Jules Janssen, 92195 Meudon, France \label{lesia}
    \and Universit\'e de Toulouse, CNRS, IRAP, 14 avenue Belin, 31400 Toulouse, France \label{irap}    
    \and Center for Astrophysics \textbar\ Harvard \& Smithsonian, 60 Garden Street, Cambridge, MA 02138, USA \label{cfa}
    \and Hazelwood Observatory, Victoria, Australia  \label{hazelwood}
    \and Department of Physics and Astronomy, Vanderbilt University, Nashville, TN 37235, USA \label{vanderbilt}   
    \and Department of Physics, Engineering and Astronomy, Stephen F. Austin State University, 1936 North St, Nacogdoches, TX 75962, USA \label{sfasu}
    \and Cerro Tololo Inter-American Observatory, Casilla 603, La Serena, Chile \label{ctio}   
    \and Department of Physics and Astronomy, The University of North Carolina at Chapel Hill, Chapel Hill, NC 27599-3255, USA \label{chapelhill} 
    \and Astrophysics Group, University of Exeter, Exeter EX4 2QL, UK \label{exeter} 
    \and Department of Astrophysical Sciences, Princeton University, Princeton, NJ 08544, USA \label{princeton} 
    \and SETI Institute, Mountain View, CA 94043 USA/NASA Ames Research Center, Moffett Field, CA 94035 USA \label{seti}
    \and NASA Exoplanet Science Institute, Caltech IPAC, 1200 E. California Blvd., Pasadena, CA 91125, USA \label{caltech}   
    \and NASA Ames Research Center, Moffett Field, CA 94035, USA \label{nasaames}     
}
    
   \date{Received; accepted}

  \abstract { 
Planetary systems around solar analogs inform us about how planets form and evolve in Solar System-like environments. We report the detection and characterization of two planetary systems around the solar analogs TOI-1736 and TOI-2141 using TESS photometry data and spectroscopic data obtained with the SOPHIE instrument on the 1.93~m telescope at the Observatoire de Haute-Provence (OHP). We performed a detailed spectroscopic analysis of these systems to obtain the precise radial velocities (RV) and physical properties of their host stars. TOI-1736 and TOI-2141 each host a transiting sub-Neptune with radii of $2.44\pm0.18$~\RE\ and $3.05\pm0.23$~\RE, orbital periods of $7.073088(7)$~d and $18.26157(6)$~d, and masses of $12.8\pm1.8$~\ME\ and $24\pm4$~\ME, respectively. TOI-1736 shows long-term RV variations that are consistent with a two-planet solution plus a linear trend of $-0.177$~m\,s$^{-1}$\,d$^{-1}$.  We measured an RV semi-amplitude of $201.1\pm0.7$~\ms\ for the outer companion, TOI-1736~c, implying a projected mass of $m_{c}\sin{i}=8.09\pm0.20$~\MJ. From the GAIA DR3 astrometric excess noise, we constrained the mass of TOI-1736~c at $8.7^{+1.5}_{-0.6}$~\MJ. This planet is in an orbit of $570.2\pm0.6$~d with an eccentricity of $0.362\pm0.003$ and a semi-major axis of $1.381\pm0.017$~au, where it receives a flux of $0.71\pm0.08$ times the bolometric flux incident on Earth, making it an interesting case of a supergiant planet that has settled into an eccentric orbit in the habitable zone of a solar analog. Our analysis of the mass-radius relation for the transiting sub-Neptunes shows that both TOI-1736~b and TOI-2141~b likely have an Earth-like dense rocky core and a water-rich envelope.
  }


   \maketitle
%
%-------------------------------------------------------------------
\section{Introduction}

Studying solar analogs is a way to understand the evolution of stars similar to our Sun. Although the physical and evolutionary stellar characteristics of solar analogs are relatively well studied, the formation and evolution of planetary systems around these stars still lack observational constraints. 

The Sun being the nearest and most studied star, where the most accurate stellar properties can be obtained, provides a way to benchmark measurements of distant stars that have similar physical properties. This fact favors a specific class of stars called solar analogs, defined as those stars that have physical properties within a certain range of solar values. \cite{Soderblom1998} defined a solar analog to be a main sequence star with an effective temperature between 5278~K and 6278~K, a metallicity within the range $\pm0.3$\,dex, and it must not have a close stellar companion (arbitrarily defined with an orbital period of less than ten days). 

The motivation to study solar analogs lies in the fact that stellar properties can be obtained with a greater precision since the Sun can be used as a reference \citep[e.g.,][]{Bedell2014,Ramirez2014}. Consequently, the uncertainties of the physical parameters of exoplanets can be significantly reduced for the solar analogs, which makes them interesting laboratories to understand the physics of planetary systems.  In addition, stellar abundances can be obtained with greater accuracy, which informs us about the chemical composition of the environment where the planets formed.  As a result, chemical clocks \citep[e.g.,][]{daSilva2012} can provide better age diagnoses that can be used to validate the isochronal ages. This can also improve, for example, the constraints in exoplanet interior structure models based on the abundance of refractory elements (Fe, Mg, and Si) of the host star \citep{Dorn2015, Adibekyan2021}.  Ultimately, studying planets around solar analogs also makes our knowledge of the conditions for the development of life less uncertain.

The Transiting Exoplanet Survey Satellite \citep[TESS,][]{tess_paper} has identified more than 6000 candidate exoplanets \citep{Guerrero2021}. The transit method is efficient to detect close-in planets. Long-term radial velocity (RV) programs, such as the one carried out with the SOPHIE instrument at the Observatoire Haute-Provence, can establish the nature of candidate planets identified by TESS and can also detect additional non-transiting companions. We have identified two exoplanet candidates detected by TESS around the solar analogs TOI-1736 (TIC 408618999) and TOI-2141 (TIC 287256467). These planet candidates were identified as a TESS object of interest (TOI), which presented recurrent transit-like events in their light curves \citep{Jenkins2002, Jenkins2010}.  Our RV follow-up with SOPHIE characterizes the nature of these two planets and detects at least two additional companions in the system TOI-1736. This paper presents the detection and characterization of both systems, where we performed a detailed analysis of their spectra to obtain a refined characterization of the host stars. The stellar parameters are summarized in Table \ref{tab:stellarparams}. We combined the TESS and RV data to characterize all planets in both systems. Finally, we present an analysis of the internal structure and possible compositions of the innermost planets.

\begin{table*}
\centering \large
\caption{Summary of the stellar parameters of TOI-1736 and TOI-2141.}
\label{tab:stellarparams}
\begin{tabular}{lccc}
\hline
Parameter & TOI-1736 & TOI-2141 & Ref. \\
\hline
ID (TYC) & 4313-01054-1 & 1540-497-1 & \\
ID (TIC) & 408618999 & 287256467 & \\
RA (hh:mm:ss.ss) & 02:53:44.4053447688 & 17:15:02.9060957472 & 1 \\
Dec (dd:mm:ss.ss) & +69:06:05.066418744 & +18:20:26.764729476 & 1 \\
Epoch (ICRS) & J2000 & J2000 & 1 \\
proper motion in RA, $\mu_{\alpha}$ (mas\,yr$^{-1}$) & $-46.468\pm0.011$ & $52.229\pm0.013$ & 1 \\
proper motion in Dec, $\mu_{\delta}$ (mas\,yr$^{-1}$) & $-14.490\pm0.015$ & $-98.286\pm0.015$ & 1 \\
parallax, $p$ (mas) & $11.343\pm0.015$ & $12.957\pm0.015$ & 1 \\
distance (pc)  & $88.9\pm0.3$ & $77.7\pm0.2$ & 1 \\
%-------------------------------------------------------
B (mag) & $9.64\pm0.03$ &  $10.13\pm0.05$  & 2 \\
V (mag) & $8.953\pm0.002$ &  $9.46\pm0.003$  & 2 \\
{\it TESS} T (mag)  & $8.330\pm0.006$ &  $8.897\pm0.006$ & 2 \\
{\it GAIA} G (mag)  & $8.77969\pm0.00018$ &  $9.34408\pm0.00021$  & 1 \\
{\it 2MASS} J (mag) & $7.69\pm0.02$ &  $8.27\pm0.02$  & 3 \\
{\it 2MASS} H (mag) & $7.42\pm0.05$ &  $7.92\pm0.03$ & 3 \\
{\it 2MASS} K (mag) & $7.28\pm0.02$ &  $7.87\pm0.02$ & 3 \\
{\it WISE} 1 (mag) & $7.23\pm0.04$ &  $7.84\pm0.03$ & 4 \\
{\it WISE} 2 (mag) & $7.30\pm0.02$ &  $7.90\pm0.02$ & 4 \\
{\it WISE} 3 (mag) & $7.28\pm0.02$ &  $7.88\pm0.02$ & 4 \\
{\it WISE} 4 (mag) & $7.24\pm0.11$ &  $7.76\pm0.14$ & 4 \\
%-------------------------------------------------------
effective temperature, $T_{\rm eff}$ (K) & $5807\pm46$ &  $5659\pm48$ & this work \\
%-------------------------------------------------------
surface gravity,$\log g$ (dex) & $4.35\pm0.04$ &  $4.42\pm0.04$ & this work \\
%-------------------------------------------------------
Fe metallicity, $[{\rm Fe}/{\rm H}]$ (dex) & $0.138\pm0.014$ &  $-0.120\pm0.013$ & this work \\
%-------------------------------------------------------
turbulence velocity, $\nu$ (km/s) & $1.10\pm0.02$ &  $0.83\pm0.02$ & this work \\
%-------------------------------------------------------
bolometric flux, $F_{\rm bol}$ ($10^{-9}$\,erg\,s$^{-1}$\,cm$^-2$) & $6.37\pm0.05$ &  $4.65\pm0.11$ & this work \\
%------------------------------------------------------- 
star mass, $M_{\star}$ (\msol) & $1.08\pm0.04$ &  $0.94\pm0.02$ & this work \\
%-------------------------------------------------------
star radius, $R_{\star}$ (\RS) & $1.15\pm0.08$ &  $0.98\pm0.06$ & this work \\
%-------------------------------------------------------
luminosity, $\log{L_{\star}/L_{\odot}}$ &  $0.25\pm0.06$ &  $-0.05\pm0.06$ & this work \\
%-------------------------------------------------------
activity index, $\log{\rm R'}_{\rm HK}$ &  $-5.01^{+0.14}_{-0.22}$ & $-4.78^{+0.10}_{-0.12}$  & this work \\ 
%-------------------------------------------------------
rotation velocity, $v_{\rm rot}\sin{i_{\star}}$ (\kms) & $4.0\pm0.6$ & $2.7\pm0.6$ & this work \\
%-------------------------------------------------------
rotation period, $P_{\rm rot}$ (d) & $28\pm5$ &  $21\pm5$ & this work (5) \\
%-------------------------------------------------------
age (Gyr) & $4.9\pm1.3$ &  $6.4\pm1.8$ & this work \\
%-------------------------------------------------------
\hline
\end{tabular}
\tablebib{
(1) \cite{GaiaEDR3Vizier};
(2) {\tt EXOFOP-TESS} website\footnote{\url{https://exofop.ipac.caltech.edu}} ;
(3) \cite{Cutri2003};
(4) \cite{Wright2010};
(5) rotation periods are estimated from activity-rotation empirical relation as detailed in Section \ref{sec:activityindicators}. 
}
\end{table*}

%--------------------------------------------------------------------
\section{Observations}
\label{sec:observations}

\subsection{TESS photometry}
\label{sec:tessphotometry}

The TESS mission observed TOI-1736 with a cadence of 2 minutes in Sectors 18, 19, and 25, and with a cadence of 20 seconds in Sectors 52, 58, and 59.  TOI-2141 was observed with a cadence of 2 minutes in Sectors 25, 26, and 52. Table \ref{tab:tessobservations} shows the log of TESS observations for these two objects. We obtained the TESS data products from the Mikulski Archive for Space Telescopes (MAST)\footnote{\url{mast.stsci.edu}}, where we used the Presearch Data Conditioning (PDC) flux time series \citep{Smith2012,Stumpe2012,Stumpe2014} processed by the TESS Science Processing Operations Center (SPOC) pipeline \citep{jenkinsSPOC2016} versions listed in Table \ref{tab:tessobservations}.

\begin{table*}
\centering
\caption{Log of TESS observations of TOI-1736 and TOI-2141.}
\label{tab:tessobservations}
\begin{tabular}{ccccccccc}
\hline
OBJECT & TSTART (UTC) & TSTOP (UTC) & Duration (d) & Cadence & Sector & Cycle & Camera & SPOC version \\
\hline
TOI-1736 & 2019-Nov-02 & 2019-Nov-27 & 24.4  & 2 min & 18 & 2 & 2 & 4.0.29-20200410 \\
TOI-1736 & 2019-Nov-27 & 2019-Dec-24 & 25.1  & 2 min & 19 & 2 & 2 & 4.0.30-20200415 \\
TOI-1736 & 2020-May-13 & 2020-Jun-08 & 25.7  & 2 min & 25 & 2 & 4 & 4.0.36-20200520 \\
TOI-1736 & 2022-May-18 & 2022-Jun-13 & 24.4  & 20 s & 52 & 4 & 4 & 5.0.72-20220608 \\
TOI-1736 & 2022-Oct-29 & 2022-Nov-26 & 27.7  & 20 s & 58 & 5 & 2 & 5.0.78-20221129 \\
TOI-1736 & 2022-Nov-26 & 2022-Dec-23 & 26.4  & 20 s & 59 & 5 & 2 & 5.0.79-20221214 \\
\hline
TOI-2141 & 2020-May-13 & 2020-Jun-08 & 25.7  & 2 min & 25 & 2 & 1 & 4.0.36-20200520 \\
TOI-2141 & 2020-Jun-08 & 2020-Jul-04 & 24.9  & 2 min & 26 & 2 & 1 & 5.0.3-20200718 \\
TOI-2141 & 2022-May-18 & 2022-Jun-13 & 24.4  & 2 min & 52 & 4 & 1 & 5.0.72-20220608 \\
\hline
\end{tabular}
\end{table*} 

The SPOC searches light curves for transiting planets with an adaptive, noise-compensating matched filter \citep{Jenkins2002,Jenkins2010,Jenkins2020}. The transit signatures of each candidate are fitted with an initial limb-darkened transit model \citep{Li:DVmodelFit2019} and subjected to a suite of diagnostic tests \citep{Twicken:DVdiagnostics2018}, all available in the TESS SPOC data validation reports (DVR).  The DVR for TOI-1736 reports a candidate planet, hereafter TOI-1736 b, with an estimated radius of R$_{\rm p}=3.0\pm1.1$~\RE\ and an orbital period of P$=7.07307(1)$~d. The difference image centroid offsets locate the source of the transit signal within $4.4\pm4.1$~arcsec of the target star. The TESS Science Office (TSO) issued an alert for TOI~1736.01 on 27 February 2020 based on the DVR associated with the combined light curve for Sectors 18-19 \citep{Guerrero2021}.

\begin{figure}
   \centering
       \includegraphics[width=0.85\hsize]{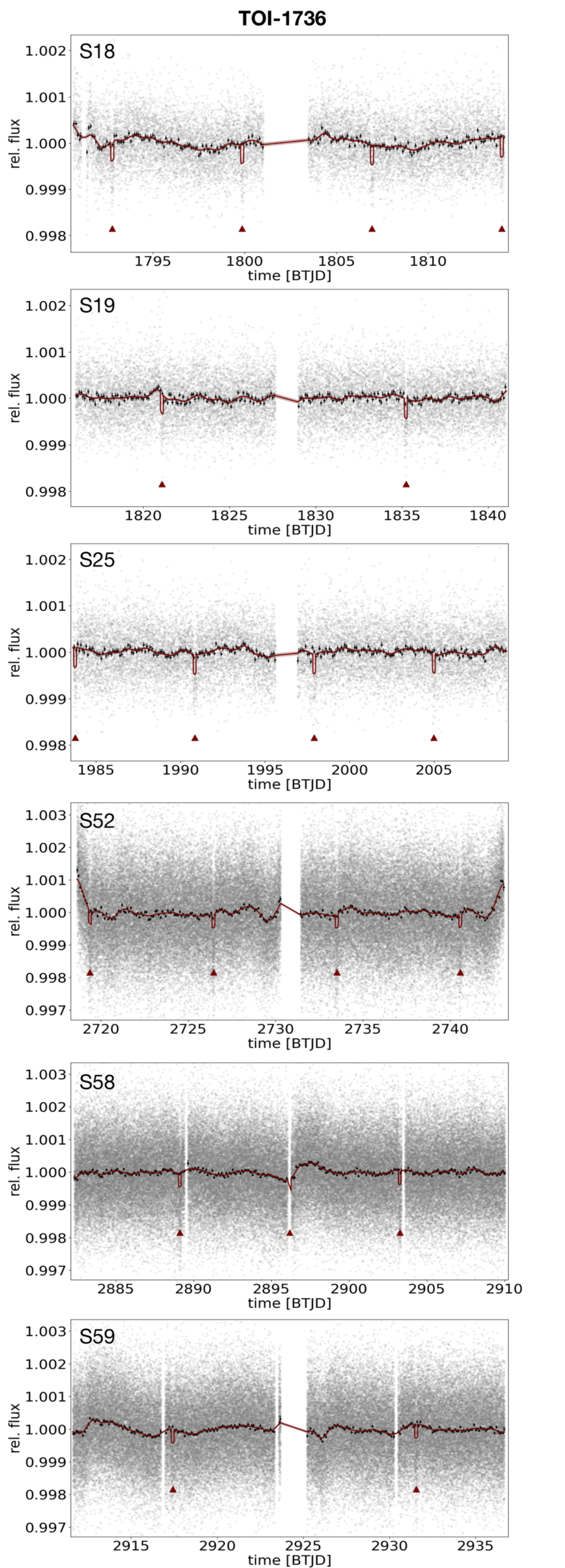}
      \caption{TESS light curve of TOI-1736. The gray points show the TESS photometry data normalized by the median flux of each sector. The black points show the same data binned by the weighted average with bin sizes of 0.1~d. The red line shows the baseline GP model that was fitted to the binned data multiplied by the best-fit transit model for TOI-1736~b. The red triangles show the predicted central time of TOI-1736~b transits.
      }
        \label{fig:toi1736_transits_fit}
  \end{figure}

\begin{figure}
   \centering
       \includegraphics[width=0.85\hsize]{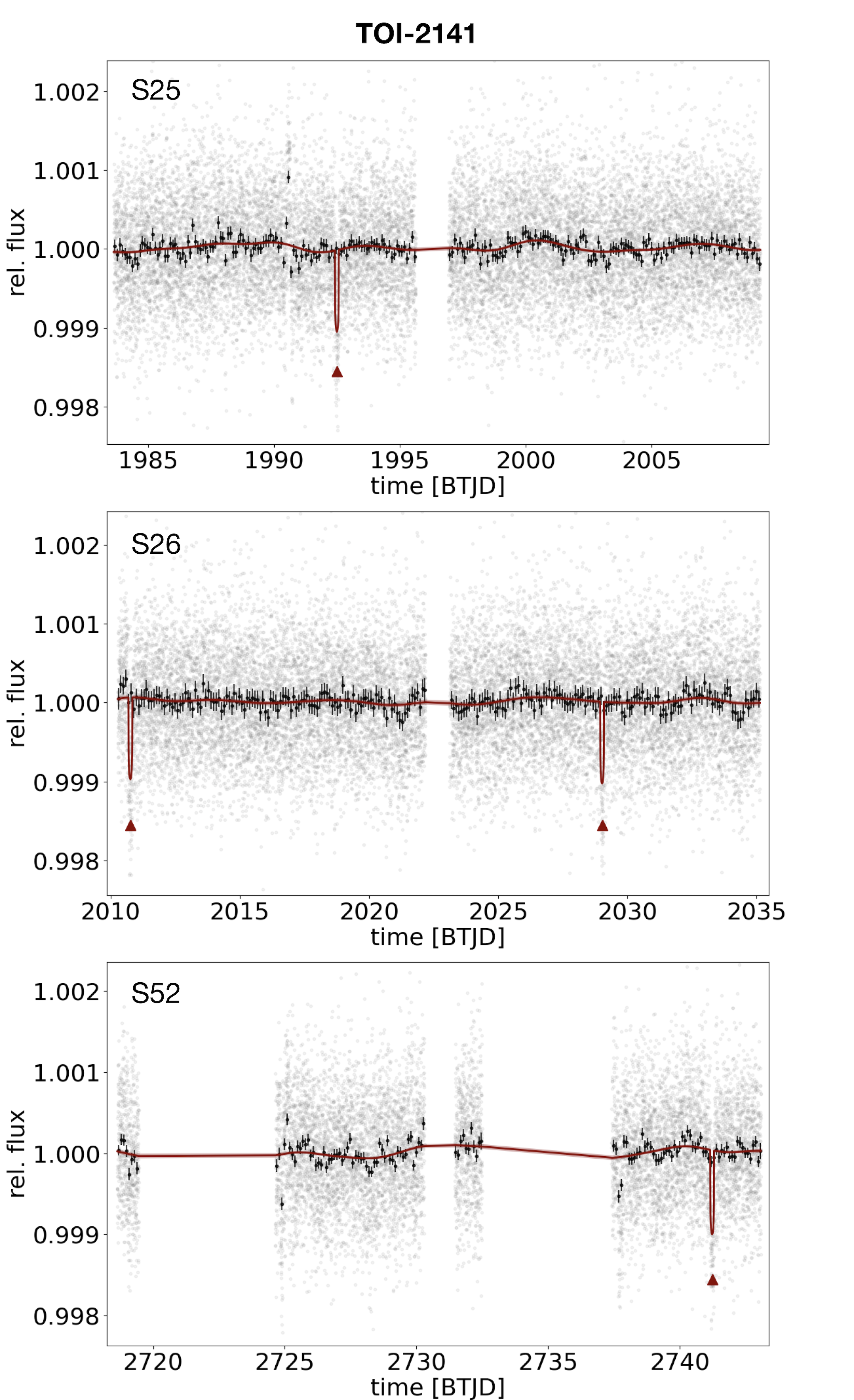}
      \caption{TESS light curve of TOI-2141. The blue points show the TESS photometry data around the transits of TOI-2141~b. The red line shows the baseline GP model that was fitted to the binned data multiplied by the best-fit transit model for TOI-2141~b. The red triangles show the predicted central time of TOI-2141~b transits.
      }
        \label{fig:toi2141_transits_fit}
  \end{figure}

The DVR for TOI-2141 reports two candidates, but only one passed validation tests. The unvalidated events have a periodicity of 30.6~d and are likely due to contamination by scattered light. We also note that the odd and even depth test supports a single strong feature folded on top of a much weaker and much less convincing transit-like feature for these events. The validated candidate planet, hereafter TOI-2141~b, has an estimated radius of R$_{\rm p}=3.2\pm0.3$~\RE\ and an orbital period of P$=18.26159(7)$~d. The centroid offsets localize the source of the transit signal within $1.7\pm6.3$~arcsec. The transit signature was first identified in the SPOC search of the Sector 26 light curve, and an alert for TOI~2141.01 was issued by TSO on 7 August 2020. Therefore, both TOI-1736 and TOI-2141 systems host at least one close-in transiting planet candidate with a size consistent with a sub-Neptune. Figures \ref{fig:toi1736_transits_fit} and \ref{fig:toi2141_transits_fit} depict the TESS photometry data for both respective objects.

%-----------------------------------------------------------------
\subsection{LCOGT 1\,m NEB Search}
\label{sec:LCOGT}

The \textit{TESS} pixel scale is $\sim 21\arcsec$ pixel$^{-1}$, and photometric apertures typically extend out to roughly 1 arcminute, which generally results in multiple stars blending in the TESS aperture. To attempt to determine the true source of the \textit{TESS} detection, we conducted ground-based photometric follow-up observations of the field around TOI-1736 as part of the {\tt TESS} Follow-up Observing Program\footnote{\url{https://tess.mit.edu/followup}} Sub Group 1 \citep[TFOP;][]{collins:2019}. 

We observed a full predicted transit window of TOI-1736~b in Pan-STARRS $z$-short band using the Las Cumbres Observatory Global Telescope \citep[LCOGT;][]{Brown:2013} 1.0\,m network node at McDonald Observatory on UTC 2020 August 28. If the event detected in the \textit{TESS} data is indeed on-target, the shallow SPOC reported depth of 380 ppm would not generally be detectable in ground-based observations. Instead, we slightly saturated TOI-1736 to enable the extraction of light curves of nearby fainter stars to attempt to rule out or identify nearby eclipsing binaries (NEBs) as potential sources of the \textit{TESS} detection. The 1\,m telescopes are equipped with $4096\times4096$ SINISTRO cameras having an image scale of $0\farcs389$ per pixel, resulting in a $26\arcmin\times26\arcmin$ field of view. The images were calibrated by the standard LCOGT {\tt BANZAI} pipeline \citep{McCully:2018}, and photometric data were extracted using {\tt AstroImageJ} \citep{Collins:2017}. 

To account for possible contamination from the wings of neighboring star point spread functions (PSFs), we searched for NEBs in all known Gaia DR3 and TICv8 nearby stars out to $2\farcm5$ from TOI-1736 that are possibly bright enough in the \textit{TESS} band to produce the \textit{TESS} detection (assuming a 100\% eclipse and 100\% contamination of the \textit{TESS} aperture). To attempt to account for possible delta-magnitude differences between the \textit{TESS} band and the follow-up filter band, we checked stars that are an extra 0.5 magnitudes fainter in \textit{TESS}-band than needed. We find that the RMS of each of the light curves of the 24 stars matching our criteria is more than a factor of 5 smaller than the expected NEB depth in the respective star. We then visually inspected each neighboring star's light curve to ensure no obvious eclipse-like signal. All of our follow-up light curves and supporting results are available on the {\tt EXOFOP-TESS} website\footnote{\url{https://exofop.ipac.caltech.edu/tess/target.php?id=408618999}}. Through our process of elimination, we find that the \textit{TESS} signal must be occurring in TOI-1736 relative to known Gaia DR3 and TICv8 stars.

\subsection{High contrast imaging}
\label{sec:highcontrastimaging}

\begin{figure}
   \centering
   \includegraphics[width=0.9\hsize]{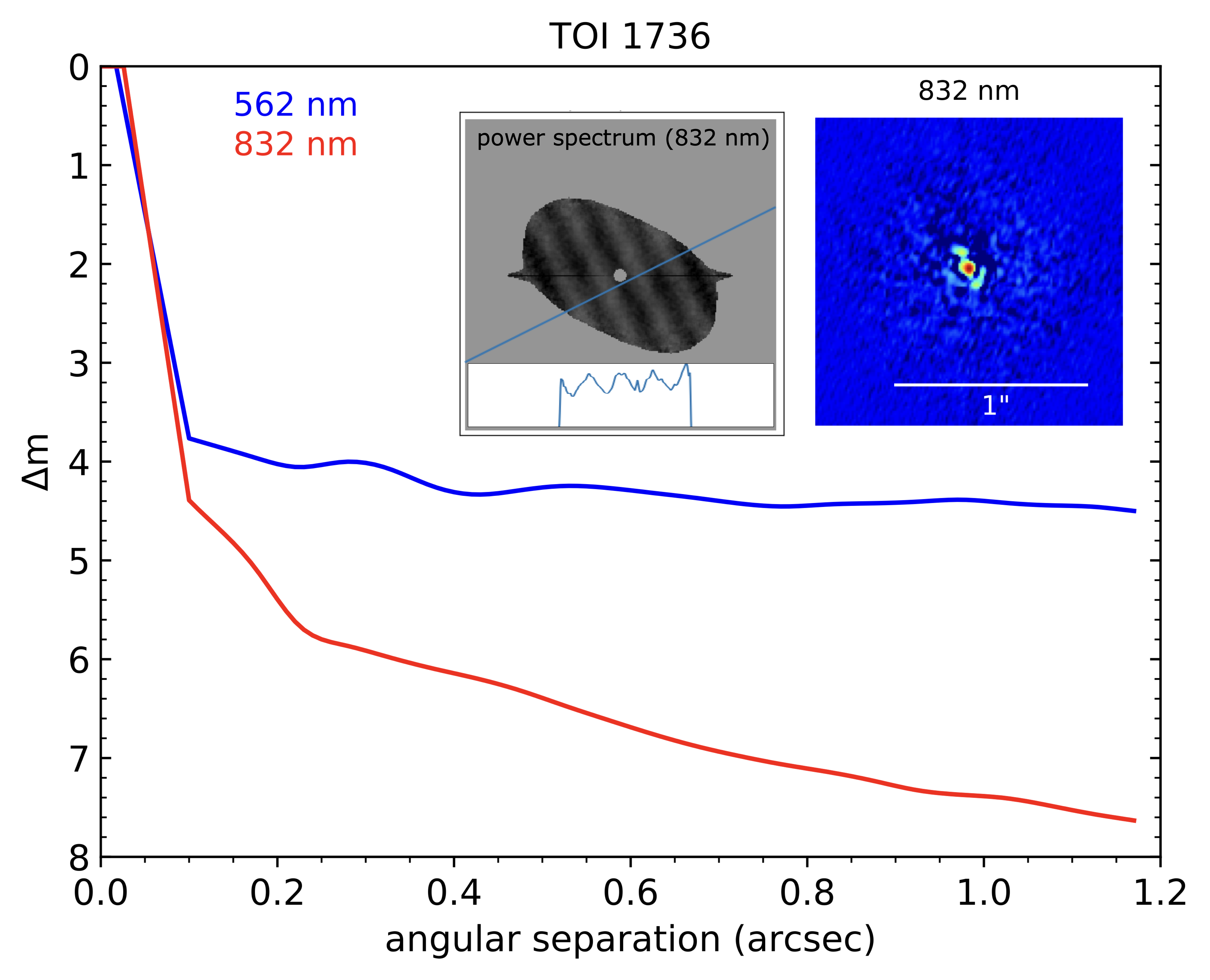}
      \caption{Contrast ratio of TOI-1736 as a function of angular separation at 562~nm (blue line) and at 832~nm (red line) obtained from the `Alopeke/Gemini speckle imaging observations. The small panels show the reconstructed speckle image (right subpanel) and its power spectrum in Fourier space (left subpanel). The power spectrum shows fringes that match a binary model.  
       }
        \label{fig:gemini_AO_constrast_TOI-1736}
  \end{figure}

High-angular-resolution observations can probe close companions within $\sim1.2$~arcsec that can create a false positive transit signal if that companion is an eclipsing binary, and which dilute the transit signal and thus yield underestimated planet radii \citep{Ciardi2015}. TOI-1736 was observed on October 19, 2021, by the `Alopeke dual-channel speckle imaging instrument on Gemini-N (PI: Howell) with a pixel scale of 0.01~arcsec/pixel and a full width at half maximum (FWHM) resolution of 0.02~arcsec. `Alopeke provided simultaneous speckle imaging at 562 and 832 nm.   The data were processed with the speckle pipeline \citep{Howell2011}, which yielded the 5-sigma sensitivity curves shown in Figure \ref{fig:gemini_AO_constrast_TOI-1736}.  These observations provide a contrast at an angular separation of 0.5~arcsec of 4.2~mag at 562~nm and 6.4~mag at 832~nm. Figure \ref{fig:gemini_AO_constrast_TOI-1736} also shows the reconstructed image at 832~nm and its Fourier transform, which  shows fringes in the power spectrum, evidence of a stellar companion to TOI-1736. The best-fit binary model for the power spectrum gives a companion at an angular separation of $0.093\pm0.002$~arcsec and position angle of $26.5\pm1.0$~deg with a magnitude difference from the primary of $\Delta{\rm mag}=2.43\pm0.15$ at 832~nm.

\begin{figure}
   \centering
   \includegraphics[width=0.9\hsize]{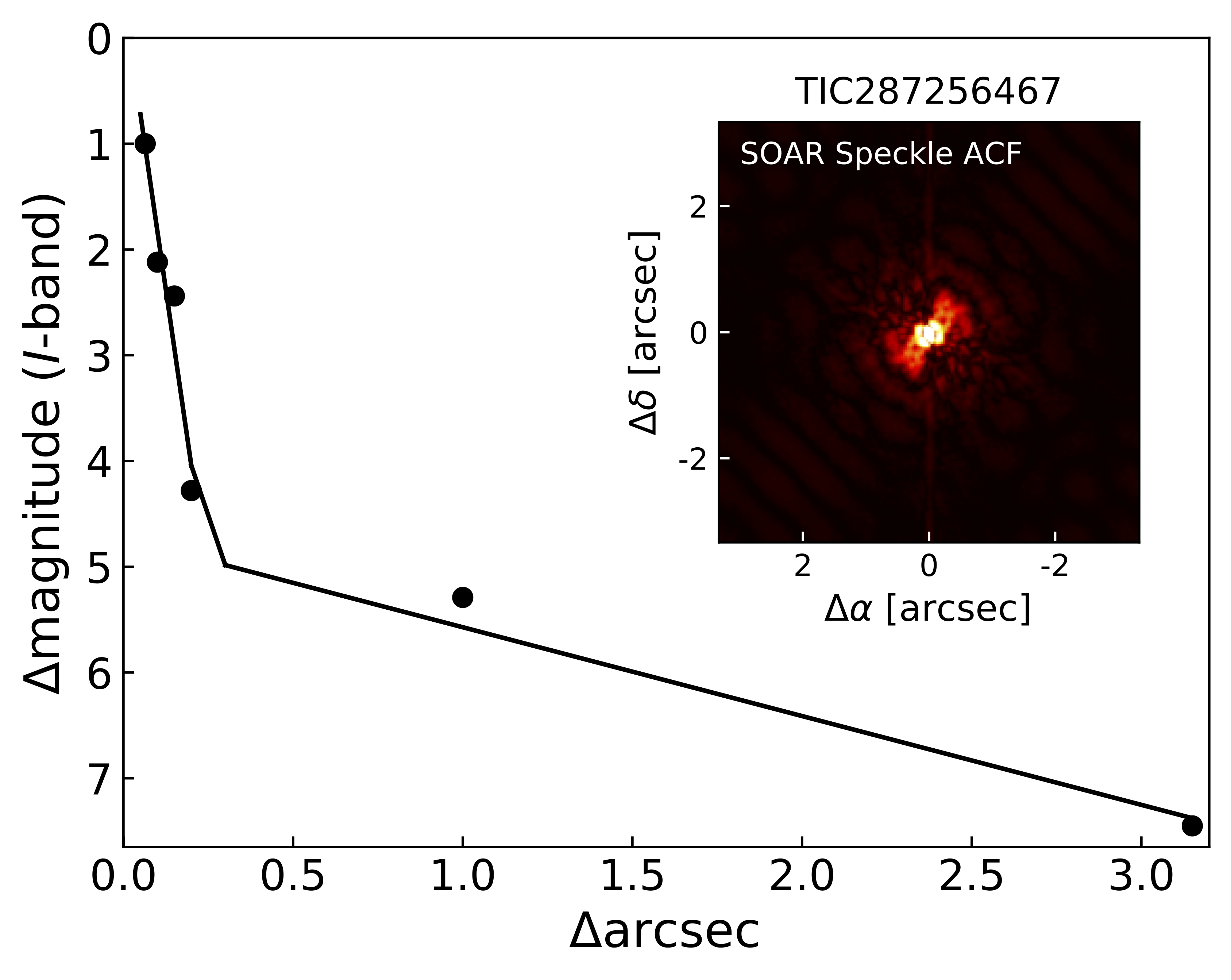}
      \caption{Contrast ratio of TOI-2141 as a function of angular separation at I-band obtained on the 4.1-m SOAR speckle imaging observations.
       }
        \label{fig:soar_AO_constrast_TOI-2141}
  \end{figure}

TOI-2141 was observed on April 25, 2021, with the 4.1-m SOAR telescope in speckle imaging.  These observations provide a contrast of 5~mag at an angular separation of 1.0~arcsec in the I-band (Figure \ref{fig:soar_AO_constrast_TOI-2141}).  These observations did not reveal any evidence of close stellar companions to TOI-2141.

%-----------------------------------------------------------------
\subsection{SOPHIE spectroscopy}
\label{sec:sophiespectroscopy}

\subsubsection{Observations}
\label{sec:sophieobservations}

SOPHIE is a high-resolution fiber-fed, cross-dispersed \'{e}chelle spectrograph mounted on the 1.93-m telescope at the Observatoire de Haute-Provence (OHP) \citep{Perruchot2008,Bouchy2013}. It covers a wavelength domain from 387.2 nm to 694.3 nm across 39 spectral orders. We observed TOI-1736 and TOI-2141 under a program dedicated to an RV follow-up of transiting candidates \citep[e.g.,][]{Konig2022,Moutou2021,Hebrard2020}, where we obtained 152 spectra of TOI-1736 between 2020-08-20 and 2023-03-11 with an average peak signal-to-noise ratio (S/N) per pixel at 550~nm of 70, and 90 spectra of TOI-2141 between 2021-02-25 and 2022-09-18 with an average peak S/N of 55. Both targets have been observed in the high spectral resolution mode (HR mode, R=75000) of SOPHIE. Tables \ref{tab:sophiervstoi1736} and \ref{tab:sophiervstoi2141} present more information about these observations.

\subsubsection{Data reduction by the DRS}
\label{sec:sophiedrs}

Our data have been reduced by the SOPHIE Data Reduction Software \citep[DRS,][]{Bouchy2009}, excluding the seven first (redder) spectral orders due to their low S/N. Five spectra of TOI-1736 were not used due to their low overall S/N (below 40). The DRS automatically extracts and calibrates the spectra and computes the RV using the cross-correlation function (CCF) between the spectra and a G2-type empirical weighted numerical mask. The DRS also uses the CCF data to deliver stellar activity indicators, such as the CCF FWHM and the bisector span (BIS).

In order to improve the accuracy of SOPHIE measurements, we used the optimized procedures presented by \cite{Heidari2022} and Heidari et al. (in prep). This includes in particular: (1) CCD charge transfer inefficiency correction \citep{Bouchy2013}; (2) correction for the moonlight contamination using the simultaneous sky spectrum obtained from the second SOPHIE fiber aperture \citep[e.g.,][]{Pollacco2008,Hebrard2008}; (3) RV constant master correction for instrumental long-term drifts \citep{Courcol2015}; and (4) correction of the instrumental short-term drifts thanks to the frequently measured drifts interpolated at the precise time of each observation.

\subsubsection{The CCF analysis by \texttt{sophie-toolkit}}
\label{sec:sophietoolkit}

In addition to the DRS reduction, we implemented an independent CCF analysis using the methodology described in \cite{Martioli2022}, where we developed the Python package \texttt{sophie-toolkit}\footnote{\url{https://github.com /edermartioli/sophie}} to obtain the RVs and other CCF quantities from a set of reduced SOPHIE spectra (both in the \texttt{e2ds} or \texttt{s1d} formats provided by the DRS). Our CCF approach is fundamentally the same as that applied by the DRS and therefore it should produce equivalent results. We compared the RVs from both CCF analyses, and despite some systematic effects in a small fraction of the data, they mostly agree within the error bars. As explained in Section \ref{sec:sophiedrs}, the DRS implements additional procedures to optimize the RV measurements in the SOPHIE data, which ultimately provide more accurate RVs. Thus, in our analysis, we adopted the DRS RVs. 

The analysis performed by \texttt{sophie-toolkit} was originally designed for near-infrared observations with the SPIRou spectrograph \citep{donati2020}, where it implements some processing steps that are aimed at mitigating the strong effects of telluric contamination and detector artifacts. Such analysis is more robust to systematics and improves some results when applied to the SOPHIE optical spectrum, where the aforementioned effects are much less significant, but still present in the data.  An important step that is applied in our tool is an iterative registration of the spectra, both in flux and wavelength domains, to match each observation to a high S/N template stacked spectrum (``template'' hereafter).  Registration is done by shifting the spectra to the same topocentric frame using the precise RV values obtained from the CCF analysis and by scaling the fluxes using an order-by-order least squares fit to a third-order polynomial \citep[see][]{Martioli2022}.  This reduces residual systematic errors between observations, providing a differential measurement of each spectrum with respect to the template. This procedure also provides the dispersion in flux for each spectral element along the time series, which allows us to estimate statistical errors and assign proper weights to the data before performing measurements such as CCFs, RVs, and spectral indices. In addition to the CCF and RV data, \texttt{sophie-toolkit} provides other important products, such as the template spectrum, obtained from the median stack of all spectra in the time series. This product is saved in FITS format to store both the template and all other registered spectra. The most relevant activity indicators are also calculated and saved in a time series product, as detailed in Section \ref{sec:activityindicators}. The Appendix \ref{app:sophiervs} presents the final RV data (from the DRS) together with the corresponding spectral quantities obtained in our analysis using the \texttt{sophie-toolkit} package, all compiled in Tables \ref{tab:sophiervstoi1736} and \ref{tab:sophiervstoi2141}.

%-----------------------------------------------------------------
\subsection{TRES spectroscopy}
\label{sec:tresspectroscopy}

For spectroscopic reconnaissance, we obtained four spectra of TOI-1736 and two spectra of TOI-2141 using the 1.5~m Tillinghast Reflector Echelle Spectrograph \citep[TRES;][]{Furesz2008} located at the Fred Lawrence Whipple Observatory (FLWO) in Arizona. The spectra for TOI-1736 and TOI-2141 were obtained between March 3 - October 4, 2020, and between August 17 - October 20, 2020. TRES is a fiber-fed echelle spectrograph with a wavelength range of 390-910 nm and a resolving power of R=44,000. Spectra were extracted and reduced as described in \cite{buchhave2010}.

%--------------------------------------------------------------------
\section{Stellar characterization} 
\label{sec:star}

We carried out an analysis of the spectral and photometric data of both systems to derive the host stars' properties as is detailed in the next sections. A summary of the final star parameters is presented in the Table \ref{tab:stellarparams}.

%-----------------------------------------------------------------
\subsection{Spectral energy distribution analysis}
\label{sec:sed}

We performed an analysis of the broadband spectral energy distribution (SED) of both stars including the {\it Gaia} Early Data Release 3 parallaxes \citep[with no systematic offset applied; see, e.g.,][]{StassunAndTorres2021} in order to determine an empirical measurement of the stellar radius, following the procedures described in \cite{Stassun2016}, \cite{Stassun2017}, and \cite{Stassun2018}.

We performed a fit using NextGen stellar atmosphere models \citep{Hauschildt1999}, with the effective temperature ($T_{\rm eff}$), metallicity ([Fe/H]), and extinction ($A_V$) as free parameters. The surface gravity ($\log g$) has little influence on the broadband SED. 

\begin{figure}
   \centering
   \includegraphics[width=0.9\hsize]{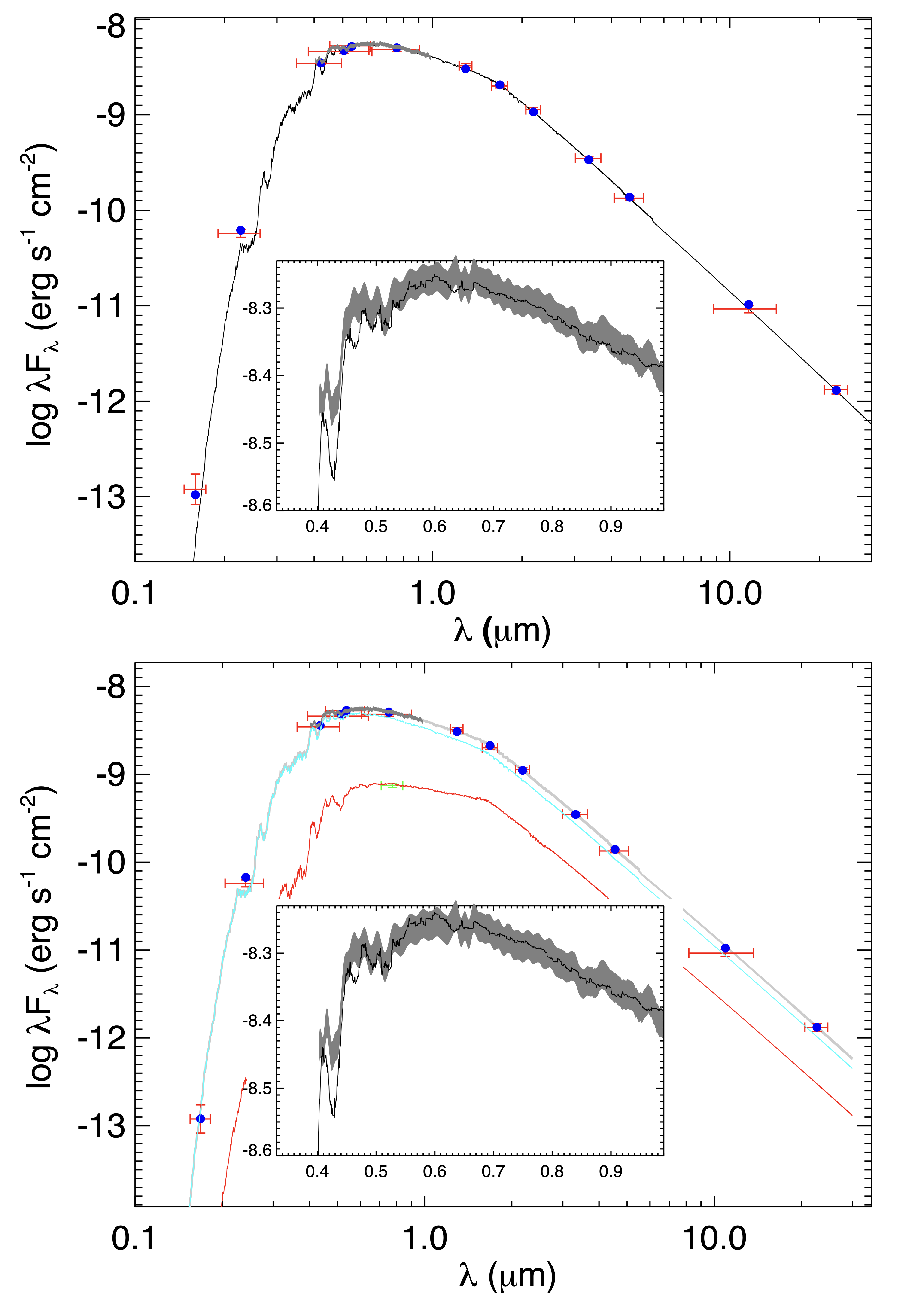}
      \caption{SED fit analysis for TOI-1736. The upper panel shows results assuming a single stellar component and the lower panel shows results assuming two stellar components. The red symbols represent the observed photometric measurements, whereas the horizontal bars represent the effective passband width. The blue symbols are the model fluxes from the best-fit NextGen atmosphere model (black and gray lines). The lower panel also shows the model for the primary component (cyan line) and the secondary component (red line), separately. The insets display our model (black curve) and the observed absolute flux-calibrated Gaia spectrum (gray swathe).}
        \label{fig:sedtoi1736}
  \end{figure}

\begin{figure}
   \centering
   \includegraphics[width=0.9\hsize]{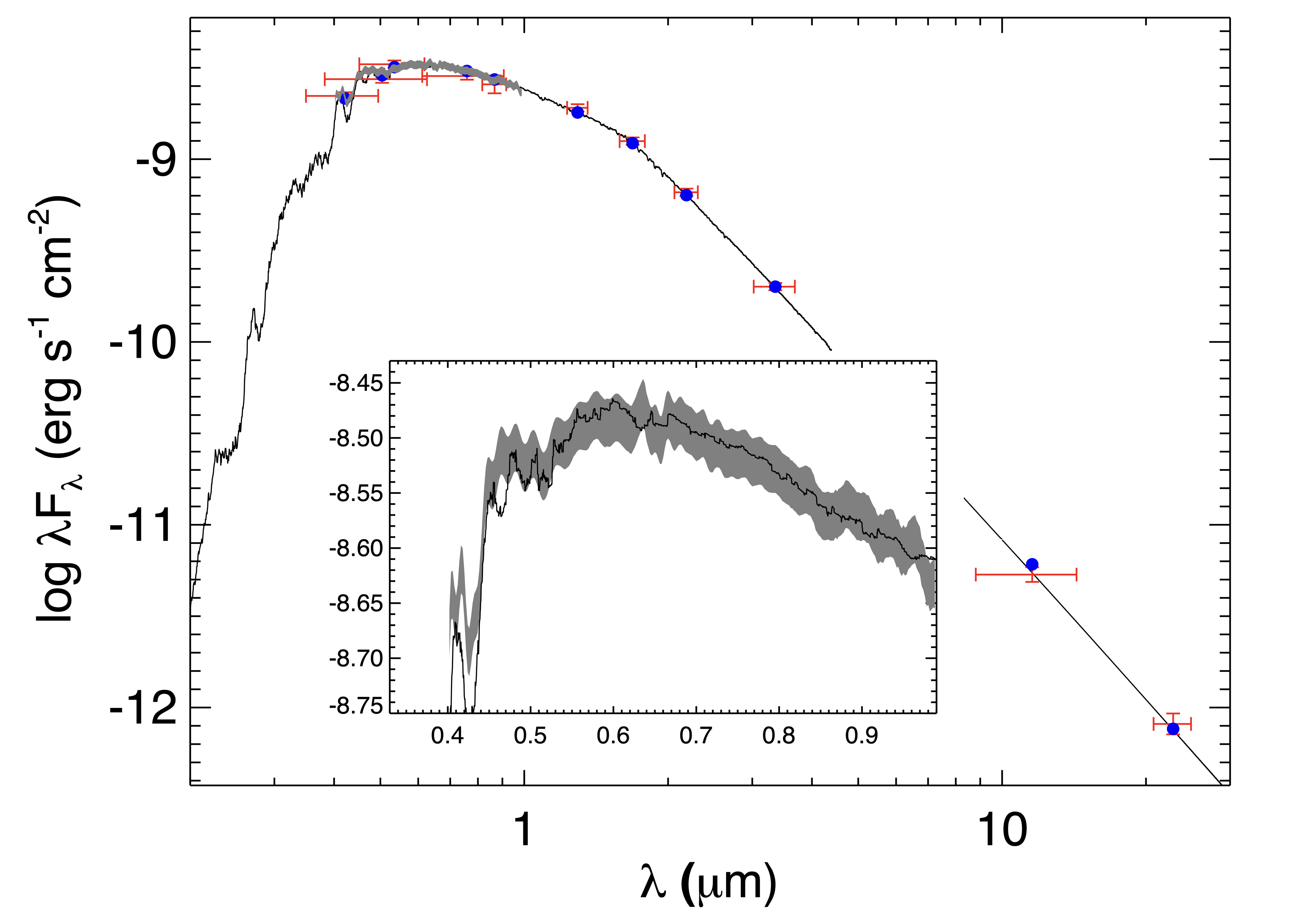}
      \caption{SED fit analysis for TOI-2141. Red symbols represent the observed photometric measurements, whereas the horizontal bars represent the effective width of the passband. Blue symbols are the model fluxes from the best-fit NextGen atmosphere model (black). The inset displays our model (black curve) alongside the observed absolute flux-calibrated Gaia spectrum (gray swathe).}
        \label{fig:sedtoi2141}
  \end{figure}

As a test to investigate whether the companion to TOI-1736 detected by the high contrast imaging results presented in Section \ref{sec:highcontrastimaging} is real, we first fit a single-component model to the SED, as shown in the top panel of Figure~\ref{fig:sedtoi1736}, where we obtained a reduced $\chi^2$ of 1.1, with the best-fit parameters $T_{\rm eff}=5800\pm75$~K, [Fe/H]=$0.2\pm0.1$, and $A_{V}=0.15\pm0.03$. According to the tool \texttt{Stilism}\footnote{\url{https://stilism.obspm.fr/}} \citep{Lallement2014}, the expected extinction for the distance of $\sim90$~pc, galactic latitude and longitude from TOI-1736 is $A_{V}=0.003$~mag, therefore an abnormal reddening (or an infrared excess flux) was detected in our model. Then we consider a two-component stellar model to fit the SED of TOI-1736, assuming that the secondary has the same extinction ($A_{V}=0.003$) and metallicity as the primary. The bottom panel of Figure \ref{fig:sedtoi1736} shows these results, which give a $\chi^2$ of 1.0 with the same fit parameters for the primary and $T_{\rm eff}=4800\pm200$~K for the secondary. Hence, the SED of TOI-1736 provides additional support to the binary hypothesis. For TOI-2141, we adopt a single-component model as illustrated in Figure~\ref{fig:sedtoi2141}, which gives a reduced $\chi^2$ of 1.3 and the best fit parameters $T_{\rm eff}=5750\pm75$~K, [Fe/H]=$-0.1\pm0.1$, and $A_{V}=0.06\pm0.04$.

Integrating the SED model gives the bolometric flux on Earth ($F_{\rm bol}$), which can be combined with the $T_{\rm eff}$ and the {\it Gaia\/} parallax to derive the stellar radius.  For TOI-1736, the single component model gives a radius of $R_\star=1.390\pm0.037$~R$_\odot$ and the two-component model gives a $3\sigma$ smaller radius of $R_\star=1.241\pm0.032$~R$_\odot$ for the primary and $R_\star=0.748\pm0.073$~R$_\odot$ for the secondary. For TOI-2141, we obtained a radius of $R_\star=0.938\pm0.027$~R$_\odot$. Using the mass-radius relations of \cite{Torres2010}, we calculate the stellar masses of $1.15\pm0.07$~M$_\odot$ for TOI-1736 and $0.98\pm0.06$~M$_\odot$ for TOI-2141.

%-----------------------------------------------------------------
\subsection{Standard analysis of SOPHIE spectra}
\label{sec:standardspectralanalysis}

To obtain the stellar parameters from the SOPHIE data, we constructed an average spectrum for both stars, using all SOPHIE spectra that were unaffected by the Moon's pollution, and performed a standard spectral analysis on them. The average spectra of TOI-1736 and TOI-2141 have an S/N at 649~nm of 1100 and 850, respectively. We use the methods presented in \cite{Santos2004} and \cite{Sousa2008} to derive the effective temperatures $T_{\rm eff}$, the surface gravities $\log g$, and the metallicities ${\rm [Fe/ H]}$. Using these spectroscopic parameters as input, the stellar masses were derived from the \cite{Torres2010} calibration with a correction following \cite{Santos2013}.  The errors were calculated from 10,000 random realizations of the stellar parameters within their error bars and assuming Gaussian distributions. The parameters obtained from this analysis are presented in Table \ref{tab:stellarparamsFromDifferentMethods} labeled as ``standard.''

%-----------------------------------------------------------------
\subsection{Spectroscopic parameters from TRES spectra}
\label{sec:tres_spectra}

We use the reduced TRES spectra to obtain the stellar parameters using the Stellar Parameter Classification tool \citep[SPC;][]{buchhave2012}. In short, SPC correlates each observed spectrum against a grid of synthetic spectra based on Kurucz \citep{kurucz1992} atmospheric models and derives the effective temperature, surface gravity, metallicity, and rotational velocity of the star. The parameters obtained from this analysis are presented in Table \ref{tab:stellarparamsFromDifferentMethods}.

%-----------------------------------------------------------------
\subsection{Strictly differential analysis of SOPHIE spectra}
\label{sec:solardifferentialanalysis}

We take advantage of the spectroscopic similarity between our targets and the Sun to apply a strictly differential analysis to determine atmospheric parameters and elemental abundances \citep[e.g.,][]{Ramirez2014,Bedell2014}. The differential analysis in such cases can potentially mitigate the Fe~I/II excitation-ionization balance internal errors caused by possible differences in the micro-physics prescriptions, in the local determination of the continuum, or in the equivalent width (EW) measurements. Our reference solar spectrum was obtained from observations of the Moon with SOPHIE on 2023-04-01 with an S/N of 245 at 649~nm, maintaining the same instrumental configuration as the one used to observe the scientific targets.   

We perform a line-by-line analysis with respect to the solar spectrum, which is based on the following steps. Given a narrow spectral range around each selected line, we normalize the stellar and solar spectrum using the same local continuum regions and then fit a Gaussian function to the solar line profiles, computing all resulting parameters (e.g., line depth, centroid, FWHM, and continuum offset). Then, we repeat the line fitting procedure for the locally normalized stellar spectrum, now fixing the continuum offset and line region information obtained from the respective solar spectrum.  We employ the ionization-excitation equilibrium method where we adopted a line list from \cite{Melendez2014}, which contains excitation potentials, oscillator strengths, laboratory $\log{gf}$ values, and hyperfine structure corrections\footnote{albeit it has a small impact on stellar abundances of solar analogs, in the Y~II lines we corrected for this effect.}.  We included iron-peak, s- and alpha-capture elements, such as Fe, Mg, Al, Si, Ti, Y, and Ca to perform this spectroscopic analysis. 

\begin{figure}
   \centering
   \includegraphics[width=1.0\hsize]{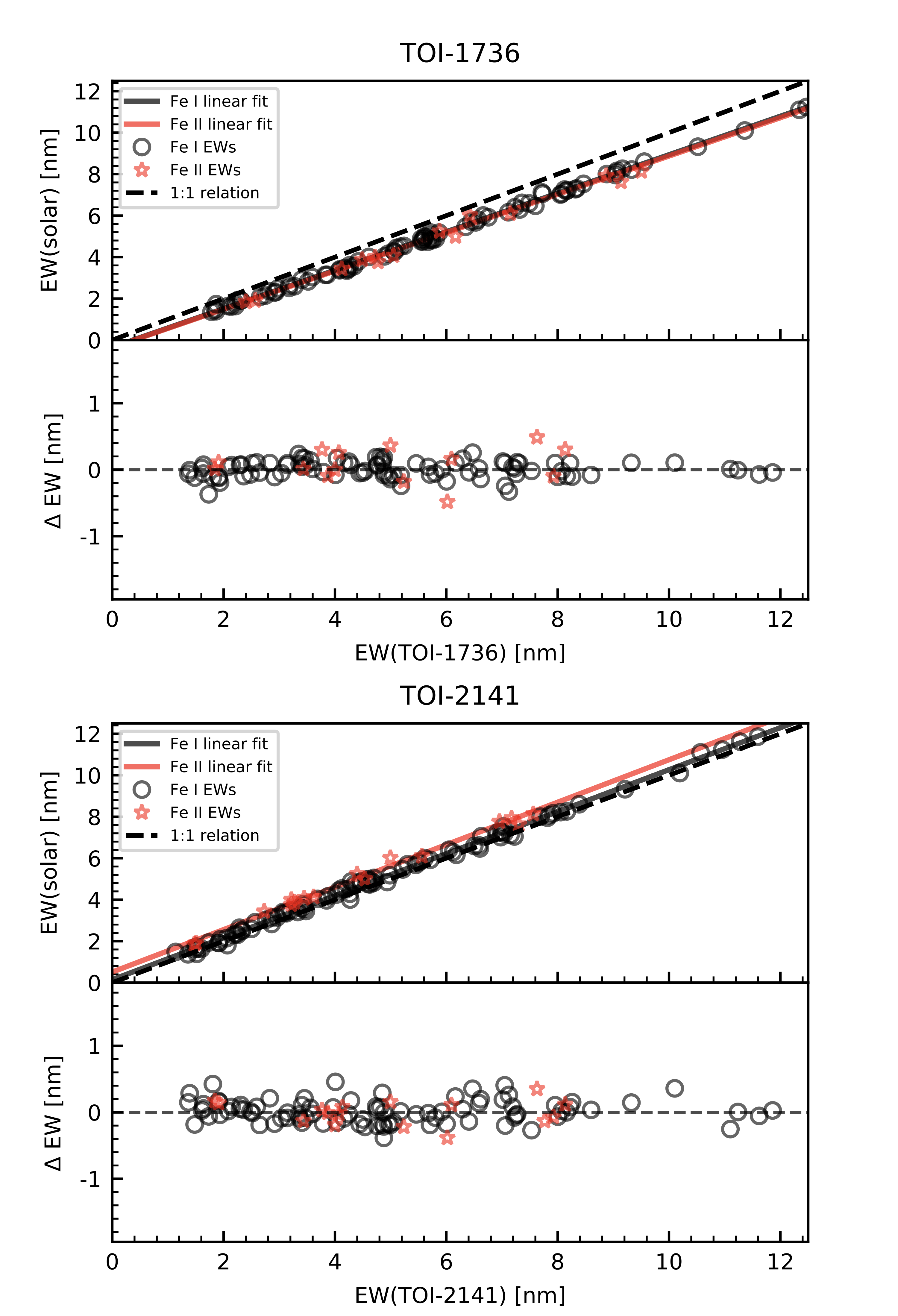}
      \caption{Comparison between solar and stellar EW measurements. TOI-1736 and TOI-2141 are shown in the top and bottom panels, respectively. The dashed black line is the 1:1 EWs ratio, and the solid black line is a linear fit based on the Fe I lines, represented by empty black circles. The red star symbols represent the Fe II lines, followed by their linear fit represented by the solid red line. The typical standard deviation of the average linear fit is around $\sim0.1-0.2$~nm, which means $\sim2$\% in relative terms.
      }
     \label{fig:solardiffew}
  \end{figure}

In Figure \ref{fig:solardiffew} we show the relationship between the solar Fe~I and Fe~II EWs and those of TOI-1736 and TOI-2141. The low scatter ($\sigma{\rm EW}\sim1-2\%$) for both stars indicates that the differential spectroscopic equilibrium analysis with respect to the Sun is sufficiently precise for the determination of the stellar atmospheric parameters. While not an exact match to our Sun, exercising caution and maintaining a conservative approach in considering uncertainties is warranted.

We use the code \texttt{q2} \citep{Ramirez2014} to model both the stellar and solar EWs of Fe~I and Fe~II from our line-by-line analysis. The input-relevant line data (e.g., oscillator strength, excitation potential, etc.) are also provided. The \texttt{q2} was configured to employ the Kurucz ODFNEW model atmospheres \citep{CastelliAndKurucz2003} and the 2019 version of the local thermodynamic equilibrium (LTE) code \texttt{MOOG} \citep{Sneden1973,Sneden2012}. The stellar parameters are obtained through an iterative convergence process, where \texttt{q2} iterates until it finds: 1) no systematic dependence of the derived Fe~I and Fe~II abundances as a function of excitation potential and reduced EW, constraining the $T_{\rm eff}$ and micro-turbulence; 2) Fe~I and Fe~II yielding the same abundance value, constraining the $\log{g}$. Once convergence is achieved, the final atmospheric parameters are estimated together with their uncertainties, which include systematic errors, as detailed in Appendix \ref{sec:validationofdifferentialanalysis}. For TOI-1736 we obtained $T_{\rm eff}=5807\pm13$~K, $\log{g}=4.35\pm0.04$~dex, $[{\rm Fe}/{\rm H}]=+0.138\pm0.014$~dex, and $v_{\rm t}=1.26\pm0.02$~\kms. For TOI-2141 we obtained $T_{\rm eff}=5659\pm12$~K, $\log{g}=4.42\pm0.04$~dex, $[{\rm Fe}/{\rm H}]=-0.120\pm0.013$~dex, and $v_{\rm t}=0.94\pm0.02$~\kms.

Stellar elemental abundances are obtained by \texttt{q2} using local thermodynamic equilibrium (LTE) model interpolations based on the previously derived atmospheric parameters. The adopted errors for each element account for the errors from atmospheric parameter measurements and the scatter from the line-to-line abundance estimates.  Table \ref{tab:abundancesfromdiffanalysis} summarizes the abundances obtained in our analysis.

\begin{table}
\centering \footnotesize
\caption{Elemental abundances from our differential analysis. }
\label{tab:abundancesfromdiffanalysis}
\begin{tabular}{cccc}
\hline
Species & TOI-1736 & TOI-2141 & Sun (Moon)\tablefootmark{a} \\
\hline
% \multicolumn{4}{l}{\it Species abundances with respect to hydrogen, [X/H]} \\
FeI & $0.135\pm0.010$ & $-0.116\pm0.0010$ & $+0.011\pm0.015$ \\ 
FeII &  $0.137\pm0.018$  & $-0.121\pm0.012$ & $+0.010\pm0.014$ \\
YII & $0.158\pm0.034$  & $-0.165\pm0.013$  & $+0.013\pm0.024$\\
AlI & $0.227\pm0.013$ & $-0.083\pm0.020$  & $+0.010\pm0.010$ \\
MgI & $0.147\pm0.023$  & $-0.051\pm0.019$  & $+0.014\pm0.013$ \\
TiI & $0.195\pm0.013$  & $-0.074\pm0.011$ & $+0.001\pm0.015$ \\
TiII & $0.177\pm0.014$  & $-0.094\pm0.013$ & $+0.002\pm0.017$ \\
SiI & $0.143\pm0.011$  & $-0.092\pm0.012$  & $+0.010\pm0.010 $ \\
CaI & $0.168\pm0.012$ & $-0.100\pm0.011$ & $ +0.008\pm0.016$ \\
\hline
\end{tabular}
\tablefoot{
\tablefoottext{a}{For comparison, we show the same abundances obtained for the solar spectrum from the observations of the Moon using the same methodology. }
}
\end{table} 

We performed an evolutionary analysis using the derived spectroscopic parameters (T$_{\rm eff}$, $\log{g}$ and abundances), where we adopted the Yonsei-Yale evolutionary tracks \citep{Yi2001,Demarque2004} that account for the contribution of the alpha enhancement elements to compute the ages, masses, and radii of TOI-1736 and TOI-2141.  The derived atmospheric parameters are compared with those predicted by stellar models using the Bayesian framework described in \cite{Grieves2018} and \cite{YanaGalarza2021}.  Table \ref{tab:stellarparamsFromDifferentMethods} shows the derived values of stellar mass and radius and the derived ages are shown in Table \ref{tab:evolutionaryparameters}.

%-----------------------------------------------------------------
\subsection{Comparison between methods for determining stellar parameters}
\label{sec:comparisonanalysis}

As we have employed different methods and used data from different instruments to obtain the stellar parameters, in this section we make a comparison between these results. At the time of writing, \cite{MacDougall2023} has also published independent results for the parameters of TOI-1736 ($T_{\rm eff}$ and $[{\rm Fe}/{\rm H}]$), which we also include here for comparison.

For TOI-1736, the mean values of $T_{\rm eff}$ range from 5636 to 5807~K, where the SED, standard, and solar diff. agree within $1\,\sigma$, whereas TRES and \cite{MacDougall2023} agree within $3\,\sigma$ with respect to the other values.  All surface gravity values also agree within $3\,\sigma$, although the solar diff. gives a higher value than other methods. The GAIA eDR3 trigonometric value is slightly more precise than the solar diff, but it can be contaminated by the close companion in the same way as the SED value. All metallicity values agree within $1\,\sigma$.  For TOI-2141, all but the SED value of $T_{\rm eff}$ agree within $1\,\sigma$, and all agree within $2\,\sigma$. All surface gravity and metallicity values also agree within $1\,\sigma$. The parameters of this star are more similiar to our Sun, being poorer in metals and more active than normal for its age.

Different instruments/methods may have different offsets for derived parameters. For this reason, we performed an absolute calibration of our differential analysis of the SOPHIE data using other stars and the Sun as a reference (see Appendix \ref{sec:validationofdifferentialanalysis}). Since the determination is more precise and since the other methods generally agree with it, we adopt the parameters of the solar differential analysis as final values.  Star masses and radii are derived from spectroscopic parameters, so they carry more or less the same level of discordance as those discussed above. For consistency, we adopted the final mass and radius values as those obtained from the solar differential analysis. As pointed out by \citep{Tayar2022}, we recognize that even the differential analysis has systematic errors in the models of stellar evolution that were not taken into account in our analysis. Therefore, we use \cite{Tayar2022}'s methods \footnote{\url{https://github.com/zclaytor/kiauhoku/blob/v1.4.0/notebooks/model_offsets.ipynb}} to calculate the additional errors in stellar parameters due to systematic differences in the models, as shown in Table \ref{tab:stellarparamsFromDifferentMethods}. We add these errors in quadrature to the uncertainties obtained by our differential analysis, giving a more realistic uncertainty in the final values.

Turbulence velocity and rotational velocity are two quantities that depend on the adopted model and can absorb some instrumental broadening. Thus, to avoid any bias toward a specific instrument/model, we adopted their mean values.

\begin{table*}
\centering
\caption{Stellar parameters of TOI-1736 and TOI-2141 obtained from different methods.}
\label{tab:stellarparamsFromDifferentMethods}
\begin{tabular}{lccc}
\hline
Parameter & TOI-1736 & TOI-2141 & Source \\
\hline
%-------------------------------------------------------
effective temperature, $T_{\rm eff}$ (K) & $5800\pm75$ &  $5750\pm75$ & SED, Sect. \ref{sec:sed} \\
effective temperature, $T_{\rm eff}$ (K) & $5750\pm70$ &  $5610\pm60$ & standard, Sect. \ref{sec:standardspectralanalysis} \\
effective temperature, $T_{\rm eff}$ (K) & $5672\pm50$ &  $5671\pm50$ & TRES spectra, Sect. \ref{sec:tres_spectra} \\
effective temperature, $T_{\rm eff}$ (K) & $5807\pm13$ &  $5659\pm12$ & solar diff., Sect. \ref{sec:solardifferentialanalysis} \\
effective temperature, $T_{\rm eff}$ (K) & $5636^{+79}_{-86}$ &  & \cite{MacDougall2023} \\
effective temperature model error, $\sigma_{T}$ (K) & $44$ & $46$ &  \cite{Tayar2022} \\
%-------------------------------------------------------
\hline
surface gravity,$\log g_{\rm dr3}$ (dex) & $4.21\pm0.03$ &  $4.42\pm0.03$ & trigonometric, GAIA eDR3 \\
surface gravity,$\log g$ (dex) & $4.10\pm0.12$ &  $4.40\pm0.10$ & standard, Sect. \ref{sec:standardspectralanalysis} \\
surface gravity,$\log g$ (dex) & $4.26\pm0.10$ &  $4.56\pm0.10$ & TRES spectra, Sect. \ref{sec:tres_spectra} \\
surface gravity,$\log g$ (dex) & $4.35\pm0.04$ &  $4.42\pm0.04$ & solar diff., Sect. \ref{sec:solardifferentialanalysis} \\
surface gravity model error, $\sigma_{\log g}$ (dex) & $0.020$ &  $0.019$ & \cite{Tayar2022} \\
\hline
%-------------------------------------------------------
Fe metallicity, $[{\rm Fe}/{\rm H}]$ (dex) & $0.20\pm0.10$ &  $-0.10\pm0.10$ & SED, Sect. \ref{sec:sed}\\
Fe metallicity, $[{\rm Fe}/{\rm H}]$ (dex) & $0.09\pm0.05$ &  $-0.17\pm0.05$ & standard, Sect. \ref{sec:standardspectralanalysis}  \\
metallicity, $[{\rm m}/{\rm H}]$ (dex) & $0.14\pm0.08$ & $-0.09\pm0.08$ & TRES spectra, Sect. \ref{sec:tres_spectra} \\
Fe metallicity, $[{\rm Fe}/{\rm H}]$ (dex) & $0.138\pm0.014$ &  $-0.120\pm0.013$ & solar diff., Sect. \ref{sec:solardifferentialanalysis} \\
Fe metallicity, $[{\rm Fe}/{\rm H}]$ (dex) & $0.15 \pm0.06$ &  &  \cite{MacDougall2023} \\
Fe metallicity model error, $\sigma_{\rm [Fe/H]}$ (dex) & $0.0004$ &  $0.0005$ & \cite{Tayar2022} \\
\hline
%------------------------------------------------------- 
star mass, $M_{\star}$ (\msol) & $1.15\pm0.07$ &  $0.98\pm0.06$ & SED, Sect. \ref{sec:sed} \\
star mass, $M_{\star}$ (\msol) & $1.07\pm0.02$ &  $0.89\pm0.01$ & standard, Sect. \ref{sec:standardspectralanalysis} \\
star mass, $M_{\star}$ (\msol) & $1.08\pm0.03$ &  $0.938\pm0.013$ & solar diff., Sect. \ref{sec:solardifferentialanalysis}\\
star mass model error, $\sigma_{M}$ (\msol) & $0.020$ &  $0.021$ & \cite{Tayar2022} \\
\hline
%-------------------------------------------------------
star radius, $R_{\star}$ (\RS) & $1.24\pm0.03$ &  $0.94\pm0.03$ & SED, Sect. \ref{sec:sed} \\
star radius, $R_{\star}$ (\RS) & $1.43\pm0.05$ &  $0.91\pm0.03$ & standard, Sect. \ref{sec:standardspectralanalysis} \\
star radius, $R_{\star}$ (\RS) & $1.15\pm0.06$ &  $0.98\pm0.04$ & solar diff., Sect. \ref{sec:solardifferentialanalysis} \\
star radius model error, $\sigma_{R}$ (\RS) & $0.048$ &  $0.041$ & \cite{Tayar2022} \\
\hline
%-------------------------------------------------------
turbulence velocity, $\nu$ (km/s) & $0.94\pm0.03$ &  $0.71\pm0.03$ & standard, Sect. \ref{sec:standardspectralanalysis}  \\
turbulence velocity, $\nu$ (km/s) & $1.26\pm0.02$ &  $0.94\pm0.02$ & solar diff., Sect. \ref{sec:solardifferentialanalysis}\\
\hline
%------------------------------------------------------- 
rotation velocity, $v_{\rm rot}\sin{i_{\star}}$ (\kms) & $4.5\pm0.5$ &  $2.7\pm0.5$ & TRES spectra, Sect. \ref{sec:tres_spectra}\\
rotation velocity, $v_{\rm rot}\sin{i_{\star}}$ (\kms) & $3.6\pm1.0$ &  $2.7\pm1.0$ & standard, Sect. \ref{sec:standardspectralanalysis}\\
%------------------------------------------------------- 
\hline

\end{tabular}
\end{table*}

%-----------------------------------------------------------------
\subsection{Stellar activity and rotation}
\label{sec:activityindicators}

The CCF analysis performed on the SOPHIE spectra provides three quantities that are sensitive to stellar activity: the FWHM, bisector span (BIS), and RVs \citep[e.g.,][]{Queloz2001,Boisse2009}, as shown in Appendix \ref{app:sophiervs}. In addition, we measured two well-known spectral indices that are proxies for chromospheric and coronal activity, the S-index, which relies on the emissions in the cores of the H and K Ca~II lines, and H$\alpha$ \citep[e.g.,][]{Cincunegui2007}. In Appendix \ref{app:activityindices} we present the details of our measurements of the S-index and H$\alpha$ from the SOPHIE spectra.  In particular, we present a calibration of the S-index to the Mount-Wilson system. The temporal variability of these activity indicators can be used to identify systematic errors in the RVs caused by spurious signals of stellar activity, thus allowing to improve the determination of planetary orbits, as will be explored in Section \ref{sec:planetcharacterization}.

We convert the Mount-Wilson S-index to the chromospheric activity index $\log{\rm R'}_{\rm HK}$ using the \cite{Czesla2019} recipe that includes the photospheric correction \citep[e.g.,][]{Mittag2013}, where we use the $B-V$ values from the magnitudes listed in Table \ref{tab:stellarparams} and T$_{\rm eff}$ from our analysis. The result was $\log{\rm R'}_{\rm HK}(B-V)=-4.94^{+0.15}_{-0.25}$ for TOI-1736 and $\log{\rm R'}_{\rm HK}(B-V)=-4.78^{+0.13}_{-0.17}$ for TOI-2141. Using the activity-rotation empirical calibration of \cite{Mamajek2008} we obtained $P_{\rm rot}=28\pm5$~d for TOI-1736 and $P_{\rm rot}=21\pm5$~d for TOI-2141. 

Also using the calibration of \cite{Mamajek2008} we obtained gyrochronological ages of $4.6^{+1.7}_{-1.3}$~Gyr for TOI-1736 and $2.7^{+1.6}_{-1.0}$~Gyr for TOI-2141.  Applying the age-chromospheric activity relation for solar analogs of \cite{LorenzoOliveira2016} we estimated the ages in $5.4^{+1.4}_{-1.1}$~Gyr for TOI-1736 and $4.2^{+1.6}_{-1.1}$~Gyr for TOI-2141, which are consistent with those from gyrochronology. We adopted the activity ages as the latter, since \cite{LorenzoOliveira2018} take mass and metallicity into account in their empirical relationship, proving to be more accurate for the range of parameters of our targets.

%-----------------------------------------------------------------
\subsection{Stellar ages} 
\label{sec:age}

Table \ref{tab:evolutionaryparameters} shows the derived age values of the evolutionary analysis of stellar parameters. As an additional check, we calculate the stellar ages from chemical clocks \citep[e.g.,][]{daSilva2012,TucciMaia2015,DelgadoMena2019}. We calculated the yttrium abundance ratios ([Y/X], where X=Si~I, Ti~I, Ti~II, or  Al~I) from the values reported in Table \ref{tab:abundancesfromdiffanalysis}, where we estimated the stellar ages using the \cite{DelgadoMena2019} relations that account for T$_{\rm eff}$ and [Fe/H] effects on chemical ages. The average calibration errors are about 1.4~Gyr and the impact of T$_{\rm eff}$-[Fe/H]-[YII/X] errors are respectively 0.5 and 0.3~Gyr for TOI-1736 and TOI-2141. The average chemical ages are shown in Table \ref{tab:evolutionaryparameters}.

\begin{figure}
   \centering
   \includegraphics[width=1.0\hsize]{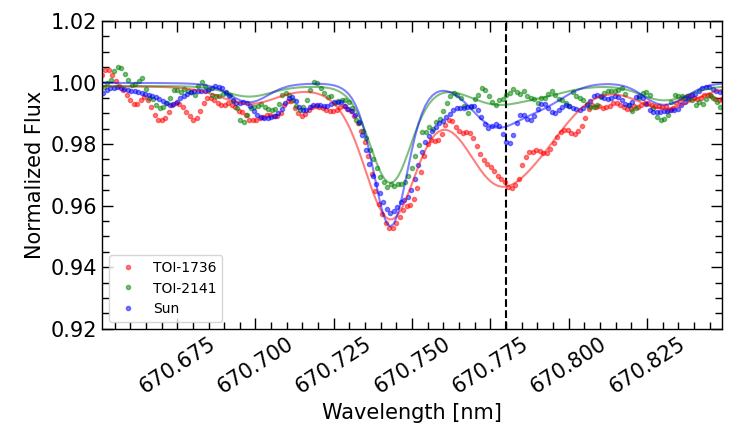}
      \caption{Spectral synthesis analysis around the Li region to obtain an estimate of the Li abundance.  The points show the SOPHIE spectra of TOI-1736 (in red), TOI-2141 (in green), and a solar spectrum (in blue) from observations of the Moon. The solid lines show the respective best-fit synthetic spectra obtained from our analysis.
      }
        \label{fig:lithiumabundances}
  \end{figure}

Lithium (Li) is also known to be a sensitive indicator of stellar evolution \cite[e.g.,][]{Carlos2016,Lyubimkov2016}. To obtain Li abundance, we performed a spectral synthesis analysis using the code \texttt{iSpec}\footnote{\url{https://www .blancocuaresma.com/s/iSpec}} \citep{BlancoCuaresma2014, BlancoCuaresma2019} and the radiative transfer code \texttt{MOOG}\citep{Sneden1973,Sneden2012}, where we use the lines in the spectral range around Li from \cite{Melendez2012}. Figure \ref{fig:lithiumabundances} illustrates our results, where we show the SOPHIE spectra and the best-fit synthetic models for TOI-1736 and TOI-2141, and also for a solar spectrum obtained from observations of the Moon.   We calculated a Li abundance for the solar spectrum of ${\rm A(Li)}=1.1\pm0.2$, which is consistent with the expected value for the Sun. For TOI-2141, our results are compatible with a null value, giving an upper limit of A(Li)$<0.5$. For TOI-1736, we obtained an abundance of ${\rm A(Li)}=1.56\pm0.16$. Using the relationship between A(Li) and age from \cite{Carlos2019} we obtained an age of $4.4\pm1.0$~Gyr for TOI-1736. 

All the different age determinations for both systems are presented in Table \ref{tab:evolutionaryparameters}. There is general agreement between the ages of TOI-1736, which are in the range of 4-6~Gyr. For TOI-2141, both the chemical and evolutionary ages agree around 7.5~Gyr, which is consistent with the absence of Li, but the activity-derived age appears younger. 
For a final determination of the ages of both stars, we adopted the average values, where we added \cite{Tayar2022}'s model uncertainties in quadrature to the measured errors, as in Section \ref{sec:comparisonanalysis}, giving an age of $4.9\pm1.3$~Gyr for TOI-1736 and $6.4\pm1.8$~Gyr for TOI-2141.

\begin{table}
\centering
\caption{Stellar ages derived from different methods in our analysis.}
\label{tab:evolutionaryparameters}
\begin{tabular}{lcc}
\hline
Parameter & TOI-1736 & TOI-2141\\
\hline
<Age> [Chem.; Gyr]  & $4.7\pm0.5$ & $7.5\pm0.3$ \\
%<Calibration errors> [Gyr]   & +-1.4 &  +-1.4 \\
Age [Iso; Gyr] & $5.1_{-0.9}^{+0.7}$ & $7.5\pm1.6$ \\
%<rms(Age)/Age>   &  +-0.4 Gyr  & +-0.3 Gyr  \\
Age [Li; Gyr] & $4.4\pm1.0$ &  \\
Age [${\rm R'}_{\rm HK}$; Gyr] & $5.4^{+1.4}_{-1.1}$ &  $4.2^{+1.6}_{-1.1}$ \\
Age model error [Gyr] \tablefootmark{a}& $1.3$ &  $1.6$ \\
\hline
\end{tabular}
\tablefoot{
\tablefoottext{a}{additional error due to systematic differences in stellar evolution models from \cite{Tayar2022}'s method.}
}
\end{table} 

%-----------------------------------------------------------------
\section{Detection and characterization of the planetary systems}
\label{sec:planetcharacterization}

\subsection{Analysis of TESS photometry data}
\label{sec:analysisoftessphotometrydata}

To further characterize the transits detected by TESS, we analyzed the PDCSAP flux data by applying the methods described in \cite{Martioli2021}, where we fit a transit model plus a baseline polynomial in selected windows around each transit in such a way that the window size is twice the duration of the transit and with a minimum of 300 data points. We selected windows for a total of 19 transits of TOI-1736~b and four transits of TOI-2141~b. Our transit model is calculated using the \texttt{BATMAN} toolkit \citep{Kreidberg2015} and the posterior distribution of transit parameters is sampled using a Bayesian Monte Carlo Markov Chain (MCMC) framework with the package \texttt{emcee} \citep{foreman2013}. We use uninformative priors for the transit parameters, as presented in Table \ref{tab:planetsparamspriors} of Appendix \ref{app:priors-and-posteriors}, with initial values given by those reported in the TESS DVRs and assuming circular orbits. 

We divided the TESS light curve by the best-fit transit model and binned the data by the weighted average with bin sizes of 0.1~d to fit a baseline flux using Gaussian processes (GP) regression with a quasi-periodic (QP) kernel as in \cite{Martioli2022}. This kernel was chosen to detect a possible variability modulated by the star's rotation period.  The priors of the GP parameters are listed in Table \ref{tab:planetsparamspriors} and the posteriors are listed in Table \ref{tab:photgpposteriors}. We opted to fix the smoothing factor and decay time values at 0.1 and 10 days, respectively, as altering these values does not impact the other parameters. Unfortunately, the periodicities obtained in our analysis are not well constrained, and their values change significantly depending on the initial guess adopted for our GP model. The low level of intrinsic photometric variability in both targets, coupled with the rotation periods expected from stellar activity (16 – 32 days) being closely aligned with the TESS sector duration, strongly biases the detection of periodicity. This bias is primarily due to systematic variations that occur during TESS observations, particularly near the edges of each sector, where the amplitude of these variations is greater.  While there may be some intrinsic stellar variability captured by TESS photometry, the GP model obtained here may account for both systematic and intrinsic stellar phenomena in unknown proportions.

Since the periodic kernel is not essential, the primary purpose of employing the GP in this analysis is for detrending.  To ensure that the choice of GP does not impact the planet parameters, we conducted experiments using both a squared exponential (SE) kernel and a model without GP. The posterior distributions of key planet parameters ($T_{c}$, $a/R_{\star}$, $P$, and $R_{p}/R_{\star}$) are consistent within 1-sigma for all three choices of baseline model. The results of our best-fit transit models and GP analysis using a QP kernel are illustrated in Figures \ref{fig:toi1736_transits_fit} and \ref{fig:toi2141_transits_fit}.

\subsection{Radial velocities}
\label{sec:rvdata}

\begin{figure}
   \centering
   \includegraphics[width=1.0\hsize]{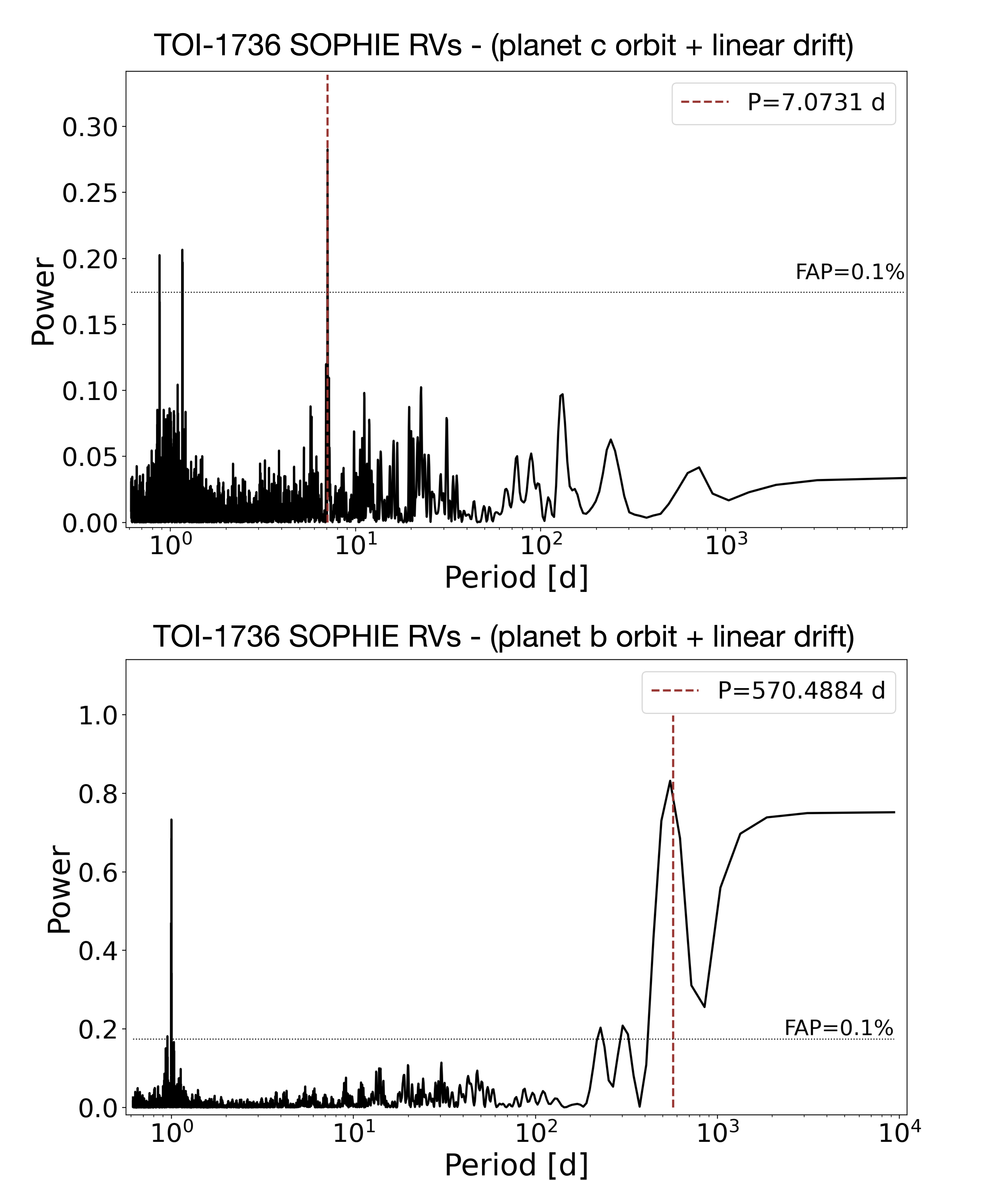}
      \caption{Generalized Lomb-Scargle periodogram for the TOI-1736 SOPHIE RVs. The top and bottom panels show the GLS for the SOPHIE RVs minus the linear trend and the fit orbits of planets c (top panel) and b (bottom panel).}
        \label{fig:toi-1736_rv_gls}
  \end{figure}

The TOI-1736 RVs have a median value of -25408.95~\ms\ with an rms of 158~\ms\ and a median individual error of 2~\ms.  These RVs present a clear signal with an amplitude of hundreds of \ms, as can be seen in Figure \ref{fig:toi-1736-rv+models}. An initial analysis of the generalized Lomb-Scargle (GLS) periodogram \citep{Zechmeister2009} of this data show that there is a dominating signal with periodicity around 570~d and a long-term linear trend was also detected by visual inspection. To detect the TESS candidate planet, TOI-1736.01, at a period of 7.073~d, we fit the SOPHIE RV data using the online tool \emph{Data \& Analysis Center for Exoplanets} (DACE\footnote{\url{https://dace.unige.ch}}), where we adopted a two-planet Keplerian model plus a linear trend.  As illustrated in Figure \ref{fig:toi-1736_rv_gls}, the GLS of the RVs after subtracting the linear trend and each planet's orbit shows a strong peak at 7.0731~d and 570.07~d for the inner and outer planet, respectively. The former agrees well with the periodicity of the transits detected by TESS, showing the RV detection of the inner planet candidate, TOI-1736~b, and an outer giant planet, TOI-1736~c. Moreover, this confirms that transits detected by TESS occur in the primary star, as the dominant flux in the SOPHIE spectra comes from the primary and not from the K3 dwarf companion. As a way to test if the SOPHIE spectra are contaminated by the flux from the companion, we performed a CCF analysis using a K0, K5, or M0 masks, which should favor the detection of spectral lines of the cooler star. Those tests resulted in slightly noisier CCF with parameters in agreement as when using the G2 mask, showing that indeed, the flux contribution of the companion to the RV measurements in the SOPHIE spectra is negligible. 

\begin{figure}
   \centering
   \includegraphics[width=1.0\hsize]{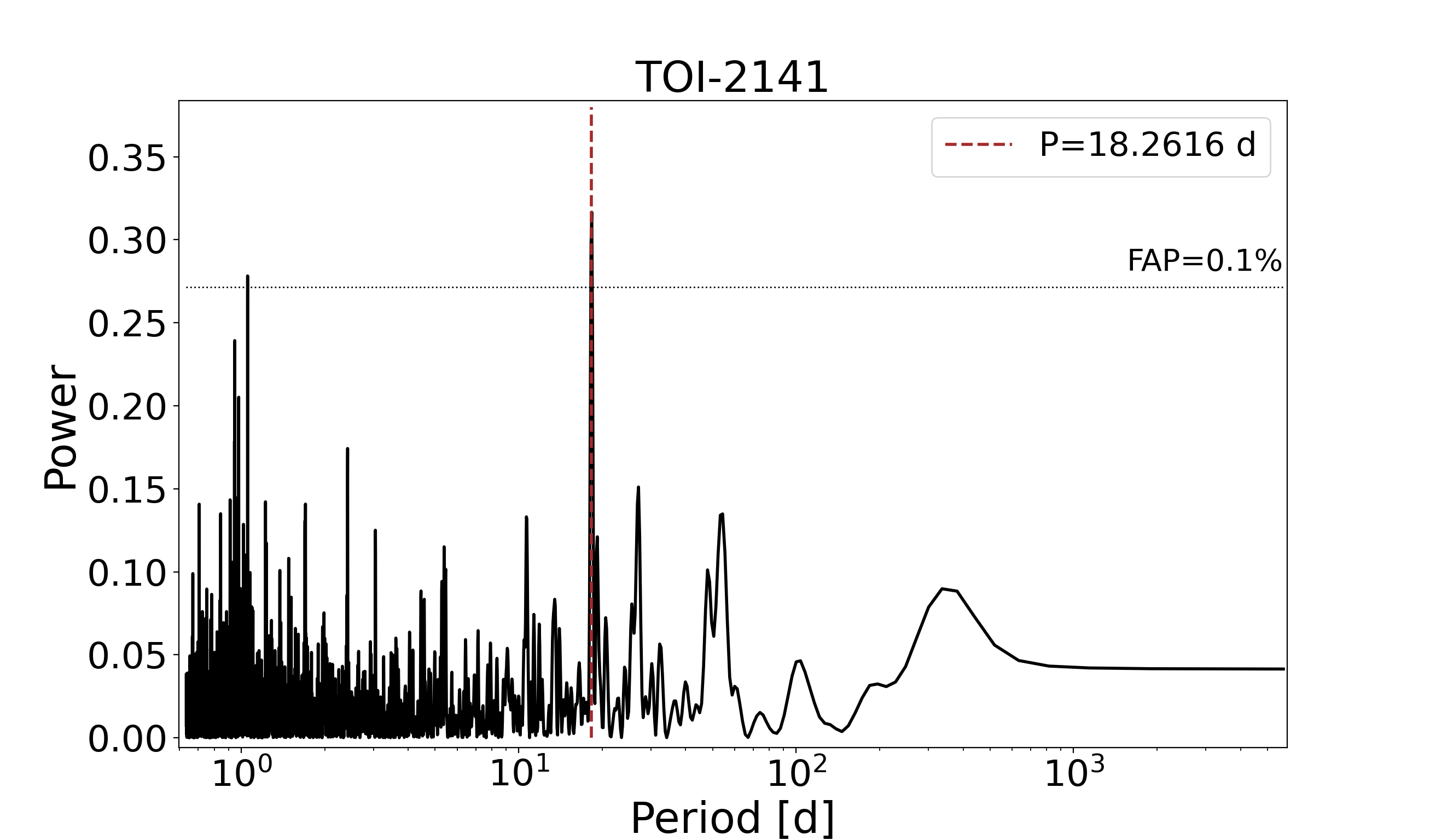}
      \caption{Generalized Lomb-Scargle periodogram for the TOI-2141 SOPHIE RVs. The blue dashed line shows the peak power at 18.3~d, which agrees with the period reported by TESS for TOI-2141~b.
      }
        \label{fig:toi-2141_rv_gls}
  \end{figure}

The TOI-2141 RVs have a median value of -19860.8~\ms\ with an rms of 7~\ms\ and a median individual error of 2.4~\ms. The RVs of this object do not appear to have any long-term trends or variability. Figure \ref{fig:toi-2141_rv_gls} shows the GLS periodogram for this data, which shows the highest peak at a period of 18.259~d, again in agreement with the periodicity of the transits detected by TESS. This shows the detection of the planet candidate TOI-2141~b in our SOPHIE data. In the following sections, we present an analysis to characterize these two planetary systems.

For both systems, we found that there is no correlation between the RVs and the four stellar activity indicators: FWHM, BIS, H$_\alpha$, and S-index. Therefore, the signals detected in RV are probably due to Doppler shift and not to scenarios that imply changes in the line profile, for example, blended binaries.  We have also inspected possible correlations between the RV residuals (after subtracting the fit models) and activity indicators. Only BIS shows a slight correlation of 0.29 and 0.15 for TOI-1736 and TOI-2141, respectively. We tried to de-correlate the RVs as in \cite{Boisse2009}, but this did not have a significant impact on the values of the fit parameters, so we preferred to keep the final analysis without applying this correction. In the Appendix \ref{app:activityindices} we show the time series of each activity indicator, an analysis of the GLS periodogram, and its correlations with the RV data.

\subsection{Joint analysis of RVs and photometry to characterize the planets}
\label{sec:jointanalysis}

To obtain the physical parameters of the planets, we perform a joint analysis of the photometry and RV data sets as in \cite{Martioli2022}. We adopt the priors listed in Table \ref{tab:planetsparamspriors} and the initial values are those obtained by the analysis performed on the SOPHIE RVs and TESS photometry data independently. 
 
We first fit the RV orbits and transit models jointly using the \texttt{scipy.optimize.leastsq} optimization tool. Then, we use the same tool to fit the RV jitter, which is a term quadratically added to the RV errors. Finally, we explore the full range covered by the priors with a Bayesian MCMC that uses 50 random walkers, with 10000 iterations, discarding the first 3000 burn-in samples. The posterior distributions of sampled parameters are illustrated in Appendix \ref{app:priors-and-posteriors} and the resulting best-fit parameters, obtained from the mode of the posterior and 1$\sigma$ uncertainties (34\% on each side of the central value) are given in Table \ref{tab:planetsparams}. Some parameters converged to values close to the edge of their physical domains. For example, it happened for the orbital inclination ($i_{b}\sim90$~deg) and the linear limb-darkening coefficient ($u_{0}\sim0$) for both transiting planets, TOI-1736~b and TOI-2141~b.  We also adopted circular orbits ($e=0$, $\omega=90$~deg) for both inner planets, as noncircular solutions resulted in higher values of the Bayesian Information Criterion (BIC), suggesting that our data cannot constrain the low values of the eccentricities of these planets. 

\begin{figure}
   \centering
   \includegraphics[width=1.0\hsize]{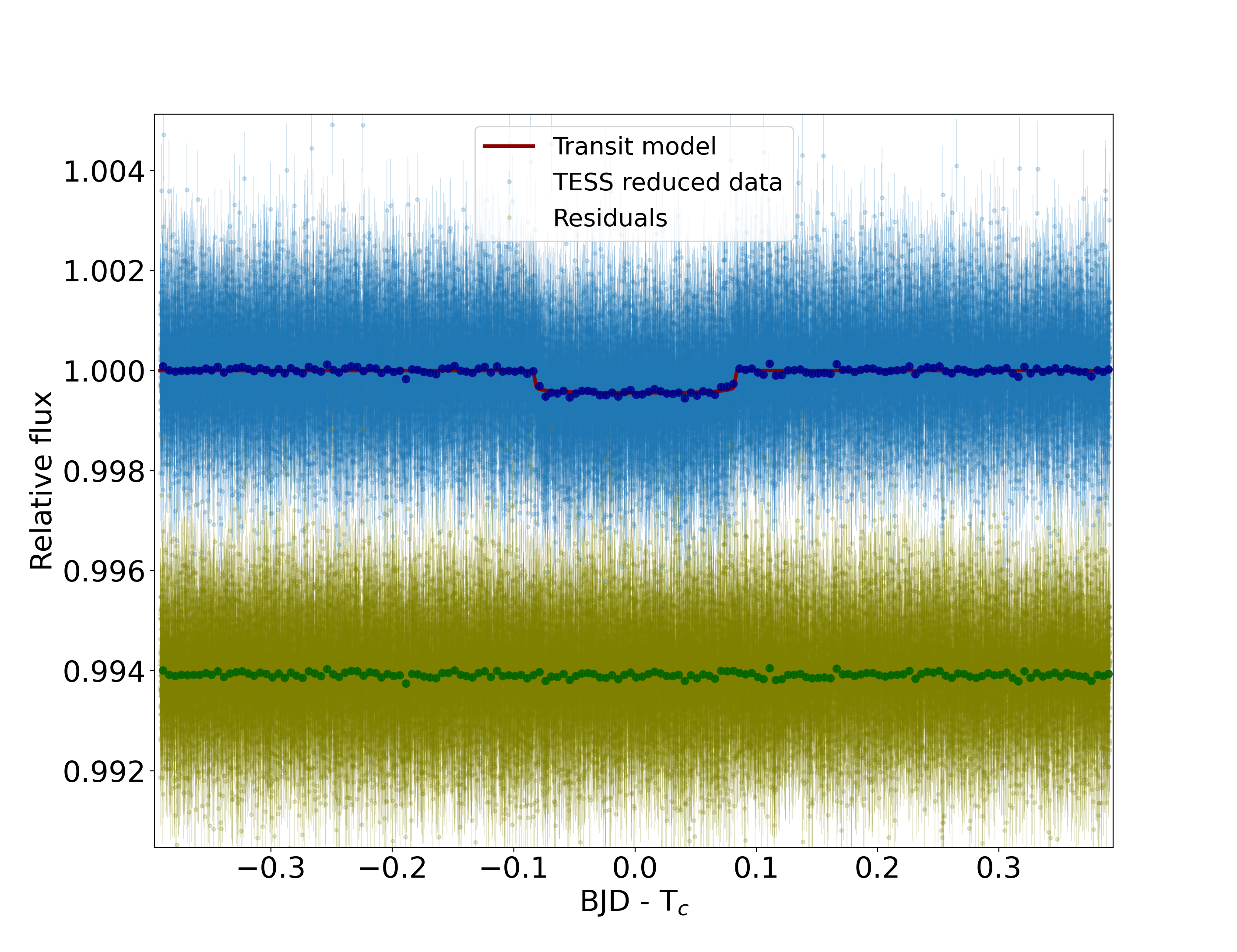}
      \caption{Transits of TOI-1736~b observed by TESS.  The blue points show the TESS photometry data around the nineteen transits of TOI-1736~b, with the times relative to the central time of each transit. The red line shows the best-fit transit model and the green points show the residuals plus an arbitrary offset for better visualization.  
      }
        \label{fig:toi-1736-tess+phase}
  \end{figure}

\begin{figure}
   \centering
   \includegraphics[width=1.0\hsize]{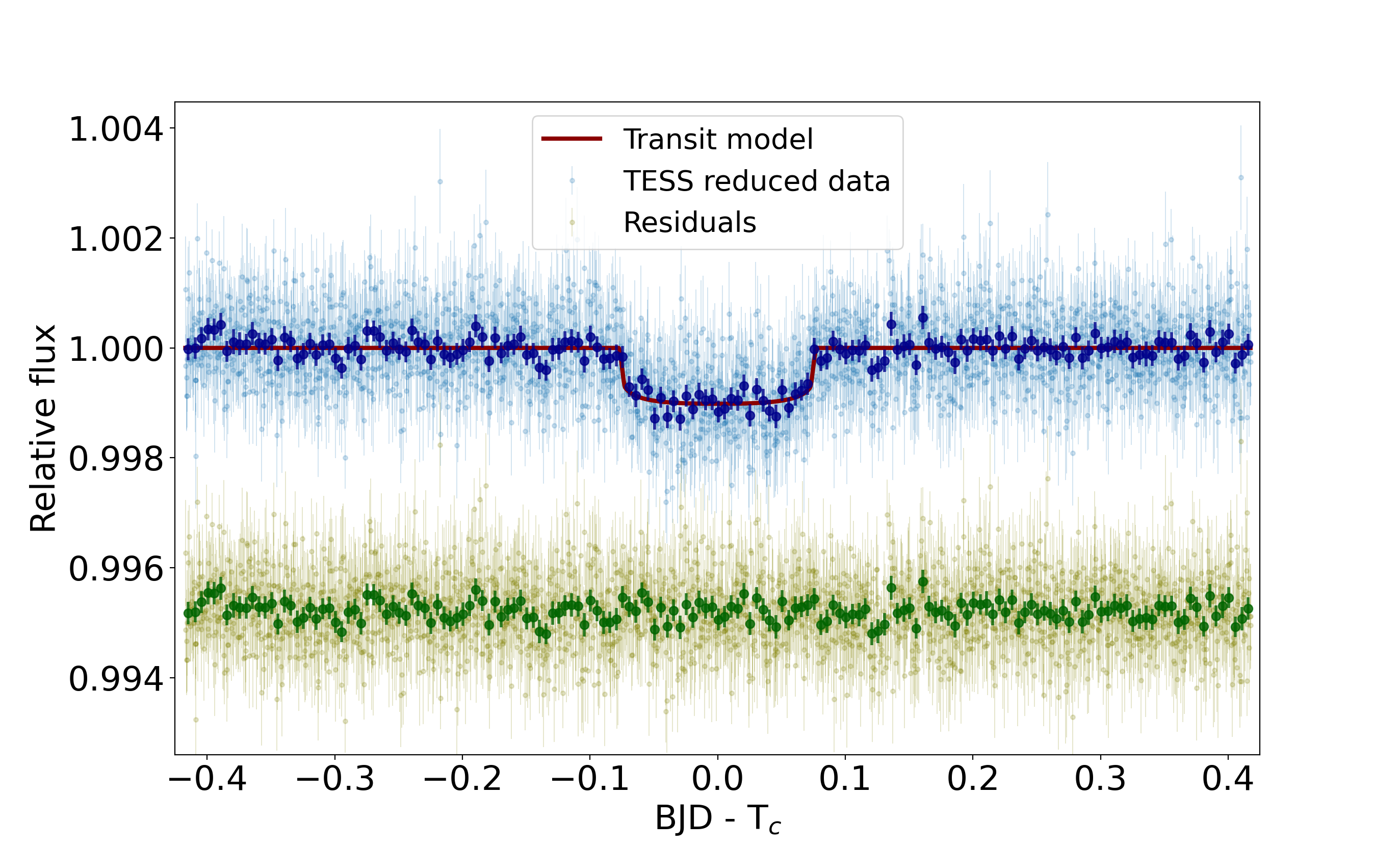}
      \caption{Transits of TOI-2141~b observed by TESS.  The blue points show the TESS photometry data around the four transits of TOI-2141~b, with the times relative to the central time of each transit. The red line shows the best-fit transit model and the green points show the residuals plus an arbitrary offset for better visualization.  
      }
        \label{fig:toi-2141-tess+phase}
  \end{figure}

\begin{figure*}
   \centering
   \includegraphics[width=1.0\hsize]{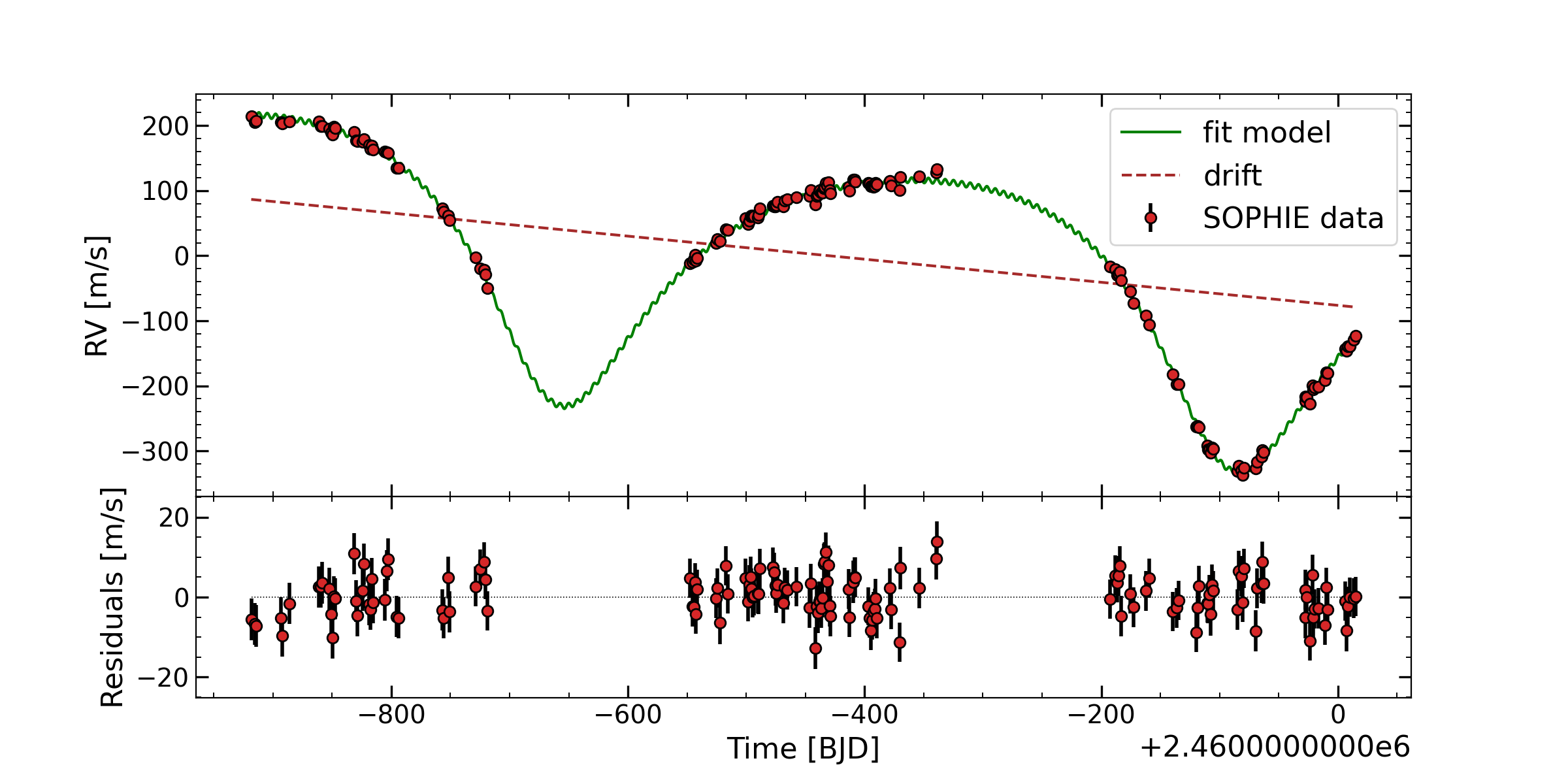}
      \caption{TOI-1736 SOPHIE RVs. In the top panel, the red points show the SOPHIE RV data and the green line shows the best-fit model, including the orbits of TOI-1736~b and TOI-1736~c, plus a linear trend, which is also represented separately by the red dashed line. The bottom panel shows the residuals. 
      }
        \label{fig:toi-1736-rv+models}
  \end{figure*}

\begin{figure*}
   \centering
   \includegraphics[width=1.0\hsize]{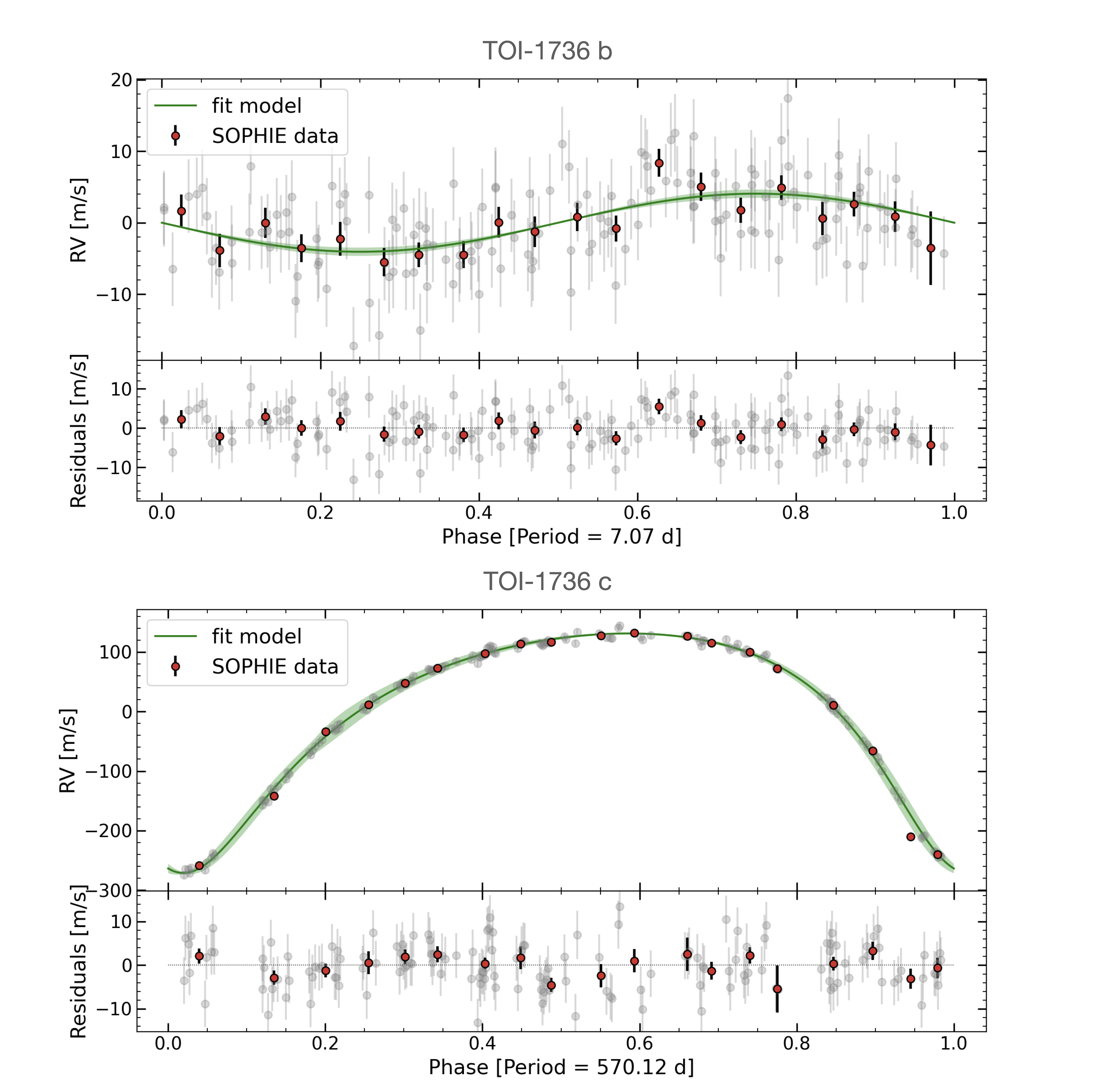}
      \caption{TOI-1736 SOPHIE RVs in phase with the orbits of the planets. In the top panel, the gray points show the RVs phase-folded in the period of planet b, where the orbit of planet c and the linear trend have been subtracted.  In the bottom panel, the gray points show the RVs phase-folded in the period of planet c, where the orbit of planet b and the linear trend has been subtracted.  The red points show the binned data, where we use a bin size of 0.05 in units of orbital phase. The green lines represent the best-fit orbit model for each respective planet.
      }
        \label{fig:toi-1736-rv+phasedmodels}
  \end{figure*}

\begin{figure*}
   \centering
   \includegraphics[width=1.0\hsize]{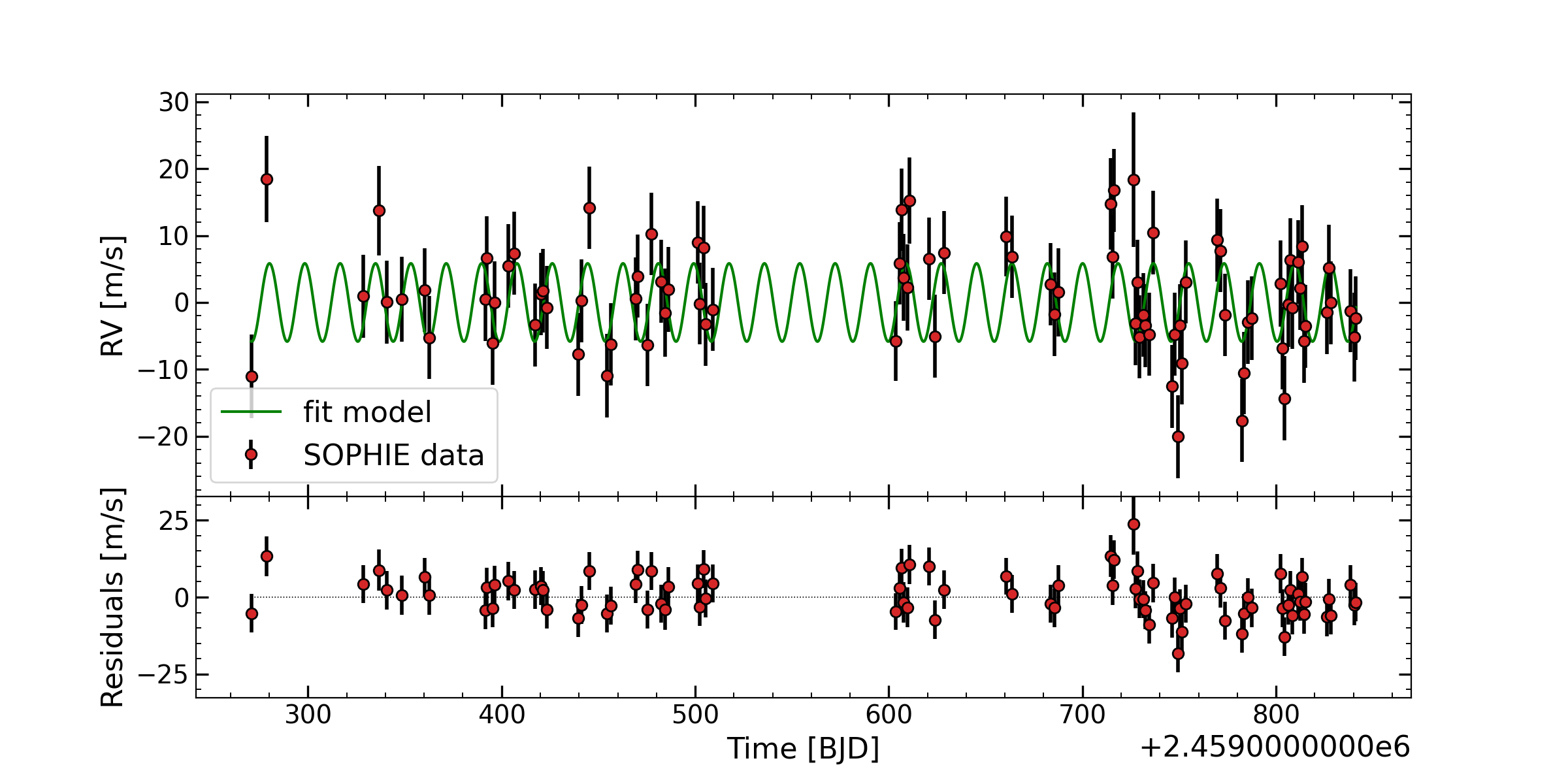}
      \caption{TOI-2141 SOPHIE RVs. In the top panel, the red points show the SOPHIE RV data and the green line shows the best-fit model for the orbit of TOI-2141~b. The bottom panel shows the residuals. 
      }
        \label{fig:toi-2141_rv+models}
  \end{figure*}

\begin{figure*}
   \centering
   \includegraphics[width=1.0\hsize]{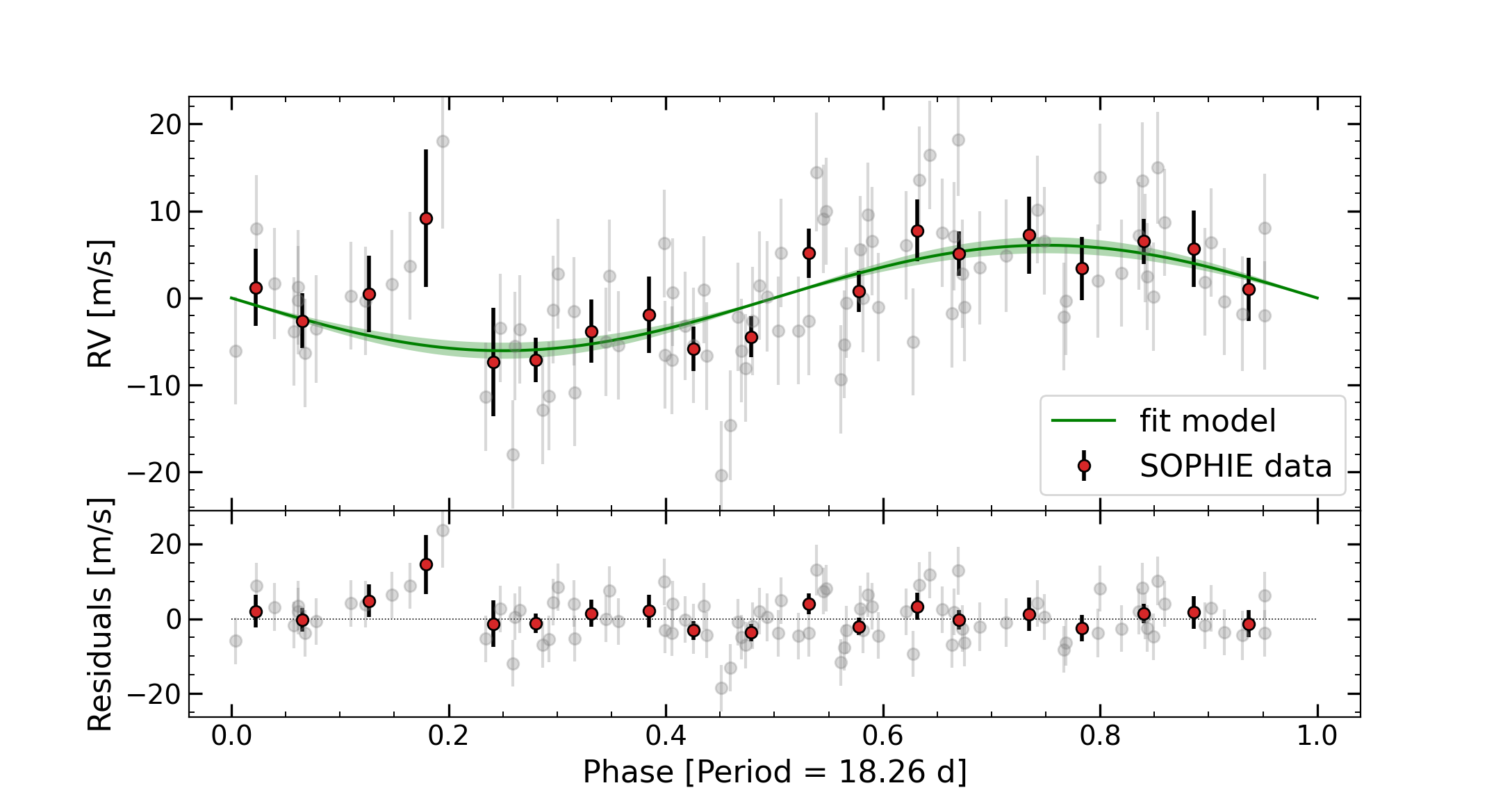}
      \caption{TOI-2141 SOPHIE RVs in phase with the orbit of the planet. The gray points show the RVs phase-folded in the period of planet TOI-2141~b, and the red points show the binned data, where we use a bin size of 0.05 in units of orbital phase. The green lines represent the best-fit orbit model for TOI-2141~b.  
      }
        \label{fig:toi-2141_rv+phasedmodels}
  \end{figure*}

Figures \ref{fig:toi-2141-tess+phase} and \ref{fig:toi-1736-tess+phase} illustrate the TESS photometry data for all transits detected by TESS and the best-fit transit models obtained by our joint analysis.  Figure \ref{fig:toi-1736-rv+models} and \ref{fig:toi-1736-rv+phasedmodels} illustrate the TOI-1736 RV data and the best-fit models that include the orbit of the two planets, TOI-1736~b and TOI-1736~c, and the linear trend. Figures \ref{fig:toi-2141_rv+models} and \ref{fig:toi-2141_rv+phasedmodels} illustrate the RV data and best-fit model for TOI-2141. 

\begin{table*}
\centering
\caption{Summary of the planetary parameters.}
\label{tab:planetsparams}
\begin{tabular}{lccc}
\hline
Parameter & TOI-1736~b & TOI-1736~c & TOI-2141~b\\
\hline
time of conjunction, $T_{c}$ (BJD) & $2458792.7947\pm0.0008$ & $2455283\pm5$ & $2458992.5033\pm0.0017$ \\
orbital period, $P$ (d) & $7.073088(7)$ & $570.1\pm0.7$ & $18.26157(6)$ \\
eccentricity, $e$  & $<0.21$  & $0.362\pm0.003$ & $<0.21$ \\
argument of periastron, $\omega$ (deg) & 90 & $164.2\pm0.6$ & 90 \\
%-------------------------------------------------------
normalized semimajor axis, $a/R_{\star}$ & $13.4^{+0.5}_{-1.8}$ & $258\pm18$ & $39^{+4}_{-9}$ \\
semimajor axis, $a_{p}$ \tablefootmark{a} (au) & $0.0740\pm0.0009$ & $1.381\pm0.017$ & $0.1330\pm0.0009$ \\
%orbital inclination, $i_{p}$  (deg) &  $89.4^{+0.9}_{-1.6}$ & $69^{+15}_{-17}$ & $89.8\pm0.7$ \\
orbital inclination, $i_{p}$  (deg) &  $>87.3$ & $69^{+15}_{-17}$ & $>89.0$ \\
transit duration, $t_{\rm dur}$  (h) & $3.95\pm0.32$ &  & $4.9\pm0.4$ \\
impact parameter, $b$  & $<0.4$ &  & $<0.4$ \\
%-------------------------------------------------------
planet-to-star radius ratio, $R_{p}/R_{\star}$  & $0.0206\pm0.0004$ &  & $0.0284\pm0.0012$ \\
planet radius, $R_{p}$ (\RJ) & $0.223\pm0.016$ &  & $0.278\pm0.021$ \\
planet radius, $R_{p}$ (\RN)  & $ 0.63\pm0.05$ &  & $0.79\pm0.06$ \\
planet radius, $R_{p}$ (\RE) & $2.44\pm0.18$ &  & $3.05\pm0.23$ \\
%-------------------------------------------------------
velocity semi-amplitude, $K_{p}$ (m\,s$^{-1}$) & $4.1\pm0.6$ & $201.1\pm0.7$ & $6.0\pm1.0$ \\
planet mass, $M_{p}$ (\MJ) & $0.040\pm0.006$ & $8.7^{+1.5}_{-0.6}$ & $0.075\pm0.012$ \\
planet mass, $M_{p}$ (\MN) & $ 0.75\pm0.11$ & $161^{+28}_{-11}$ & $1.38\pm0.22$ \\
planet mass, $M_{p}$ (\ME) & $12.8\pm1.8$ & $2.8^{+0.5}_{-0.2}\times10^{3}$ & $24\pm4$ \\
planet min. mass, $M_{p}\sin{i_{p}}$ (\MJ) &  & $8.09\pm0.20$ &  \\
planet density, $\rho_{p}$ (g\,cm$^{-3}$) & $4.9\pm1.3$ &  & $4.6\pm1.3$ \\
%-------------------------------------------------------
equilibrium temperature, $T_{\rm eq}$ \tablefootmark{b} (K) & $1076\pm39$ & $249\pm9$ & $722\pm23$ \\
%-------------------------------------------------------
linear limb dark. coef., $u_{0}$  & $0.01^{+0.12}_{-0.07}$ &  & $0.15^{+0.13}_{-0.09}$ \\
quadratic limb dark. coef., $u_{1}$ & $0.23^{+0.20}_{-0.17}$ &  & $0.42^{+0.21}_{-0.17}$ \\
%-------------------------------------------------------
RV linear trend, $\alpha$ (m\,s$^{-1}$\,d$^{-1}$) &  \multicolumn{2}{c}{$-0.17741424(18)$}  & 0 \\
RV jitter, $\sigma_{\rm jitter}$ (m\,s$^{-1}$) & $5\pm3$ & $5\pm3$ & $6\pm4$ \\
RMS of RV residuals (m\,s$^{-1}$) & 5.0 & 5.0 & 6.5 \\
RMS of flux residuals (ppm) & 941 &  & 630 \\
%-------------------------------------------------------
\hline
\end{tabular}
\tablefoot{
\tablefoottext{a}{semi-major axis derived from the fit period and Kepler's law.}
\tablefoottext{b}{assuming a uniform heat redistribution and an arbitrary geometric albedo of 0.1.}
}
\end{table*}

%-----------------------------------------------------------------
\section{GAIA astrometry to constrain the mass of TOI-1736~c}
\label{sec:gaiaastrometry}

\begin{figure}
   \centering
   \includegraphics[width=1.0\hsize]{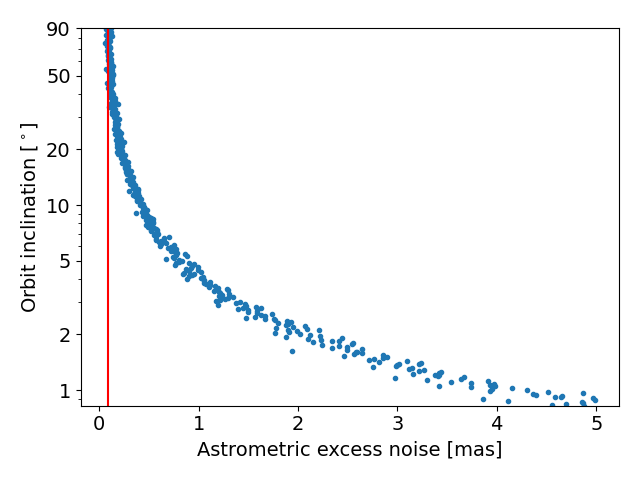}
      \caption{Astrometric constraint to the orbital inclination of TOI-1736~c. The blue points show the excess noise data obtained from GASTON simulations, assuming the RV-derived parameters of TOI-1736~c. The red line shows the GAIA DR3 excess noise of TOI-1736, $\varepsilon_{\rm DR3}=0.09221$~mas.      
      }
        \label{fig:toi-1736_gaston_inclination}
  \end{figure}

The expected astrometric signal due to the orbit of the star TOI-1736 around the center of mass of the system can be estimated by the following equation:

\begin{equation}
 a_{\star} = p a_{p} \left( \frac{m_{p}}{M_{\star}} \right),  
\end{equation}

where $p$ is the parallax in milli-arcseconds, $a_{p}$ is the semi-major axis of the planet's orbit in au, $m_{p}$ is the planet's mass and $M_{\star }$ is the stellar mass. For TOI-1736~c, we estimate $a_{\star}=0.112\pm0.005$~mas.  Here, we use the method Gaia Astrometric noise Simulation To derive Orbit iNclination \citep[GASTON;][]{Kiefer2019a,Kiefer2019b} to constrain, from GAIA DR3 \citep{Gaia2021} astrometric excess noise and RV-derived orbital parameters, the orbital inclination and true mass of TOI-1736~c.  The GAIA DR3 astrometric excess noise of TOI-1736 is $\varepsilon_{\rm DR3}=0.09221$~mas. Figure \ref{fig:toi-1736_gaston_inclination} shows the results of our GASTON simulations of astrometric excess noise as a function of the orbital inclination, assuming the RV orbit of TOI-1736~c.  We obtained an orbital inclination of $i_{p}=69^{+15}_{-17}$~deg, which gives a semi-major axis of the star photocenter of $a_{\star}=0.119_{-0.007}^{+0.022}$~mas. 

\begin{figure}
   \centering
   \includegraphics[width=1.0\hsize]{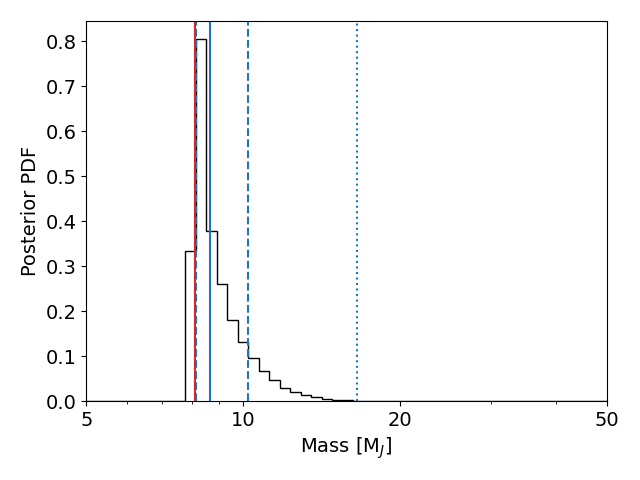}
      \caption{TOI-1736~c mass determination. The black line shows the mass posterior distribution obtained from RV-derived parameters and GASTON simulations of astrometric excess noise. The solid and dashed blue lines show the median and the $\pm1\sigma$ range of the distribution. The red line shows the minimum mass of $M_{p}\sin{i_{p}}=8.09$~\MJ\ obtained from our RV analysis and the dotted blue line shows the $3\sigma$ upper limit of $<16.54$~\MJ.  
      }
        \label{fig:toi-1736_gaston_mass}
  \end{figure}

These results show that the orbit of the TOI-1736~c seems to be tilted with respect to the inner planet, but is not inconsistent, within $2\sigma$, with a coplanar orientation. Assuming coplanarity, we computed the prospective transit epochs for TOI-1736~c; however, none of the predicted events align with TESS observations. Finally, we derived the mass of TOI-1736~c by considering our measurements of $M_{p}\sin{i_{p}}$ combined with the astrometric constraint. Figure \ref{fig:toi-1736_gaston_mass} shows the posterior distribution for the mass of TOI-1736~c, where we obtained $M_{p}=8.7^{+1.5}_{-0.6}$ at $1\sigma$ with a maximum value of $M_{p}<16.54$~\MJ\ at $3\sigma$.

%-----------------------------------------------------------------
\section{Discussions}
\label{sec:discussions}

The SOPHIE RV data show periodic variations at the same period as the transits detected by TESS, where the orbital fit gives a semi-amplitude of $K_{p}=4.1\pm0.6$~m\,s$^{-1}$ for TOI-1736~b and $K_{p}=6.0\pm1.0$~m\,s$^{-1}$ for TOI-2141~b. The TOI-1736 RVs also revealed an outer giant planet TOI-1736~c with an RV semi-amplitude of $K_{p}=201.1\pm0.7$~m\,s$^{-1}$, in addition to a long term linear trend of $-0.177$~m\,s$^{-1}$\,d$^{-1}$. This trend could be due to the perturbation caused by the orbital motion of an outer planetary or stellar companion. We note that this deviation could be caused by the K-dwarf stellar companion that we detected with photometry, although we did not detect its signature in GAIA's excess astrometric noise. Current SOPHIE data are not sufficient to distinguish at what periodicity and amplitude this variability occurs. Therefore, we have not yet inferred any characterization for this possible companion, and we will continue to monitor this star in the coming years to discover the nature of this signal in future work.

The inner planets, TOI-1736~b and TOI-2141~b, with radii of $2.44\pm0.18$~\RE\ and $3.05\pm0.23$~\RE\ and masses of $12.8\pm1.8$~\ME\ and $24\pm4$~\ME\, have similar mean densities of $4.9\pm1.3$~g\,cm$^{-3}$ and $4.6\pm1.3$~g\,cm$^{-3}$, respectively. The giant planet TOI-1736~c has a mass of $8.7^{+1.5}_{-0.6}$~\MJ\ and orbits the star with a period of $570.1\pm0.7$~d and an orbital eccentricity of $e=0.362\pm0.003$. With a semi-major axis of $1.381\pm0.017$~au, this giant resides in a temperate orbit around the star, which is in the habitable zone of that star as detailed in Section \ref{sec:hzs}.  TOI-1736 is therefore similar to the system HD~137496 (K2-364) \citep{Silva2022}, which is also a solar analog hosting a hot, dense inner planet and an outer giant with about the same mass as TOI-1736~c and with similar orbital properties.  Both are nearby and bright systems, making them interesting cases to study the evolution of planetary systems around solar analogs and, in particular, studying the occupation of their habitable zones by giant planets.

%-----------------------------------------------------------------
\subsection{The exoplanets' radiation environment}
\label{sec:hzs}

We consider the best-fit parameters from our analysis to calculate the habitable zone for these solar analogs, using the equations and data from \cite{Kopparapu2014}. We obtained an optimistic lower limit (recent Venus) at 0.87~au for TOI-1736 and 0.71~au for TOI-2141, and an upper limit (early Mars) at 2.05~au for TOI-1736 and 1.68~au for TOI-2141. The runaway greenhouse limits ($M_{\rm p}=1$~\ME) ranges between 1.10~au and 1.94~au for TOI-1736, and between 0.90~au and 1.59~au for TOI-2141. 

\begin{figure}
   \includegraphics[width=1.0\hsize]{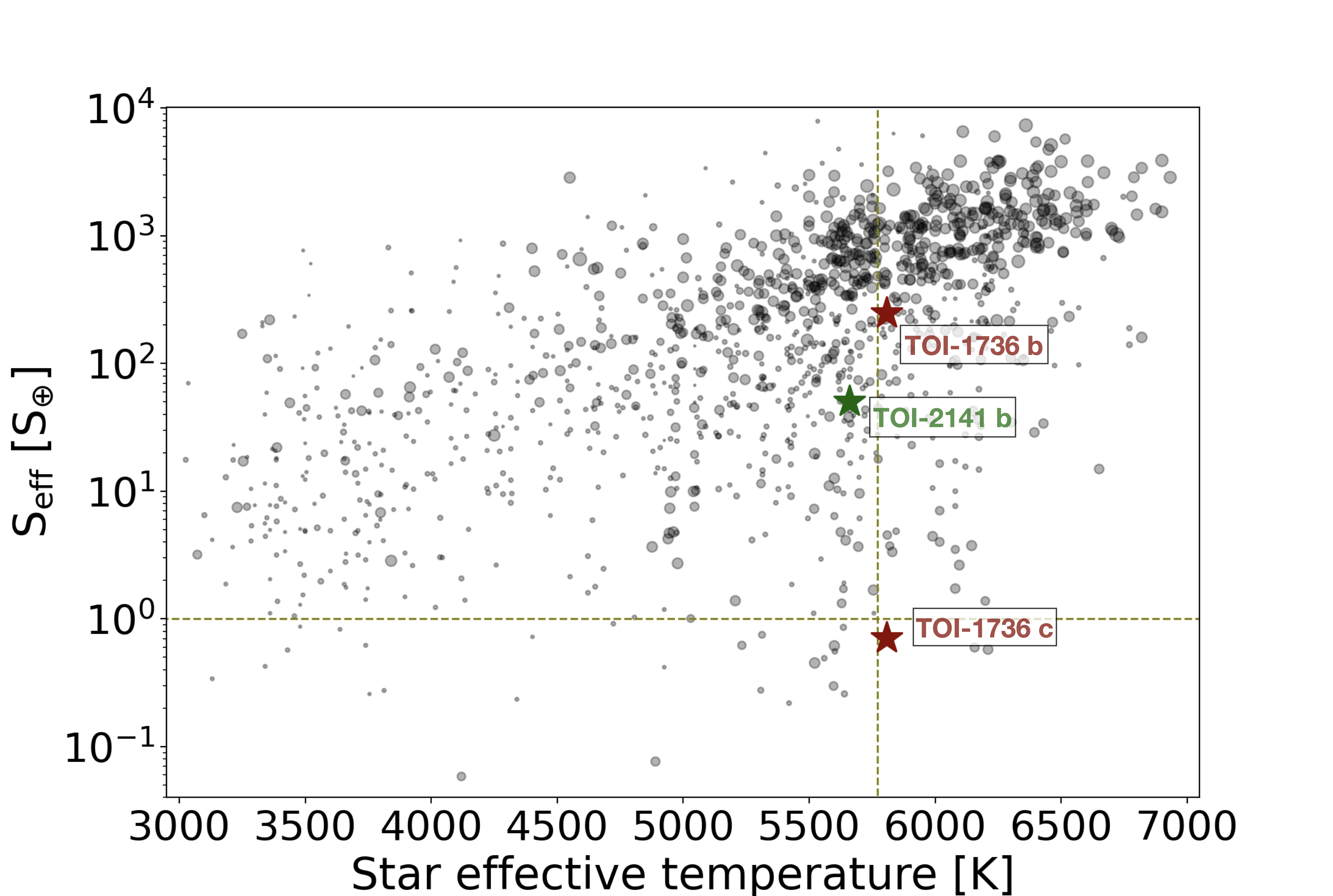}
      \caption{Exoplanet insolation compared to the incident flux on Earth. The black circles show the exoplanet data from \texttt{exoplanets.eu} where the sizes are proportional to the planet's radius, the red stars show the two planets in the TOI-1736 system, and the green star shows planet TOI-2141~b. Light green dashed lines show Earth values. The insolation for both transiting planets is very high compared to Earth, resulting in an equilibrium temperature of 1076~K and 722~K for TOI-1736~b and TOI-2141~b, respectively. The giant planet TOI-1736~c receives insolation very close to Earth values, placing it in the habitable zone around the star.      
      }
        \label{fig:exoplanetinsolation}
  \end{figure}

To estimate the level of radiation that each planet receives from its host star, we calculate the effective insolation flux $S_{\rm eff}$ relative to the flux incident on Earth $S_\oplus$ as $S_{\rm eff}/S_\oplus = L_{\star}/a_{p}^{2}$, where $L_{\star}$ is the star luminosity in solar units and $a_{p}$ is the semi-major axis of the planet's orbit in au.  Figure \ref{fig:exoplanetinsolation} shows the insolation for exoplanets as a function of star effective temperature. TOI-1736~b and TOI-2141~b receive an insolation of $246\pm26\,S_\oplus$ and $50\pm4\,S_\oplus$, respectively, showing that these planets reside in a hot environment. We estimate the equilibrium temperature for the planets as in \cite{heng2013}, assuming a uniform heat redistribution and an arbitrary geometric albedo of 0.1, which gives $T_{\rm eq}=1076\pm39$~K and $T_{\rm eq}=722\pm23$~K for TOI-1736~b and TOI-2141~b, respectively.

On the other hand, TOI-1736~c resides at a larger distance from the star, in an eccentric orbit with $e_{c}=0.362\pm0.003$, and a semi-minor axis of $1.287\pm0.016$~au and a semi-major axis of $1.381\pm0.017$~au.   At these distances, TOI-1736~c receives an insolation of $0.81\pm0.09\,S_\oplus$ and $0.71\pm0.08\,S_\oplus$ for which we estimate an equilibrium temperature of $T_{\rm eq}=258\pm9$~K and $T_{\rm eq}=249\pm9$~K, respectively. This planet resides within the conservative habitable zone around the star.

%-----------------------------------------------------------------
\subsection{Mass-radius relation for the inner planets}
\label{sec:massradius}

\begin{figure*}
  \sidecaption
   \includegraphics[width=1.0\hsize]{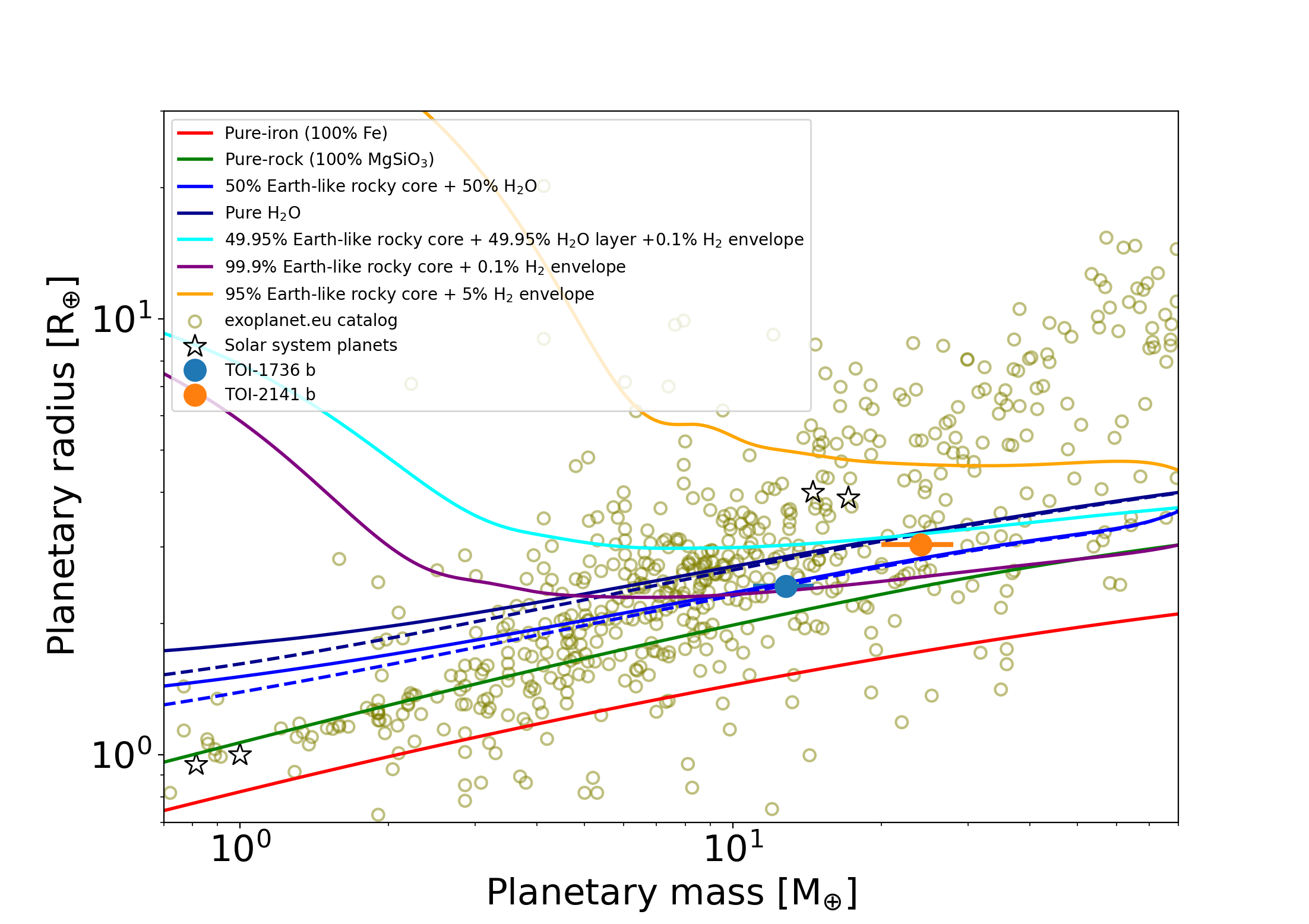}
      \caption{ Mass-radius diagram.  The green circles show the exoplanet data from \texttt{exoplanets.eu}, the black stars show the Solar System planets and the blue and orange points show the measured values for TOI-1736~b and TOI-2141~b, respectively.  Several models from \cite{Zeng2019} are also plotted for comparison. 
      }
    \label{fig:massradiusdiagram}
  \end{figure*}

From mass and radius measurements of the transiting planets TOI-1736~b and TOI~2141~b, we derive their mean densities of $4.9\pm1.3$~g\,cm$^{-3}$ and $4.6\pm1.3$~g\,cm$^{-3}$, respectively, showing that these sub-Neptune sized planets have densities similar to the rocky planets of the Solar System. To explore further the possible chemical composition and structure of these planets we inspect the mass-radius (M-R) diagram illustrated in Figure \ref{fig:massradiusdiagram}, which shows the exoplanet data from \texttt{exoplanets.eu}, the data from the two hot sub-Neptunes studied here, and a set of models from \cite{Zeng2019}, for comparison.  The M-R location of TOI-1736~b matches two models: (1) a 50\% Earth-like rocky core plus a 50\% H$_2$O layer; (2) an almost pure Earth-like rocky core with a thin 0.1\% H$_2$ envelop. Therefore, this planet probably has a large dense rocky core surrounded by either a thick water-rich atmosphere or a thin H$_2$ atmosphere.  TOI-2141~b is compatible with three scenarios: (1) a 50\% Earth-like rocky core with a 50\% H$_2$O layer; (2) a 49.95\% Earth-like rocky core + 49.95\% H$_2$O layer and a thin 0.1\% H$_2$ envelop; and (3) a pure H$_2$O. Thus, TOI-2141~b is likely a water-rich planet. A further investigation of the internal structure of these planets can be done by including simulations constrained by the star metallicity and abundances of volatiles, which is out of the scope of this paper.

%-----------------------------------------------------------------
\section{Conclusions}
\label{sec:conclusions}

We reported the discovery of two new planetary systems around the solar analogs TOI-1736 and TOI-2141.  We monitored these systems spectroscopically with the SOPHIE instrument at OHP to obtain high-precision RV time series. These data show periodic variations at the same period as the transiting candidates detected by TESS, establishing the planetary nature of the sub-Neptunes TOI-1736~b and TOI-2141~b.  Our RV data also revealed an outer giant planet in the TOI-1736 system, in addition to a long-term linear trend. This drift could be due to an additional companion that will be investigated further in future observations.

We characterized the two host stars by analyzing their high-resolution spectra combined with photometry and high-contrast imaging data. We found that TOI-1736 likely has a K3 dwarf stellar companion, which is evidenced both by the power spectrum of the speckle imaging data at 832~nm and by an infrared excess in the SED. The presence of this companion has an important impact on the determination of the star's radius, which affects the derivation of the planet's radius and other physical properties of the system. The flux of this possible companion has a negligible contribution to our high-resolution SOPHIE spectra.

We obtained the stellar parameters from several methods, including a SED fit analysis of the photometry data, a spectral analysis of the high-resolution SOPHIE and TRES spectra, and a differential analysis of the SOPHIE spectra using a solar spectrum from observations of the Moon as a reference. The latter provides the most accurate stellar parameters, which also improves the determination of the physical parameters of the planets. This shows the importance of studying planetary systems around solar analogs, where the solar spectrum can benchmark measurements of stellar properties and therefore provide more accurate values. We also determined elemental abundances for several species, including refractory elements (Fe, Mg, and Si) that provide constraints for the planet's interior structure models. We found that both systems exhibit low levels of stellar activity from their $\log{\rm R'}_{\rm HK}$, which are consistent with their non-variable time series of activity indices, and with the mature ages of about 5~Gyr for TOI-1736 and 7~Gyr for TOI-2141, found from chemical clocks and evolutionary analyses. The inferred radii and stellar masses for these two stars have values that differ by about $\pm10\%$ of solar values, showing that they are indeed solar analogs, but not solar twins. 

Therefore, these two new systems represent perspectives on how planetary systems can form and evolve around solar analogs. On the one hand, we have TOI-2141, an evolved star that hosts a warm sub-Neptune with a likely dense rocky core and a possible thick, water-rich envelope. On the other hand, we have TOI-1736, a hierarchical star system with a solar analog that hosts a hot, dense sub-Neptune and an outer giant planet in an eccentric orbit in the habitable zone. 

\begin{acknowledgements}

    E.M. acknowledges funding from \emph{Funda\c{c}\~{a}o de Amparo \`{a} Pesquisa do Estado de Minas Gerais} (FAPEMIG) under project number APQ-02493-22 and research productivity grant (PQ) number 309829/2022-4 awarded by the \emph{National Council for Scientific and Technological Development} (CNPq), Brazil. We thank the Observatoire de Haute-Provence (CNRS) staff for their support. This work was supported by the ``Programme National de Plan\'etologie'' (PNP) of CNRS/INSU and CNES. This paper includes data collected with the TESS mission, obtained from the MAST data archive at the Space Telescope Science Institute (STScI). Funding for the TESS mission is provided by NASA's Science Mission Directorate. We acknowledge the use of public TESS data from pipelines at the TESS Science Office and at the TESS Science Processing Operations Center. Resources supporting this work were provided by the NASA High-End Computing (HEC) Program through the NASA Advanced Supercomputing (NAS) Division at Ames Research Center for the production of the SPOC data products. This publication makes use of The Data \& Analysis Center for Exoplanets (DACE), which is a facility based at the University of Geneva (CH) dedicated to extrasolar planets data visualization, exchange, and analysis. DACE is a platform of the Swiss National Centre of Competence in Research (NCCR) PlanetS, federating the Swiss expertise in Exoplanet research. The DACE platform is available at \url{https://dace.unige.ch}. We acknowledge funding from the French ANR under contract number ANR18CE310019 (SPlaSH). X.D. and T.F. acknoweldge support by the French National Research Agency in the framework of the Investissement d’Avenir program (ANR-15-IDEX-02), through the funding of the "Origin of Life" project of the Grenoble-Alpes University. This work was supported by FCT - Fundação para a Ciência e a Tecnologia through national funds and by FEDER through COMPETE2020 - Programa Operacional Competitividade e Internacionalizacão by these grants: UID/FIS/04434/2019, UIDB/04434/2020, UIDP/04434/2020, PTDC/FIS-AST/32113/2017 \& POCI-01-0145-FEDER- 032113, PTDC/FIS-AST/28953/2017 \& POCI-01-0145-FEDER-028953, PTDC/FIS-AST/28987/2017 \& POCI-01-0145-FEDER-028987. NCS further acknowledges funding by the European Union (ERC, FIERCE, 101052347). Views and opinions expressed are however those of the author(s) only and do not necessarily reflect those of the European Union or the European Research Council. Neither the European Union nor the granting authority can be held responsible for them. This work makes use of observations from the LCOGT network. Part of the LCOGT telescope time was granted by NOIRLab through the Mid-Scale Innovations Program (MSIP). MSIP is funded by NSF. KAC acknowledges support from the TESS mission via subaward s3449 from MIT. S.D. is funded by the UK Science and Technology Facilities Council (grant number ST/V004735/1). S.G.S acknowledges the support from FCT through the contract nr.CEECIND/00826/2018 and POPH/FSE (EC).

\end{acknowledgements}


\bibliographystyle{aa}

\bibliography{bibliography}

\begin{appendix}
%-------------------------------------------------------------------

\section{SOPHIE RVs and spectroscopic quantities}
\label{app:sophiervs}

This appendix presents a compilation of all quantities derived from the SOPHIE spectra. Tables \ref{tab:sophiervstoi1736} and \ref{tab:sophiervstoi2141} show these quantities obtained for the spectra of TOI-1736 and TOI-2141, respectively.

\onecolumn

\begin{landscape}
\begingroup\small
\begin{longtable}{ccccccccccccccc}
\caption{Log of SOPHIE spectroscopic observations of TOI-1736.} \\
\hline\hline
\label{tab:sophiervstoi1736}
Obs  & Time & RV & $\sigma_{\rm RV}$ & BERV & Exptime & S/N & FWHM & $\sigma_{\rm FWHM}$ & Bis & $\sigma_{\rm Bis}$ & S$_{\rm MW}$ & $\sigma_{\rm S_{\rm MW}}$ & H$_\alpha$ & $\sigma_{H\alpha}$  \\
index & BJD & \kms & \kms & \kms & s & @649~nm & \kms & \kms & \kms & \kms & & & &\\
\hline
\endfirsthead
\caption{continued.}\\
\hline\hline
Obs  & Time & RV & $\sigma_{\rm RV}$ & BERV & Exptime & S/N & FWHM & $\sigma_{\rm FWHM}$ & Bis & $\sigma_{\rm Bis}$ & S$_{\rm MW}$ & $\sigma_{\rm S_{\rm MW}}$ & H$_\alpha$ & $\sigma_{H\alpha}$\\
index & BJD & \kms & \kms & \kms & s & @649~nm & \kms & \kms & \kms & \kms & & & & \\
\hline
\endhead
\hline
\endfoot
\hline
\endlastfoot
1 & 2459081.61496 & -25.2590 & 0.0025 & 18.9856 & 285 & 55 & 8.207 & 0.022 & 0.001 & 0.013 & 0.136 & 0.036 & 0.2481 & 0.0023 \\
2 & 2459084.64339 & -25.2680 & 0.0026 & 19.0918 & 742 & 54 & 8.188 & 0.022 & -0.021 & 0.010 & 0.161 & 0.038 & 0.2405 & 0.0016 \\
3 & 2459085.62038 & -25.2669 & 0.0026 & 19.1373 & 580 & 55 & 8.184 & 0.022 & -0.013 & 0.005 & 0.158 & 0.038 & 0.2411 & 0.0018 \\
4 & 2459106.60980 & -25.2690 & 0.0025 & 18.5376 & 297 & 56 & 8.203 & 0.022 & -0.023 & 0.012 & 0.119 & 0.037 & 0.2425 & 0.0018 \\
5 & 2459107.65934 & -25.2704 & 0.0025 & 18.4075 & 402 & 57 & 8.175 & 0.022 & -0.051 & 0.009 & 0.121 & 0.036 & 0.2484 & 0.0018 \\
6 & 2459113.61605 & -25.2674 & 0.0025 & 17.7880 & 520 & 56 & 8.207 & 0.022 & -0.022 & 0.009 & 0.136 & 0.038 & 0.2443 & 0.0021 \\
7 & 2459138.53335 & -25.2670 & 0.0025 & 13.1954 & 454 & 55 & 8.218 & 0.022 & 0.006 & 0.010 & 0.165 & 0.038 & 0.2364 & 0.0015 \\
8 & 2459140.50963 & -25.2742 & 0.0025 & 12.7191 & 291 & 54 & 8.198 & 0.022 & -0.004 & 0.009 & 0.178 & 0.040 & 0.2418 & 0.0020 \\
9 & 2459141.47077 & -25.2741 & 0.0025 & 12.5008 & 287 & 55 & 8.212 & 0.021 & -0.018 & 0.005 & 0.146 & 0.037 & 0.2294 & 0.0019 \\
10 & 2459147.51262 & -25.2783 & 0.0026 & 10.8427 & 670 & 54 & 8.199 & 0.022 & -0.011 & 0.009 & 0.126 & 0.037 & 0.2375 & 0.0016 \\
11 & 2459149.47203 & -25.2843 & 0.0025 & 10.3192 & 474 & 55 & 8.175 & 0.021 & -0.011 & 0.004 & 0.152 & 0.038 & 0.2377 & 0.0026 \\
12 & 2459150.49656 & -25.2873 & 0.0025 & 10.0069 & 500 & 55 & 8.200 & 0.022 & 0.000 & 0.009 & 0.112 & 0.035 & 0.2420 & 0.0015 \\
13 & 2459151.39709 & -25.2755 & 0.0025 & 9.8144 & 374 & 55 & 8.196 & 0.022 & -0.007 & 0.004 & 0.160 & 0.038 & 0.2299 & 0.0030 \\
14 & 2459152.51351 & -25.2774 & 0.0025 & 9.4063 & 686 & 55 & 8.179 & 0.022 & -0.011 & 0.008 & 0.143 & 0.037 & 0.2420 & 0.0028 \\
15 & 2459168.46094 & -25.2833 & 0.0025 & 4.4145 & 419 & 55 & 8.189 & 0.022 & -0.009 & 0.008 & 0.137 & 0.037 & 0.2323 & 0.0014 \\
16 & 2459170.42527 & -25.2968 & 0.0027 & 3.7804 & 1109 & 51 & 8.171 & 0.022 & -0.024 & 0.010 & 0.137 & 0.037 & 0.2382 & 0.0023 \\
17 & 2459171.48232 & -25.2978 & 0.0025 & 3.3803 & 443 & 54 & 8.189 & 0.021 & -0.021 & 0.006 & 0.165 & 0.039 & 0.2425 & 0.0024 \\
18 & 2459175.52276 & -25.2986 & 0.0025 & 1.9766 & 339 & 55 & 8.191 & 0.022 & -0.015 & 0.007 & 0.154 & 0.038 & 0.2376 & 0.0018 \\
19 & 2459176.59152 & -25.2944 & 0.0025 & 1.5743 & 469 & 56 & 8.200 & 0.022 & -0.015 & 0.004 & 0.125 & 0.037 & 0.2439 & 0.0020 \\
20 & 2459181.44213 & -25.3038 & 0.0026 & 0.0059 & 608 & 54 & 8.204 & 0.021 & -0.023 & 0.004 & 0.143 & 0.038 & 0.2327 & 0.0020 \\
21 & 2459182.44156 & -25.3095 & 0.0025 & -0.3365 & 331 & 54 & 8.184 & 0.023 & -0.010 & 0.008 & 0.130 & 0.037 & 0.2285 & 0.0019 \\
22 & 2459183.46451 & -25.3047 & 0.0025 & -0.7041 & 347 & 55 & 8.187 & 0.022 & -0.013 & 0.018 & 0.161 & 0.039 & 0.2295 & 0.0021 \\
23 & 2459184.47339 & -25.3105 & 0.0025 & -1.0562 & 387 & 55 & 8.193 & 0.022 & -0.017 & 0.014 & 0.170 & 0.039 & 0.2274 & 0.0020 \\
24 & 2459194.42648 & -25.3141 & 0.0025 & -4.4200 & 485 & 55 & 8.203 & 0.022 & -0.022 & 0.005 & 0.135 & 0.037 & 0.2380 & 0.0017 \\
25 & 2459196.32307 & -25.3144 & 0.0025 & -4.9851 & 446 & 55 & 8.207 & 0.022 & -0.020 & 0.010 & 0.167 & 0.038 & 0.2385 & 0.0023 \\
26 & 2459197.47794 & -25.3156 & 0.0026 & -5.4837 & 1439 & 54 & 8.204 & 0.022 & -0.014 & 0.011 & 0.151 & 0.038 & 0.2334 & 0.0022 \\
27 & 2459204.50298 & -25.3392 & 0.0025 & -7.7975 & 326 & 55 & 8.200 & 0.022 & -0.038 & 0.009 & 0.146 & 0.038 & 0.2404 & 0.0017 \\
28 & 2459206.33359 & -25.3390 & 0.0025 & -8.2648 & 416 & 54 & 8.205 & 0.022 & -0.011 & 0.007 & 0.148 & 0.038 & 0.2349 & 0.0020 \\
29 & 2459243.33563 & -25.4015 & 0.0025 & -17.4102 & 616 & 54 & 8.247 & 0.022 & -0.016 & 0.009 & 0.131 & 0.037 & 0.2222 & 0.0018 \\
30 & 2459244.32973 & -25.4063 & 0.0025 & -17.5633 & 840 & 54 & 8.262 & 0.022 & -0.001 & 0.019 & 0.110 & 0.036 & 0.2228 & 0.0026 \\
31 & 2459248.32999 & -25.4119 & 0.0025 & -18.1445 & 352 & 55 & 8.200 & 0.021 & -0.027 & 0.012 & 0.171 & 0.039 & 0.2326 & 0.0023 \\
32 & 2459249.30840 & -25.4196 & 0.0024 & -18.2600 & 379 & 55 & 8.219 & 0.022 & -0.029 & 0.004 & 0.116 & 0.036 & 0.2313 & 0.0019 \\
33 & 2459271.29327 & -25.4768 & 0.0020 & -19.6493 & 538 & 75 & 8.236 & 0.022 & -0.002 & 0.002 & 0.151 & 0.037 & 0.2308 & 0.0015 \\
34 & 2459275.35142 & -25.4938 & 0.0019 & -19.6159 & 876 & 78 & 8.218 & 0.022 & -0.018 & 0.011 & 0.162 & 0.038 & 0.2327 & 0.0011 \\
35 & 2459278.30685 & -25.4954 & 0.0020 & -19.5006 & 1200 & 76 & 8.200 & 0.022 & -0.008 & 0.004 & 0.147 & 0.037 & 0.2410 & 0.0011 \\
36 & 2459279.35578 & -25.5029 & 0.0020 & -19.4690 & 1070 & 81 & 8.224 & 0.022 & -0.014 & 0.004 & 0.177 & 0.040 & 0.2331 & 0.0014 \\
37 & 2459281.28440 & -25.5233 & 0.0020 & -19.3403 & 1070 & 78 & 8.214 & 0.022 & -0.018 & 0.005 & 0.147 & 0.037 & 0.2430 & 0.0010 \\
38 & 2459452.64790 & -25.4850 & 0.0020 & 19.1468 & 779 & 77 & 8.223 & 0.022 & -0.010 & 0.005 & 0.170 & 0.038 & 0.2495 & 0.0014 \\
39 & 2459454.60297 & -25.4832 & 0.0020 & 19.2081 & 1152 & 77 & 8.210 & 0.021 & -0.009 & 0.004 & 0.149 & 0.037 & 0.2387 & 0.0013 \\
40 & 2459455.62707 & -25.4793 & 0.0020 & 19.1985 & 887 & 77 & 8.209 & 0.021 & -0.011 & 0.008 & 0.131 & 0.036 & 0.2494 & 0.0021 \\
41 & 2459456.63136 & -25.4722 & 0.0017 & 19.1969 & 1200 & 89 & 8.219 & 0.022 & -0.004 & 0.001 & 0.159 & 0.037 & 0.2473 & 0.0011 \\
42 & 2459457.57059 & -25.4815 & 0.0020 & 19.2352 & 608 & 77 & 8.224 & 0.022 & -0.023 & 0.007 & 0.112 & 0.035 & 0.2449 & 0.0016 \\
43 & 2459458.55691 & -25.4770 & 0.0019 & 19.2350 & 648 & 77 & 8.225 & 0.022 & -0.012 & 0.005 & 0.140 & 0.037 & 0.2468 & 0.0010 \\
44 & 2459474.63564 & -25.4548 & 0.0020 & 18.2417 & 1774 & 75 & 8.207 & 0.022 & -0.006 & 0.006 & 0.126 & 0.035 & 0.2469 & 0.0011 \\
45 & 2459475.52288 & -25.4480 & 0.0020 & 18.2289 & 668 & 77 & 8.219 & 0.022 & -0.004 & 0.005 & 0.143 & 0.037 & 0.2448 & 0.0012 \\
46 & 2459477.65307 & -25.4509 & 0.0028 & 17.8954 & 1200 & 50 & 8.220 & 0.021 & -0.023 & 0.010 & 0.133 & 0.038 & 0.2434 & 0.0024 \\
47 & 2459482.51629 & -25.4336 & 0.0020 & 17.3625 & 1200 & 73 & 8.199 & 0.021 & -0.023 & 0.005 & 0.159 & 0.037 & 0.2405 & 0.0017 \\
48 & 2459484.52336 & -25.4346 & 0.0020 & 17.0625 & 642 & 76 & 8.220 & 0.022 & -0.016 & 0.011 & 0.145 & 0.037 & 0.2459 & 0.0013 \\
49 & 2459499.58244 & -25.4157 & 0.0020 & 14.1379 & 989 & 76 & 8.225 & 0.021 & -0.021 & 0.006 & 0.136 & 0.036 & 0.2417 & 0.0015 \\
50 & 2459501.50241 & -25.4253 & 0.0020 & 13.7513 & 797 & 77 & 8.221 & 0.022 & -0.012 & 0.005 & 0.148 & 0.037 & 0.2384 & 0.0014 \\
51 & 2459502.45601 & -25.4197 & 0.0019 & 13.5564 & 735 & 77 & 8.206 & 0.022 & -0.010 & 0.005 & 0.142 & 0.037 & 0.2400 & 0.0010 \\
52 & 2459503.56633 & -25.4133 & 0.0020 & 13.2085 & 922 & 76 & 8.209 & 0.022 & -0.011 & 0.006 & 0.138 & 0.037 & 0.2450 & 0.0020 \\
53 & 2459504.46508 & -25.4124 & 0.0020 & 13.0616 & 558 & 76 & 8.200 & 0.022 & -0.008 & 0.010 & 0.135 & 0.036 & 0.2388 & 0.0012 \\
54 & 2459505.49056 & -25.4123 & 0.0020 & 12.7884 & 645 & 76 & 8.217 & 0.022 & -0.017 & 0.009 & 0.141 & 0.037 & 0.2434 & 0.0018 \\
55 & 2459506.51220 & -25.4127 & 0.0020 & 12.5142 & 550 & 75 & 8.208 & 0.022 & -0.010 & 0.006 & 0.152 & 0.037 & 0.2381 & 0.0012 \\
56 & 2459509.54265 & -25.4150 & 0.0020 & 11.7028 & 1103 & 76 & 8.214 & 0.021 & -0.006 & 0.008 & 0.149 & 0.037 & 0.2448 & 0.0014 \\
57 & 2459510.48874 & -25.4116 & 0.0020 & 11.4901 & 726 & 76 & 8.227 & 0.022 & -0.005 & 0.005 & 0.174 & 0.039 & 0.2385 & 0.0014 \\
58 & 2459511.49134 & -25.4008 & 0.0020 & 11.2166 & 564 & 75 & 8.208 & 0.022 & -0.014 & 0.004 & 0.179 & 0.039 & 0.2468 & 0.0012 \\
59 & 2459522.48363 & -25.3972 & 0.0020 & 8.0144 & 1200 & 75 & 8.220 & 0.023 & -0.012 & 0.004 & 0.157 & 0.038 & 0.2406 & 0.0009 \\
60 & 2459523.48010 & -25.3980 & 0.0020 & 7.7062 & 907 & 76 & 8.209 & 0.022 & -0.011 & 0.004 & 0.161 & 0.037 & 0.2343 & 0.0012 \\
61 & 2459524.44250 & -25.3982 & 0.0020 & 7.4312 & 605 & 76 & 8.206 & 0.021 & -0.016 & 0.003 & 0.155 & 0.037 & 0.2387 & 0.0008 \\
62 & 2459525.51092 & -25.3956 & 0.0020 & 7.0413 & 881 & 77 & 8.208 & 0.022 & -0.012 & 0.003 & 0.148 & 0.036 & 0.2403 & 0.0015 \\
63 & 2459526.61447 & -25.3908 & 0.0021 & 6.6227 & 1200 & 72 & 8.237 & 0.022 & -0.025 & 0.008 & 0.151 & 0.037 & 0.2409 & 0.0011 \\
64 & 2459531.33958 & -25.3980 & 0.0020 & 5.2700 & 1200 & 74 & 8.218 & 0.022 & -0.013 & 0.006 & 0.117 & 0.035 & 0.2436 & 0.0010 \\
65 & 2459532.40862 & -25.3895 & 0.0020 & 4.8744 & 953 & 76 & 8.221 & 0.021 & -0.028 & 0.010 & 0.166 & 0.038 & 0.2406 & 0.0012 \\
66 & 2459534.62096 & -25.3875 & 0.0019 & 4.0030 & 774 & 78 & 8.220 & 0.022 & -0.015 & 0.008 & 0.150 & 0.037 & 0.2411 & 0.0010 \\
67 & 2459542.56269 & -25.3843 & 0.0020 & 1.3556 & 907 & 77 & 8.221 & 0.022 & -0.007 & 0.004 & 0.151 & 0.037 & 0.2393 & 0.0013 \\
68 & 2459553.56087 & -25.3816 & 0.0021 & -2.4250 & 1200 & 72 & 8.242 & 0.022 & -0.024 & 0.011 & 0.151 & 0.037 & 0.2332 & 0.0017 \\
69 & 2459554.55368 & -25.3725 & 0.0020 & -2.7635 & 1078 & 78 & 8.206 & 0.023 & -0.010 & 0.005 & 0.118 & 0.036 & 0.2402 & 0.0018 \\
70 & 2459558.39681 & -25.3946 & 0.0026 & -3.9773 & 1464 & 55 & 8.224 & 0.022 & -0.028 & 0.013 & 0.137 & 0.037 & 0.2301 & 0.0018 \\
71 & 2459559.45330 & -25.3822 & 0.0028 & -4.3759 & 1400 & 52 & 8.227 & 0.022 & -0.027 & 0.014 & 0.128 & 0.037 & 0.2296 & 0.0027 \\
72 & 2459560.33748 & -25.3803 & 0.0025 & -4.5890 & 1200 & 58 & 8.212 & 0.022 & -0.013 & 0.004 & 0.122 & 0.036 & 0.2316 & 0.0018 \\
73 & 2459561.42902 & -25.3737 & 0.0020 & -5.0226 & 922 & 76 & 8.216 & 0.022 & -0.013 & 0.008 & 0.161 & 0.038 & 0.2367 & 0.0014 \\
74 & 2459562.35541 & -25.3732 & 0.0020 & -5.2770 & 608 & 76 & 8.210 & 0.022 & -0.016 & 0.004 & 0.162 & 0.038 & 0.2303 & 0.0018 \\
75 & 2459563.37261 & -25.3767 & 0.0020 & -5.6274 & 552 & 76 & 8.227 & 0.022 & -0.024 & 0.004 & 0.151 & 0.037 & 0.2371 & 0.0013 \\
76 & 2459564.38234 & -25.3772 & 0.0020 & -5.9678 & 813 & 76 & 8.207 & 0.022 & -0.007 & 0.006 & 0.160 & 0.038 & 0.2349 & 0.0011 \\
77 & 2459565.39268 & -25.3700 & 0.0020 & -6.3070 & 694 & 76 & 8.235 & 0.022 & -0.014 & 0.010 & 0.163 & 0.038 & 0.2356 & 0.0014 \\
78 & 2459566.36453 & -25.3681 & 0.0020 & -6.6029 & 663 & 76 & 8.221 & 0.022 & -0.017 & 0.004 & 0.173 & 0.039 & 0.2339 & 0.0012 \\
79 & 2459567.33342 & -25.3621 & 0.0020 & -6.8939 & 550 & 75 & 8.206 & 0.022 & -0.012 & 0.006 & 0.185 & 0.039 & 0.2316 & 0.0015 \\
80 & 2459568.47918 & -25.3656 & 0.0020 & -7.3645 & 819 & 77 & 8.214 & 0.022 & -0.003 & 0.013 & 0.171 & 0.039 & 0.2310 & 0.0012 \\
81 & 2459569.29039 & -25.3607 & 0.0020 & -7.4920 & 621 & 76 & 8.215 & 0.021 & -0.009 & 0.004 & 0.166 & 0.038 & 0.2286 & 0.0011 \\
82 & 2459570.30873 & -25.3725 & 0.0026 & -7.8287 & 726 & 55 & 8.197 & 0.023 & -0.010 & 0.009 & 0.130 & 0.038 & 0.2286 & 0.0033 \\
83 & 2459571.34632 & -25.3783 & 0.0023 & -8.1841 & 2005 & 63 & 8.215 & 0.023 & -0.008 & 0.009 & 0.170 & 0.039 & 0.2366 & 0.0012 \\
84 & 2459586.35956 & -25.3680 & 0.0020 & -12.6133 & 1104 & 77 & 8.210 & 0.022 & -0.004 & 0.004 & 0.172 & 0.039 & 0.2304 & 0.0012 \\
85 & 2459587.34825 & -25.3744 & 0.0020 & -12.8673 & 1200 & 74 & 8.221 & 0.022 & -0.019 & 0.008 & 0.170 & 0.038 & 0.2307 & 0.0011 \\
86 & 2459590.24201 & -25.3571 & 0.0027 & -13.5317 & 1200 & 53 & 8.236 & 0.021 & -0.009 & 0.007 & 0.108 & 0.036 & 0.2341 & 0.0027 \\
87 & 2459591.28806 & -25.3571 & 0.0020 & -13.8243 & 792 & 76 & 8.213 & 0.022 & -0.006 & 0.007 & 0.159 & 0.037 & 0.2318 & 0.0013 \\
88 & 2459592.29227 & -25.3603 & 0.0020 & -14.0717 & 572 & 76 & 8.230 & 0.022 & -0.007 & 0.008 & 0.160 & 0.037 & 0.2237 & 0.0013 \\
89 & 2459603.37588 & -25.3615 & 0.0015 & -16.5208 & 1200 & 102 & 8.217 & 0.021 & -0.023 & 0.006 & 0.150 & 0.036 & 0.2413 & 0.0008 \\
90 & 2459604.29341 & -25.3626 & 0.0020 & -16.6379 & 618 & 76 & 8.213 & 0.021 & -0.020 & 0.009 & 0.148 & 0.037 & 0.2322 & 0.0010 \\
91 & 2459605.38067 & -25.3668 & 0.0019 & -16.8944 & 786 & 78 & 8.213 & 0.022 & -0.007 & 0.007 & 0.138 & 0.036 & 0.2356 & 0.0009 \\
92 & 2459606.29699 & -25.3671 & 0.0020 & -17.0033 & 677 & 76 & 8.218 & 0.022 & -0.015 & 0.002 & 0.148 & 0.037 & 0.2290 & 0.0012 \\
93 & 2459607.40746 & -25.3679 & 0.0019 & -17.2618 & 1194 & 78 & 8.221 & 0.022 & -0.019 & 0.004 & 0.124 & 0.036 & 0.2370 & 0.0011 \\
94 & 2459608.37671 & -25.3673 & 0.0020 & -17.4088 & 1200 & 75 & 8.206 & 0.022 & -0.011 & 0.002 & 0.159 & 0.038 & 0.2316 & 0.0014 \\
95 & 2459609.34453 & -25.3621 & 0.0020 & -17.5469 & 1200 & 76 & 8.211 & 0.021 & -0.020 & 0.006 & 0.164 & 0.038 & 0.2315 & 0.0010 \\
96 & 2459610.33301 & -25.3635 & 0.0020 & -17.6943 & 1014 & 77 & 8.219 & 0.022 & -0.016 & 0.009 & 0.144 & 0.037 & 0.2315 & 0.0010 \\
97 & 2459621.28804 & -25.3594 & 0.0020 & -18.9846 & 832 & 77 & 8.218 & 0.022 & -0.015 & 0.004 & 0.151 & 0.037 & 0.2376 & 0.0011 \\
98 & 2459622.37203 & -25.3657 & 0.0019 & -19.1223 & 1200 & 77 & 8.222 & 0.022 & -0.017 & 0.009 & 0.153 & 0.037 & 0.2313 & 0.0014 \\
99 & 2459629.35554 & -25.3733 & 0.0019 & -19.5399 & 942 & 78 & 8.210 & 0.022 & -0.016 & 0.007 & 0.161 & 0.038 & 0.2354 & 0.0015 \\
100 & 2459630.39834 & -25.3525 & 0.0028 & -19.5922 & 1500 & 52 & 8.233 & 0.022 & -0.009 & 0.010 & 0.136 & 0.038 & 0.2373 & 0.0021 \\
101 & 2459646.28753 & -25.3520 & 0.0022 & -19.3589 & 800 & 68 & 8.225 & 0.022 & -0.004 & 0.006 & 0.138 & 0.038 & 0.2323 & 0.0024 \\
102 & 2459660.29145 & -25.3460 & 0.0027 & -17.9657 & 1200 & 56 & 8.172 & 0.021 & -0.018 & 0.007 & 0.120 & 0.037 & 0.2341 & 0.0018 \\
103 & 2459661.29920 & -25.3413 & 0.0023 & -17.8286 & 1200 & 66 & 8.242 & 0.024 & -0.015 & 0.007 & 0.135 & 0.037 & 0.2269 & 0.0020 \\
104 & 2459807.61069 & -25.4909 & 0.0020 & 18.6878 & 731 & 77 & 8.228 & 0.021 & -0.003 & 0.006 & 0.148 & 0.038 & 0.2524 & 0.0011 \\
105 & 2459811.63296 & -25.4946 & 0.0020 & 18.9366 & 1121 & 76 & 8.235 & 0.022 & -0.007 & 0.005 & 0.141 & 0.037 & 0.2449 & 0.0013 \\
106 & 2459813.61785 & -25.5036 & 0.0020 & 19.0453 & 964 & 76 & 8.240 & 0.022 & -0.010 & 0.007 & 0.153 & 0.037 & 0.2476 & 0.0013 \\
107 & 2459814.60411 & -25.5016 & 0.0020 & 19.0954 & 969 & 76 & 8.219 & 0.022 & -0.016 & 0.011 & 0.140 & 0.036 & 0.2401 & 0.0012 \\
108 & 2459815.59752 & -25.4982 & 0.0020 & 19.1353 & 690 & 76 & 8.239 & 0.022 & -0.014 & 0.004 & 0.146 & 0.037 & 0.2481 & 0.0016 \\
109 & 2459816.57868 & -25.5116 & 0.0023 & 19.1769 & 1200 & 62 & 8.240 & 0.021 & -0.009 & 0.005 & 0.141 & 0.038 & 0.2474 & 0.0022 \\
110 & 2459824.60025 & -25.5284 & 0.0020 & 19.1951 & 910 & 77 & 8.244 & 0.022 & -0.003 & 0.002 & 0.154 & 0.037 & 0.2449 & 0.0008 \\
111 & 2459827.52971 & -25.5468 & 0.0020 & 19.1583 & 1012 & 77 & 8.240 & 0.022 & -0.016 & 0.002 & 0.159 & 0.038 & 0.2516 & 0.0010 \\
112 & 2459837.66052 & -25.5659 & 0.0020 & 18.4402 & 1200 & 74 & 8.237 & 0.021 & -0.012 & 0.008 & 0.155 & 0.038 & 0.2486 & 0.0013 \\
113 & 2459840.63718 & -25.5798 & 0.0020 & 18.1713 & 1069 & 76 & 8.241 & 0.021 & -0.012 & 0.008 & 0.156 & 0.037 & 0.2454 & 0.0011 \\
114 & 2459860.50050 & -25.6567 & 0.0020 & 15.1320 & 649 & 77 & 8.233 & 0.021 & -0.012 & 0.003 & 0.164 & 0.038 & 0.2415 & 0.0013 \\
115 & 2459863.48808 & -25.6716 & 0.0020 & 14.5056 & 816 & 77 & 8.245 & 0.021 & -0.017 & 0.011 & 0.149 & 0.037 & 0.2464 & 0.0011 \\
116 & 2459865.58128 & -25.6716 & 0.0020 & 13.9698 & 644 & 76 & 8.228 & 0.021 & -0.011 & 0.007 & 0.148 & 0.037 & 0.2445 & 0.0017 \\
117 & 2459880.48259 & -25.7369 & 0.0020 & 10.1756 & 694 & 75 & 8.229 & 0.022 & -0.020 & 0.006 & 0.143 & 0.036 & 0.2407 & 0.0013 \\
118 & 2459881.42661 & -25.7360 & 0.0020 & 9.9406 & 584 & 75 & 8.253 & 0.022 & -0.010 & 0.012 & 0.151 & 0.037 & 0.2498 & 0.0018 \\
119 & 2459882.45022 & -25.7379 & 0.0024 & 9.6238 & 1200 & 59 & 8.239 & 0.022 & -0.002 & 0.006 & 0.146 & 0.037 & 0.2439 & 0.0017 \\
120 & 2459889.60802 & -25.7662 & 0.0020 & 7.3360 & 973 & 78 & 8.246 & 0.021 & -0.009 & 0.005 & 0.142 & 0.037 & 0.2407 & 0.0011 \\
121 & 2459890.51346 & -25.7713 & 0.0020 & 7.1139 & 718 & 76 & 8.241 & 0.021 & -0.010 & 0.009 & 0.131 & 0.036 & 0.2347 & 0.0011 \\
122 & 2459891.54302 & -25.7716 & 0.0021 & 6.7679 & 1404 & 74 & 8.216 & 0.021 & -0.003 & 0.003 & 0.140 & 0.036 & 0.2346 & 0.0013 \\
123 & 2459892.43560 & -25.7764 & 0.0029 & 6.5632 & 3009 & 55 & 8.079 & 0.056 & -0.032 & 0.007 & 0.061 & 0.035 & 0.2196 & 0.0021 \\
124 & 2459893.62259 & -25.7684 & 0.0020 & 6.0597 & 1031 & 77 & 8.255 & 0.022 & -0.006 & 0.003 & 0.143 & 0.037 & 0.2417 & 0.0014 \\
125 & 2459894.38557 & -25.7710 & 0.0020 & 5.9727 & 577 & 76 & 8.237 & 0.021 & -0.015 & 0.003 & 0.174 & 0.038 & 0.2448 & 0.0011 \\
126 & 2459915.33130 & -25.8054 & 0.0024 & -1.0811 & 1200 & 60 & 8.267 & 0.021 & -0.002 & 0.005 & 0.118 & 0.036 & 0.2392 & 0.0020 \\
127 & 2459916.38555 & -25.7964 & 0.0020 & -1.4813 & 816 & 77 & 8.239 & 0.022 & -0.010 & 0.003 & 0.185 & 0.040 & 0.2387 & 0.0014 \\
128 & 2459918.52501 & -25.8041 & 0.0022 & -2.3102 & 1200 & 68 & 8.250 & 0.022 & -0.019 & 0.003 & 0.153 & 0.037 & 0.2329 & 0.0017 \\
129 & 2459919.40221 & -25.8110 & 0.0020 & -2.5259 & 1070 & 76 & 8.246 & 0.022 & 0.006 & 0.006 & 0.171 & 0.038 & 0.2323 & 0.0008 \\
130 & 2459920.39359 & -25.8000 & 0.0020 & -2.8572 & 559 & 76 & 8.253 & 0.022 & -0.006 & 0.003 & 0.163 & 0.037 & 0.2396 & 0.0013 \\
131 & 2459930.60199 & -25.8004 & 0.0024 & -6.3835 & 1200 & 63 & 8.262 & 0.021 & -0.011 & 0.007 & 0.124 & 0.036 & 0.2283 & 0.0022 \\
132 & 2459931.57908 & -25.7913 & 0.0020 & -6.6998 & 1105 & 79 & 8.254 & 0.022 & -0.010 & 0.002 & 0.169 & 0.038 & 0.2385 & 0.0009 \\
133 & 2459935.35577 & -25.7827 & 0.0020 & -7.8008 & 888 & 77 & 8.255 & 0.022 & -0.005 & 0.005 & 0.163 & 0.038 & 0.2404 & 0.0010 \\
134 & 2459936.30712 & -25.7728 & 0.0021 & -8.0692 & 1200 & 73 & 8.247 & 0.022 & 0.001 & 0.006 & 0.181 & 0.039 & 0.2351 & 0.0013 \\
135 & 2459937.32833 & -25.7762 & 0.0020 & -8.4086 & 841 & 76 & 8.245 & 0.022 & -0.004 & 0.007 & 0.191 & 0.040 & 0.2405 & 0.0014 \\
136 & 2459972.34432 & -25.6976 & 0.0026 & -17.1785 & 383 & 55 & 8.246 & 0.022 & -0.003 & 0.013 & 0.104 & 0.035 & 0.2304 & 0.0019 \\
137 & 2459972.35432 & -25.6906 & 0.0020 & -17.1864 & 796 & 78 & 8.243 & 0.022 & -0.016 & 0.010 & 0.150 & 0.037 & 0.2307 & 0.0012 \\
138 & 2459973.40495 & -25.6918 & 0.0020 & -17.3862 & 1003 & 79 & 8.255 & 0.022 & -0.018 & 0.006 & 0.147 & 0.037 & 0.2362 & 0.0014 \\
139 & 2459976.30450 & -25.7010 & 0.0019 & -17.7744 & 830 & 77 & 8.246 & 0.022 & -0.003 & 0.003 & 0.154 & 0.037 & 0.2374 & 0.0012 \\
140 & 2459978.32918 & -25.6730 & 0.0020 & -18.0758 & 762 & 77 & 8.245 & 0.022 & -0.007 & 0.004 & 0.164 & 0.037 & 0.2364 & 0.0009 \\
141 & 2459979.29409 & -25.6798 & 0.0019 & -18.1803 & 1045 & 76 & 8.259 & 0.022 & -0.020 & 0.006 & 0.172 & 0.038 & 0.2351 & 0.0010 \\
142 & 2459980.29695 & -25.6767 & 0.0020 & -18.3100 & 1710 & 72 & 8.271 & 0.022 & -0.010 & 0.003 & 0.177 & 0.038 & 0.2304 & 0.0016 \\
143 & 2459983.32323 & -25.6754 & 0.0023 & -18.6793 & 1200 & 61 & 8.263 & 0.022 & -0.023 & 0.005 & 0.167 & 0.039 & 0.2357 & 0.0017 \\
144 & 2459989.33814 & -25.6651 & 0.0020 & -19.2376 & 635 & 78 & 8.238 & 0.022 & -0.021 & 0.005 & 0.172 & 0.039 & 0.2402 & 0.0015 \\
145 & 2459990.39263 & -25.6534 & 0.0019 & -19.3346 & 649 & 77 & 8.232 & 0.022 & -0.013 & 0.005 & 0.173 & 0.038 & 0.2378 & 0.0011 \\
146 & 2459991.41345 & -25.6538 & 0.0020 & -19.4053 & 1200 & 77 & 8.209 & 0.023 & -0.013 & 0.011 & 0.085 & 0.035 & 0.2276 & 0.0010 \\
147 & 2460006.27843 & -25.6171 & 0.0019 & -19.5753 & 982 & 79 & 8.272 & 0.021 & -0.010 & 0.011 & 0.169 & 0.038 & 0.2381 & 0.0012 \\
148 & 2460007.27996 & -25.6202 & 0.0026 & -19.5434 & 1200 & 54 & 8.321 & 0.022 & 0.032 & 0.012 & 0.126 & 0.037 & 0.2346 & 0.0024 \\
149 & 2460008.31062 & -25.6127 & 0.0020 & -19.5185 & 1200 & 77 & 8.255 & 0.022 & -0.007 & 0.011 & 0.179 & 0.039 & 0.2366 & 0.0010 \\
150 & 2460010.33038 & -25.6127 & 0.0019 & -19.4302 & 1083 & 78 & 8.272 & 0.022 & -0.004 & 0.004 & 0.174 & 0.039 & 0.2266 & 0.0014 \\
151 & 2460013.27481 & -25.6030 & 0.0020 & -19.2289 & 699 & 78 & 8.247 & 0.022 & -0.014 & 0.003 & 0.158 & 0.038 & 0.2356 & 0.0013 \\
152 & 2460015.28434 & -25.5966 & 0.0019 & -19.0851 & 868 & 78 & 8.242 & 0.022 & -0.006 & 0.005 & 0.171 & 0.038 & 0.2438 & 0.0010 \\
\hline
\end{longtable}
\endgroup

\begingroup\small
\begin{longtable}{ccccccccccccccccc}
\caption{Log of SOPHIE spectroscopic observations of TOI-2141.} \\
\hline\hline
\label{tab:sophiervstoi2141}
Obs  & Time & RV & $\sigma_{\rm RV}$ & BERV & Exptime & S/N & FWHM & $\sigma_{\rm FWHM}$ & Bis & $\sigma_{\rm Bis}$ & S$_{\rm MW}$ & $\sigma_{\rm S_{\rm MW}}$ & H$_\alpha$ & $\sigma_{H\alpha}$  \\
index & BJD & \kms & \kms & \kms & s & @649~nm & \kms & \kms & \kms & \kms & & & &\\
\hline
\endfirsthead
\caption{continued.}\\
\hline\hline
Obs  & Time & RV & $\sigma_{\rm RV}$ & BERV & Exptime & S/N & FWHM & $\sigma_{\rm FWHM}$ & Bis & $\sigma_{\rm Bis}$ & S$_{\rm MW}$ & $\sigma_{\rm S_{\rm MW}}$ & H$_\alpha$ & $\sigma_{H\alpha}$\\
index & BJD & \kms & \kms & \kms & s & @649~nm & \kms & \kms & \kms & \kms & & & & \\
\hline
\endhead
\hline
\endfoot
\hline
\endlastfoot
1 & 2459270.70771 & -19.8722 & 0.0025 & 22.4143 & 627 & 55 & 7.521 & 0.023 & -0.016 & 0.007 & 0.165 & 0.041 & 0.2505 & 0.0018 \\
2 & 2459278.65511 & -19.8427 & 0.0030 & 22.7614 & 1477 & 47 & 7.484 & 0.022 & -0.031 & 0.011 & 0.149 & 0.040 & 0.2491 & 0.0031 \\
3 & 2459328.63428 & -19.8602 & 0.0024 & 15.0975 & 861 & 59 & 7.501 & 0.024 & -0.017 & 0.012 & 0.211 & 0.042 & 0.2514 & 0.0019 \\
4 & 2459336.53217 & -19.8474 & 0.0035 & 12.9076 & 1800 & 38 & 7.509 & 0.024 & 0.015 & 0.011 & 0.198 & 0.044 & 0.2498 & 0.0034 \\
5 & 2459340.58941 & -19.8611 & 0.0025 & 11.4775 & 902 & 55 & 7.518 & 0.023 & -0.034 & 0.014 & 0.237 & 0.044 & 0.2501 & 0.0015 \\
6 & 2459348.48389 & -19.8607 & 0.0028 & 8.9598 & 1800 & 48 & 7.505 & 0.023 & -0.025 & 0.004 & 0.262 & 0.047 & 0.2421 & 0.0018 \\
7 & 2459360.43069 & -19.8593 & 0.0025 & 4.6977 & 1265 & 56 & 7.504 & 0.023 & -0.027 & 0.005 & 0.185 & 0.040 & 0.2338 & 0.0018 \\
8 & 2459362.50511 & -19.8664 & 0.0025 & 3.7888 & 593 & 55 & 7.539 & 0.024 & -0.027 & 0.006 & 0.206 & 0.041 & 0.2459 & 0.0018 \\
9 & 2459391.50390 & -19.8607 & 0.0025 & -7.0632 & 905 & 56 & 7.506 & 0.024 & -0.031 & 0.008 & 0.179 & 0.040 & 0.2487 & 0.0017 \\
10 & 2459392.47824 & -19.8545 & 0.0025 & -7.3612 & 1187 & 55 & 7.531 & 0.025 & -0.013 & 0.006 & 0.184 & 0.040 & 0.2520 & 0.0014 \\
11 & 2459395.49367 & -19.8672 & 0.0025 & -8.4551 & 1738 & 55 & 7.519 & 0.023 & -0.024 & 0.014 & 0.181 & 0.039 & 0.2445 & 0.0019 \\
12 & 2459396.51007 & -19.8612 & 0.0025 & -8.8397 & 1200 & 56 & 7.511 & 0.023 & -0.018 & 0.004 & 0.228 & 0.044 & 0.2457 & 0.0020 \\
13 & 2459403.50659 & -19.8557 & 0.0025 & -11.2073 & 892 & 56 & 7.518 & 0.023 & -0.009 & 0.008 & 0.184 & 0.043 & 0.2481 & 0.0018 \\
14 & 2459406.41548 & -19.8538 & 0.0025 & -11.9824 & 1007 & 57 & 7.514 & 0.024 & -0.023 & 0.014 & 0.182 & 0.042 & 0.2488 & 0.0015 \\
15 & 2459417.37289 & -19.8645 & 0.0025 & -15.1647 & 718 & 57 & 7.533 & 0.022 & -0.008 & 0.008 & 0.143 & 0.038 & 0.2514 & 0.0016 \\
16 & 2459420.46780 & -19.8599 & 0.0023 & -16.1737 & 1170 & 58 & 7.533 & 0.024 & -0.031 & 0.007 & 0.187 & 0.041 & 0.2497 & 0.0016 \\
17 & 2459421.40228 & -19.8594 & 0.0025 & -16.2998 & 766 & 57 & 7.516 & 0.023 & -0.024 & 0.009 & 0.192 & 0.041 & 0.2582 & 0.0019 \\
18 & 2459423.40030 & -19.8619 & 0.0025 & -16.8058 & 1800 & 55 & 7.523 & 0.024 & -0.011 & 0.010 & 0.122 & 0.037 & 0.2510 & 0.0025 \\
19 & 2459439.42347 & -19.8689 & 0.0024 & -20.2294 & 871 & 58 & 7.525 & 0.023 & -0.031 & 0.017 & 0.220 & 0.043 & 0.2445 & 0.0016 \\
20 & 2459441.39841 & -19.8609 & 0.0024 & -20.5047 & 717 & 56 & 7.518 & 0.023 & -0.039 & 0.009 & 0.188 & 0.040 & 0.2419 & 0.0027 \\
21 & 2459445.38616 & -19.8470 & 0.0023 & -21.0516 & 725 & 57 & 7.521 & 0.023 & -0.026 & 0.005 & 0.230 & 0.044 & 0.2443 & 0.0014 \\
22 & 2459454.38684 & -19.8721 & 0.0025 & -22.0027 & 1641 & 57 & 7.506 & 0.024 & -0.021 & 0.008 & 0.195 & 0.044 & 0.2420 & 0.0017 \\
23 & 2459456.33576 & -19.8674 & 0.0024 & -22.0624 & 808 & 56 & 7.513 & 0.023 & -0.023 & 0.004 & 0.211 & 0.043 & 0.2492 & 0.0022 \\
24 & 2459469.30862 & -19.8606 & 0.0024 & -22.3470 & 616 & 57 & 7.511 & 0.024 & -0.036 & 0.007 & 0.161 & 0.039 & 0.2425 & 0.0012 \\
25 & 2459470.30193 & -19.8572 & 0.0024 & -22.3135 & 893 & 57 & 7.523 & 0.024 & -0.021 & 0.006 & 0.106 & 0.036 & 0.2424 & 0.0021 \\
26 & 2459475.29197 & -19.8675 & 0.0024 & -22.0924 & 611 & 57 & 7.531 & 0.023 & -0.025 & 0.012 & 0.173 & 0.044 & 0.2533 & 0.0026 \\
27 & 2459477.30701 & -19.8509 & 0.0024 & -21.9923 & 1244 & 56 & 7.526 & 0.024 & -0.025 & 0.006 & 0.190 & 0.040 & 0.2454 & 0.0020 \\
28 & 2459482.27556 & -19.8580 & 0.0024 & -21.5264 & 623 & 56 & 7.524 & 0.023 & -0.025 & 0.008 & 0.191 & 0.042 & 0.2522 & 0.0027 \\
29 & 2459484.30844 & -19.8627 & 0.0033 & -21.3654 & 1800 & 44 & 7.522 & 0.024 & -0.046 & 0.010 & 0.147 & 0.042 & 0.2543 & 0.0031 \\
30 & 2459486.28811 & -19.8592 & 0.0028 & -21.1018 & 1800 & 50 & 7.499 & 0.024 & -0.029 & 0.007 & 0.168 & 0.043 & 0.2502 & 0.0034 \\
31 & 2459501.25879 & -19.8522 & 0.0023 & -18.4803 & 1013 & 57 & 7.533 & 0.023 & -0.019 & 0.004 & 0.173 & 0.042 & 0.2553 & 0.0023 \\
32 & 2459502.26110 & -19.8613 & 0.0023 & -18.2632 & 833 & 59 & 7.524 & 0.023 & -0.022 & 0.011 & 0.128 & 0.039 & 0.2521 & 0.0020 \\
33 & 2459504.25263 & -19.8529 & 0.0024 & -17.7981 & 608 & 59 & 7.508 & 0.023 & -0.004 & 0.007 & 0.146 & 0.042 & 0.2521 & 0.0022 \\
34 & 2459505.25776 & -19.8644 & 0.0024 & -17.5690 & 728 & 59 & 7.518 & 0.022 & -0.008 & 0.004 & 0.181 & 0.042 & 0.2460 & 0.0020 \\
35 & 2459509.23883 & -19.8622 & 0.0024 & -16.5576 & 996 & 57 & 7.545 & 0.023 & -0.019 & 0.011 & 0.223 & 0.044 & 0.2582 & 0.0016 \\
36 & 2459603.70734 & -19.8669 & 0.0018 & 17.1155 & 1800 & 82 & 7.520 & 0.024 & -0.021 & 0.013 & 0.102 & 0.039 & 0.2478 & 0.0012 \\
37 & 2459605.71392 & -19.8553 & 0.0023 & 17.6262 & 1057 & 61 & 7.527 & 0.023 & 0.002 & 0.007 & 0.130 & 0.041 & 0.2464 & 0.0018 \\
38 & 2459606.70224 & -19.8473 & 0.0023 & 17.8886 & 1171 & 60 & 7.501 & 0.024 & -0.013 & 0.005 & 0.104 & 0.039 & 0.2470 & 0.0020 \\
39 & 2459607.71998 & -19.8574 & 0.0032 & 18.1152 & 1800 & 44 & 7.499 & 0.024 & -0.006 & 0.013 & 0.241 & 0.048 & 0.2425 & 0.0036 \\
40 & 2459609.71220 & -19.8589 & 0.0031 & 18.5972 & 1994 & 45 & 7.497 & 0.024 & -0.039 & 0.007 & 0.192 & 0.044 & 0.2494 & 0.0032 \\
41 & 2459610.71164 & -19.8459 & 0.0030 & 18.8253 & 1843 & 47 & 7.517 & 0.024 & -0.035 & 0.007 & 0.167 & 0.043 & 0.2562 & 0.0033 \\
42 & 2459620.67858 & -19.8546 & 0.0024 & 20.7845 & 1460 & 58 & 7.524 & 0.023 & -0.027 & 0.005 & 0.210 & 0.044 & 0.2443 & 0.0014 \\
43 & 2459623.69925 & -19.8662 & 0.0024 & 21.2066 & 764 & 58 & 7.508 & 0.023 & -0.028 & 0.013 & 0.122 & 0.040 & 0.2466 & 0.0022 \\
44 & 2459628.65568 & -19.8537 & 0.0024 & 21.8874 & 1190 & 59 & 7.479 & 0.022 & -0.020 & 0.019 & 0.229 & 0.044 & 0.2443 & 0.0019 \\
45 & 2459660.62852 & -19.8513 & 0.0018 & 21.9324 & 1800 & 83 & 7.523 & 0.023 & -0.036 & 0.009 & 0.176 & 0.039 & 0.2559 & 0.0010 \\
46 & 2459663.59201 & -19.8543 & 0.0024 & 21.6532 & 795 & 58 & 7.514 & 0.023 & -0.014 & 0.008 & 0.144 & 0.042 & 0.2498 & 0.0015 \\
47 & 2459683.57818 & -19.8584 & 0.0024 & 17.9520 & 612 & 56 & 7.525 & 0.022 & -0.022 & 0.008 & 0.134 & 0.040 & 0.2543 & 0.0047 \\
48 & 2459685.55853 & -19.8629 & 0.0025 & 17.5017 & 886 & 56 & 7.536 & 0.022 & -0.014 & 0.008 & 0.244 & 0.047 & 0.2533 & 0.0025 \\
49 & 2459687.55852 & -19.8596 & 0.0033 & 16.9941 & 900 & 42 & 7.507 & 0.023 & -0.017 & 0.014 & 0.121 & 0.042 & 0.2522 & 0.0043 \\
50 & 2459714.54196 & -19.8464 & 0.0037 & 8.5630 & 1800 & 39 & 7.473 & 0.027 & -0.025 & 0.021 & 0.111 & 0.042 & 0.2357 & 0.0037 \\
51 & 2459715.47570 & -19.8543 & 0.0025 & 8.3573 & 1419 & 57 & 7.493 & 0.024 & -0.007 & 0.010 & 0.129 & 0.040 & 0.2445 & 0.0016 \\
52 & 2459716.44585 & -19.8444 & 0.0025 & 8.0616 & 1058 & 57 & 7.505 & 0.023 & -0.012 & 0.007 & 0.132 & 0.038 & 0.2434 & 0.0018 \\
53 & 2459726.52254 & -19.8428 & 0.0083 & 4.2235 & 1800 & 33 & 7.577 & 0.028 & -0.017 & 0.011 & 0.195 & 0.043 & 0.2291 & 0.0029 \\
54 & 2459727.48916 & -19.8643 & 0.0025 & 3.9254 & 1501 & 56 & 7.510 & 0.023 & -0.006 & 0.011 & 0.181 & 0.045 & 0.2450 & 0.0014 \\
55 & 2459728.45605 & -19.8581 & 0.0027 & 3.6230 & 1800 & 51 & 7.533 & 0.024 & -0.010 & 0.008 & 0.127 & 0.040 & 0.2454 & 0.0016 \\
56 & 2459729.47627 & -19.8663 & 0.0025 & 3.1989 & 1141 & 56 & 7.517 & 0.024 & -0.020 & 0.007 & 0.187 & 0.043 & 0.2479 & 0.0018 \\
57 & 2459731.48177 & -19.8630 & 0.0025 & 2.4267 & 1007 & 56 & 7.521 & 0.024 & -0.021 & 0.010 & 0.138 & 0.040 & 0.2459 & 0.0028 \\
58 & 2459732.50174 & -19.8646 & 0.0024 & 1.9991 & 1072 & 57 & 7.512 & 0.024 & -0.026 & 0.008 & 0.173 & 0.040 & 0.2568 & 0.0019 \\
59 & 2459734.42702 & -19.8659 & 0.0024 & 1.4088 & 899 & 56 & 7.514 & 0.023 & -0.024 & 0.013 & 0.148 & 0.040 & 0.2598 & 0.0033 \\
60 & 2459736.52112 & -19.8507 & 0.0025 & 0.4303 & 1110 & 55 & 7.516 & 0.023 & -0.007 & 0.011 & 0.246 & 0.047 & 0.2412 & 0.0029 \\
61 & 2459746.46050 & -19.8737 & 0.0025 & -3.1924 & 1801 & 55 & 7.509 & 0.025 & -0.001 & 0.016 & 0.206 & 0.046 & 0.2500 & 0.0024 \\
62 & 2459747.53184 & -19.8659 & 0.0025 & -3.7311 & 1211 & 56 & 7.524 & 0.024 & -0.030 & 0.011 & 0.169 & 0.039 & 0.2572 & 0.0026 \\
63 & 2459749.46244 & -19.8812 & 0.0025 & -4.3122 & 1266 & 57 & 7.517 & 0.023 & -0.030 & 0.005 & 0.210 & 0.043 & 0.2550 & 0.0021 \\
64 & 2459750.41988 & -19.8646 & 0.0025 & -4.5827 & 1005 & 56 & 7.513 & 0.023 & -0.034 & 0.015 & 0.221 & 0.045 & 0.2490 & 0.0025 \\
65 & 2459751.48063 & -19.8702 & 0.0025 & -5.0950 & 1337 & 56 & 7.527 & 0.024 & -0.023 & 0.009 & 0.190 & 0.041 & 0.2497 & 0.0019 \\
66 & 2459753.52354 & -19.8581 & 0.0025 & -5.9280 & 1313 & 57 & 7.537 & 0.023 & -0.031 & 0.005 & 0.179 & 0.043 & 0.2535 & 0.0029 \\
67 & 2459769.44974 & -19.8518 & 0.0025 & -11.3311 & 1053 & 56 & 7.519 & 0.024 & -0.010 & 0.002 & 0.217 & 0.043 & 0.2487 & 0.0020 \\
68 & 2459771.44841 & -19.8534 & 0.0025 & -11.9695 & 1375 & 55 & 7.532 & 0.023 & -0.030 & 0.011 & 0.123 & 0.041 & 0.2471 & 0.0023 \\
69 & 2459773.49163 & -19.8630 & 0.0024 & -12.6885 & 1437 & 56 & 7.570 & 0.023 & 0.002 & 0.008 & 0.154 & 0.040 & 0.2388 & 0.0024 \\
70 & 2459782.48405 & -19.8788 & 0.0025 & -15.3286 & 1056 & 57 & 7.532 & 0.023 & -0.037 & 0.011 & 0.141 & 0.038 & 0.2557 & 0.0010 \\
71 & 2459783.52357 & -19.8717 & 0.0024 & -15.6691 & 1800 & 63 & 7.536 & 0.024 & -0.036 & 0.015 & 0.079 & 0.040 & 0.2491 & 0.0015 \\
72 & 2459785.38372 & -19.8641 & 0.0025 & -15.9387 & 724 & 56 & 7.509 & 0.024 & -0.014 & 0.009 & 0.189 & 0.042 & 0.2558 & 0.0019 \\
73 & 2459787.45927 & -19.8635 & 0.0025 & -16.6227 & 1783 & 56 & 7.513 & 0.024 & -0.017 & 0.009 & 0.236 & 0.044 & 0.2499 & 0.0019 \\
74 & 2459802.36596 & -19.8583 & 0.0030 & -19.7113 & 1800 & 47 & 7.509 & 0.023 & -0.023 & 0.015 & 0.161 & 0.042 & 0.2486 & 0.0020 \\
75 & 2459803.41882 & -19.8680 & 0.0024 & -19.9905 & 928 & 58 & 7.514 & 0.023 & -0.017 & 0.014 & 0.187 & 0.043 & 0.2461 & 0.0013 \\
76 & 2459804.39858 & -19.8755 & 0.0027 & -20.1243 & 1475 & 51 & 7.497 & 0.023 & -0.031 & 0.012 & 0.224 & 0.045 & 0.2420 & 0.0018 \\
77 & 2459806.35127 & -19.8614 & 0.0028 & -20.3575 & 1800 & 50 & 7.520 & 0.023 & -0.028 & 0.013 & 0.215 & 0.043 & 0.2436 & 0.0015 \\
78 & 2459807.36005 & -19.8548 & 0.0025 & -20.5321 & 780 & 56 & 7.520 & 0.022 & -0.020 & 0.005 & 0.216 & 0.043 & 0.2423 & 0.0021 \\
79 & 2459808.33717 & -19.8619 & 0.0025 & -20.6358 & 976 & 57 & 7.494 & 0.024 & -0.034 & 0.004 & 0.189 & 0.042 & 0.2550 & 0.0022 \\
80 & 2459811.37954 & -19.8551 & 0.0025 & -21.1393 & 1800 & 55 & 7.521 & 0.023 & -0.030 & 0.012 & 0.157 & 0.040 & 0.2524 & 0.0026 \\
81 & 2459812.37860 & -19.8590 & 0.0025 & -21.2651 & 1040 & 56 & 7.506 & 0.023 & -0.023 & 0.008 & 0.191 & 0.042 & 0.2486 & 0.0009 \\
82 & 2459813.39003 & -19.8528 & 0.0025 & -21.4068 & 1365 & 56 & 7.540 & 0.023 & -0.005 & 0.005 & 0.170 & 0.040 & 0.2528 & 0.0019 \\
83 & 2459814.34520 & -19.8669 & 0.0025 & -21.4392 & 962 & 56 & 7.523 & 0.024 & -0.010 & 0.003 & 0.157 & 0.040 & 0.2483 & 0.0015 \\
84 & 2459815.32563 & -19.8647 & 0.0025 & -21.5099 & 827 & 56 & 7.530 & 0.023 & -0.025 & 0.008 & 0.213 & 0.043 & 0.2533 & 0.0016 \\
85 & 2459826.38784 & -19.8626 & 0.0025 & -22.3888 & 1800 & 56 & 7.522 & 0.023 & -0.019 & 0.016 & 0.124 & 0.042 & 0.2630 & 0.0019 \\
86 & 2459827.30702 & -19.8560 & 0.0030 & -22.2859 & 1800 & 45 & 7.542 & 0.023 & -0.032 & 0.006 & 0.177 & 0.044 & 0.2475 & 0.0020 \\
87 & 2459828.30875 & -19.8612 & 0.0025 & -22.3144 & 754 & 56 & 7.524 & 0.023 & -0.021 & 0.008 & 0.227 & 0.044 & 0.2507 & 0.0011 \\
88 & 2459838.29509 & -19.8624 & 0.0025 & -22.2108 & 787 & 56 & 7.533 & 0.023 & -0.009 & 0.004 & 0.186 & 0.040 & 0.2547 & 0.0020 \\
89 & 2459840.30256 & -19.8663 & 0.0034 & -22.1330 & 1800 & 42 & 7.538 & 0.023 & -0.016 & 0.007 & 0.258 & 0.048 & 0.2513 & 0.0038 \\
90 & 2459841.29660 & -19.8635 & 0.0025 & -22.0682 & 1083 & 56 & 7.494 & 0.023 & -0.035 & 0.015 & 0.197 & 0.041 & 0.2446 & 0.0020 \\
\hline
\end{longtable}
\endgroup

\end{landscape}

\twocolumn
%-----------------------------------------------------------------
\section{Calibration of our spectral differential analysis}
\label{sec:validationofdifferentialanalysis}

\begin{figure}
   \centering
   \includegraphics[width=1.0\hsize]{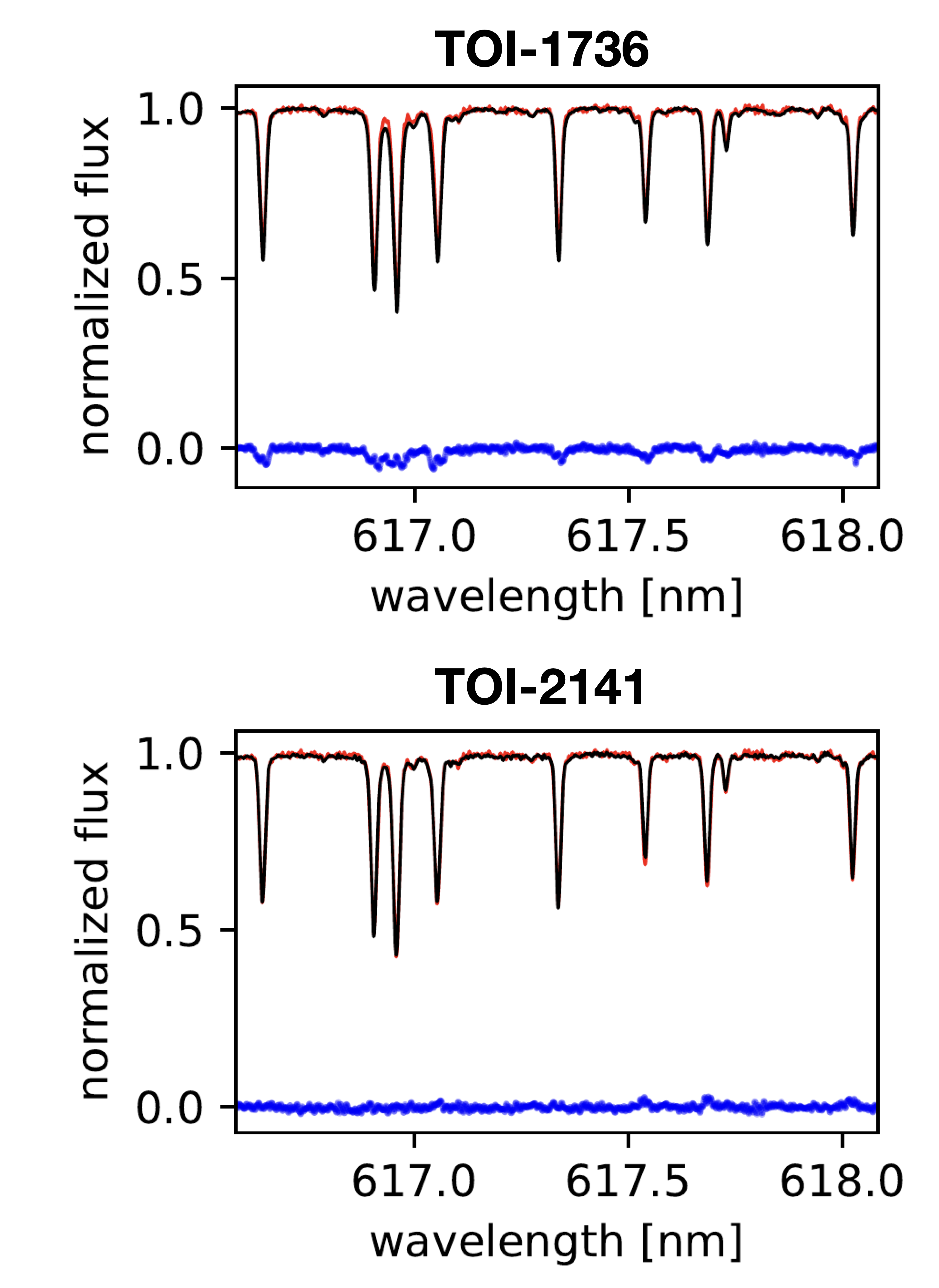}
      \caption{Visual spectral differences between stellar and solar spectra. Solid red lines show the solar spectrum used in this work as the reference spectrum for the strictly differential analysis. The black solid lines show the stellar spectrum for TOI-1736 (upper panel) and TOI-2141 (bottom panel), and the blue points show the respective spectral differences between stellar and solar normalized fluxes.
      }
        \label{fig:solardiff}
  \end{figure}

 The resulting stellar parameters obtained by our strictly differential analysis presented in Section \ref{sec:solardifferentialanalysis} have typical internal precision of $\sigma(T_{\rm eff})=8$~K, $\sigma(\log{g})=0.01$~dex, $\sigma({\rm [Fe/H]})=0.01$~dex, and $\sigma(v_{\rm t})=0.01$~\kms. To obtain the accuracy of these parameters, we conducted a series of validation tests, which are presented as supplemental material in this appendix.  As a first sanity check, we visually inspect the similarity between the stellar and solar spectra for our targets, which is illustrated in Figure \ref{fig:solardiff}.

 To estimate the errors introduced by instrumental instabilities, we calculated a stacked spectrum from the mean of a randomly selected subsample containing half of the available spectra, where we performed the differential analysis. We repeat this procedure 30 times. As a result, our EWs measurements are stable at $1.5\pm0.5$\% level, which translates into 7~K, 0.02~dex and 0.01~dex in $T_{\rm eff}$, $\log{g}$ and [Fe/H]. These are considered the minimum uncertainties achievable by our automatic differential technique applied to the SOPHIE data. Thus, we added these errors quadratically to the internal errors of our measurements.

 To estimate possible offsets in our spectroscopic parameters and validate the error analysis, we applied the same spectroscopic analysis to observations of the Moon and to a small subset of 5 previously characterized solar analogs bracketing our sample stars' parameters, namely: HIP~7585, HIP~28066, HIP~29432, HIP~77052, and HIP~79672. The comparison stars’ literature parameters were obtained from \cite{Spina2018} (hereafter S18) which was based on the same line list of this work and used a similar spectroscopic technique. The differences between our analysis and those from S18 are the higher S/N (500-1000) and the higher spectral resolution (R=120000) from their HARPS data. In addition, the authors performed manual EW line measurements, in contrast to our automatic procedure. Therefore, in terms of internal achievable error levels, we can assume their measurements as \textit{noise-free} in comparison to ours. In Table \ref{tab:firstcheck} we report the comparison between SOPHIE and S18 atmospheric parameters. The typical precision achieved by S18 are $\sigma(T_{\rm eff})=4$~K, $\sigma(\log{g})=0.012$~dex, and $\sigma({\rm [Fe/H]})=0.004$~dex. Our results agree with their estimates with an average offset of $\Delta(T_{\rm eff})=5\pm10$~K, $\Delta(\log{g})=+0.00\pm0.02$~dex, and $\Delta({\rm [Fe/H]})=+0.00\pm0.02$~dex. \cite{YanaGalarza2019} also validated their measurements with those in the literature using the same technique as in our work and using instrumentation with similar spectral resolution and higher signal-to-noise ratio (R=60000, S/N$\sim$400) compared to our SOPHIE data. The authors reported an rms of 12~K, 0.03~dex and 0.02~dex for $T_{\rm eff}$, $\log{g}$, and [Fe/H], respectively. In a similar fashion, S18 found an rms of $\sigma(T_{\rm eff})=15$~K, $\sigma(\log{g})=0.03$~dex, and $\sigma({\rm [Fe/H]})=0.017$~dex. These results are in line with our differential atmospheric parameters, highlighting the robustness of our methodology applied to similar Sun-like stars.   

\begin{table*}
\centering
\caption{Consistency check of our derived stellar parameters ($T_{\rm eff}$, $\log g$, and $[{\rm Fe}/{\rm H}]$).
}
\label{tab:firstcheck}
\begin{tabular}{cccccccc}
\hline
Star & $T_{\rm eff}$  & \multicolumn{2}{c}{$T_{\rm eff}$}& \multicolumn{2}{c}{$\log g$} & \multicolumn{2}{c}{$[{\rm Fe}/{\rm H}]$} \\
ID & (interf)\tablefootmark{a} & SOPHIE\tablefootmark{b} & S18\tablefootmark{c} & SOPHIE\tablefootmark{b} & S18\tablefootmark{c} & SOPHIE\tablefootmark{b} & S18\tablefootmark{c} \\
\hline
HIP~7585 & & $5821\pm10$ & $5822\pm3$ & $4.455\pm0.027$ & $4.445\pm0.008$ & $+0.059 \pm 0.011$ &  $+0.083\pm0.003$ \\
HIP~28066 & & $5757\pm13$ & $5742\pm4$ & $4.303\pm0.035$  & $4.300\pm0.011$ & $-0.147 \pm 0.014$  & $-0.147\pm0.003$ \\
HIP~29432 & & $5747\pm15$  & $5762\pm3$ & $4.378\pm0.045$ & $4.45\pm0.01$ & $-0.137 \pm 0.015$ & $-0.112\pm0.003$ \\
HIP~77052 & $5692\pm74$ & $5695\pm11$ & $5687\pm3$ & $4.460\pm0.036$  & $4.450\pm0.012$ & $+0.074 \pm 0.015$ & $+0.051\pm0.003$ \\
HIP~79672 & $5811\pm28$ & $5819\pm11$  & $5808\pm3$ & $4.429\pm0.034$ & $4.440\pm0.009$ & $+0.082 \pm 0.013$ & $+0.041\pm0.003$ \\
\hline
\multicolumn{2}{c}{$\Delta = {\rm SOPHIE} - {\rm S18}$} & \multicolumn{2}{c}{$-4\pm11$} & \multicolumn{2}{c}{$-0.012\pm0.031$} &  \multicolumn{2}{c}{$+0.003\pm0.026$} \\
\hline
\end{tabular}
\tablefoot{
\tablefoottext{a}{Effective temperatures for the available sources obtained through interferometry.}
\tablefoottext{b}{Spectroscopic parameters from our differential analysis.}
\tablefoottext{c}{Spectroscopic parameters from \cite{Spina2018} (S18).}
}
\end{table*} 

 Finally, to evaluate the accuracy of our methodology, we derived the solar atmospheric parameters from another moonlight spectrum obtained with the same instrumental setup reported in this work. Solar observations were analyzed following the same steps used to determine the parameters of solar analogs. We show in Table \ref{tab:finaldiffanalysisresults} the accuracy test using moonlight observations and final differential atmospheric parameters for TOI-1736 and TOI-2141. The agreement with our solar reference values used in \texttt{q2} confirms the differential analysis as accurate and precise, assuming there is negligible unaccounted-for residual systematics from the reduction process, for example.

\begin{table}
\centering
\caption{Atmospheric parameters obtained through our differential analysis.  
}
\label{tab:finaldiffanalysisresults}
\begin{tabular}{cccc}
\hline
Parameter & TOI-1736 & TOI-2141 & Sun (Moon)\\
\hline
$T_{\rm eff}$ & $5804\pm11$ & $5660\pm11$ &  $5778\pm7$ \\
$\log g$ & $4.33\pm0.03$ & $4.41\pm0.02$ & $4.426\pm0.021$  \\
$[{\rm Fe}/{\rm H}]$ & $0.14\pm0.03$ & $-0.10\pm0.02$ &  $-0.004\pm0.012$ \\
%$v_{\rm t}$ & $1.05\pm0.02$ & $0.72\pm0.02$ &   \\
\hline
\end{tabular}
\tablefoot{
As a second consistency check, we present our results for TOI-1736 and TOI-2141, as well as for a solar spectrum obtained from observations of the Moon.
}
\end{table} 

\section{Activity indices}
\label{app:activityindices}
This appendix details our S-index and H$_\alpha$ measurements. It also shows the time series analysis and RV correlation of the activity indices obtained from the SOPHIE spectra of TOI-1736 and TOI-2141.

\subsection{S-index}
\label{app:sindex}

 We computed the SOPHIE's instrumental S-index as follows:

\begin{equation}
S_{\rm SOPHIE} = \frac{N_H + N_K}{N_V + N_R},
\end{equation}

where $N_H$ and $N_K$ are the integrated fluxes of two triangular band-passes in a 0.109~nm wide window centered on the H (396.85 nm) and K (393.37 nm) Ca~II emission lines, and $N_V$ and $N_R$ are two continuum regions 2~nm wide centered at 390.107~nm and 400.107~nm, as illustrated in Figures \ref{fig:toi1736_sindex_halpha} and  \ref{fig:toi2141_sindex_halpha}. For the computation of $S_{\rm SOPHIE}$, we performed flux integration using spectral values generated via Monte Carlo (MC) sampling from a normal distribution, taking into account the central value and errors associated with each spectral element. Figures \ref{fig:toi1736_sindex_halpha} and  \ref{fig:toi2141_sindex_halpha} depict the posterior distributions for the MC samples.

\begin{figure*}
\centering
\includegraphics[width=0.5\hsize]{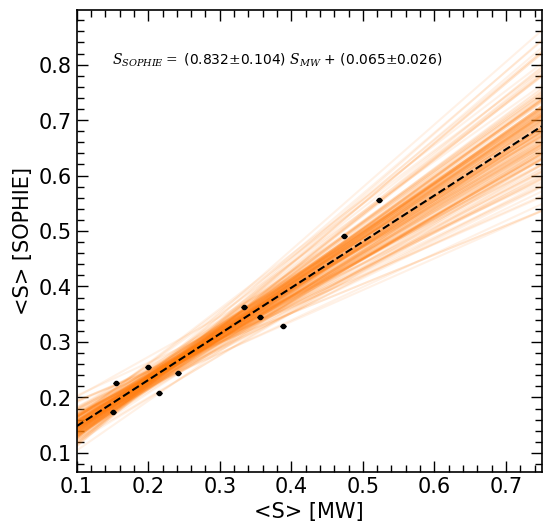}
\caption{Calibration of S-index from the SOPHIE HR spectra. Black points show the S-index for several targets (see Table  measured from the SOPHIE HR spectrum as a function of the reference MW values. The orange lines show a randomly selected sample of the fit models obtained from a posterior distribution of the model parameters, and the dashed black line show the median linear model.}
\label{fig:sophie-sindex-cal}
\end{figure*}

\begin{figure*}
\centering
\includegraphics[width=0.5\hsize]{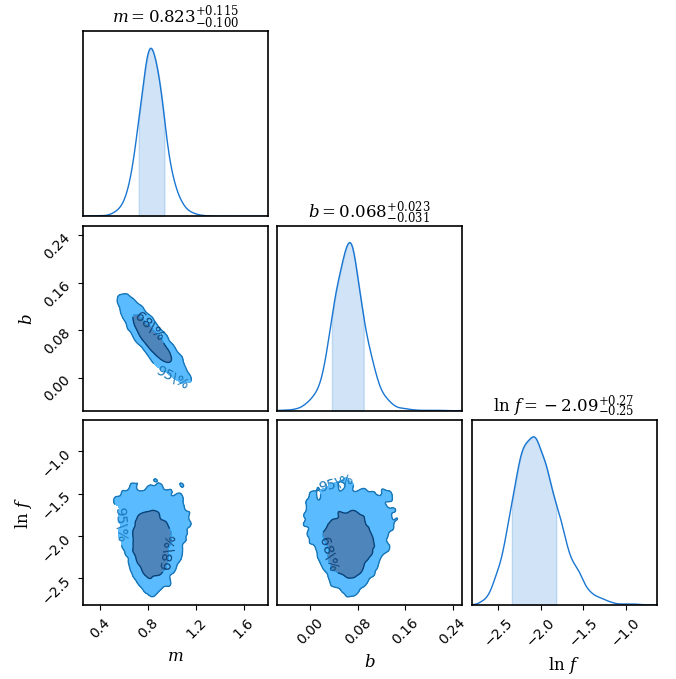}
\caption{Pairs plot showing the MCMC samples and posterior distributions of the coefficients for the SOPHIE S-index calibration.}
\label{fig:sophie-sindex_pairsplot}
\end{figure*}

To calibrate SOPHIE's instrumental S-index to the Mt. Wilson Observatory (MWO) system \citep[$S_{\rm MW}$,][]{Wilson1968,Egeland2017}, we downloaded archival spectra of ten selected stars that are known to have $S_{\rm MW}$ measurements in the range of 0.15 to 0.36. We only used observations in HR mode and with a peak S/N$>$100 in the first order, and we computed the S-index for each calibrator in the same way as described above. We least-square fit a linear relationship between $S_{\rm SOPHIE}$ and the $S_{\rm MW}$ values from \cite{Duncan1991}, and then we sampled the posterior distribution of the linear coefficients using a Bayesian MCMC framework with the package \texttt{emcee} \citep {foreman2013} as shown in figures \ref{fig:sophie-sindex-cal} and \ref{fig:sophie-sindex_pairsplot}. The best solution is expressed as follows:

\begin{equation}
    S_{\rm SOPHIE} = 0.832\pm0.104 \times S_{\rm MW} + 0.065\pm0.026.
    \label{eq:sindexcalibration}
\end{equation}

This calibration is consistent with the previous calibration obtained by \cite{Boisse2010}.  The $S_{\rm MW}$ values from \cite{Duncan1991} and our measurements of $S_{\rm SOPHIE}$ for our sample of calibrators are listed in Table \ref{tab:sindexcalibrationtable}.  Solving Equation \ref{eq:sindexcalibration} to get the $S_{\rm MW}$ for our two stars, we obtained $S_{\rm MW} = 0.16_{-0.03}^{+0.04}$ for TOI-1736 and $S_{\rm MW} = 0.21_{-0.03}^{+0.04}$ for TOI-2141, where the $S_{\rm SOPHIE}$ values were measured in the template spectrum of each target.  We have also measured the $S_{\rm MW}$ for each individual exposure, providing the S-index time series for both stars as presented in Tables \ref{tab:sophiervstoi1736} and \ref{tab:sophiervstoi2141}. 

\begin{table}
\caption{S-index calibrators observed with SOPHIE in HR mode.}
\label{tab:sindexcalibrationtable}
\begin{tabular}{ccccc}
\hline
ID & $S_{\rm SOPHIE}$ &$\sigma$(SOPHIE) & $S_{\rm MW}$ & $\sigma$(MW) \\ 
\hline
HD097334 & 0.3449 & 0.0022 & 0.356 & 0.012 \\
HD101501 & 0.3635 & 0.0023 & 0.334 & 0.002 \\
HD114378 & 0.2438 & 0.0015 & 0.241 & 0.005 \\
HD114710 & 0.2549 & 0.0016 & 0.200 & 0.010 \\
HD115404 & 0.5558 & 0.0040 & 0.523 & 0.044 \\
HD131156 & 0.4912 & 0.0035 & 0.473 & 0.010 \\
HD141004 & 0.2254 & 0.0014 & 0.155 & 0.021 \\
%HD149661 & 0.3836 & 0.000 & 0.327 & 0.029 \\
HD152391 & 0.3288 & 0.0020 & 0.389 & 0.036 \\
HD182101 & 0.2080 & 0.0012 & 0.215 & 0.013 \\
HD187013 & 0.1745 & 0.0010 & 0.151 & 0.005 \\ 
\hline
\end{tabular}
\end{table}

\subsection{H$\alpha$}
\label{app:halpha}

\begin{figure*}
\centering
\includegraphics[width=1.0\hsize]{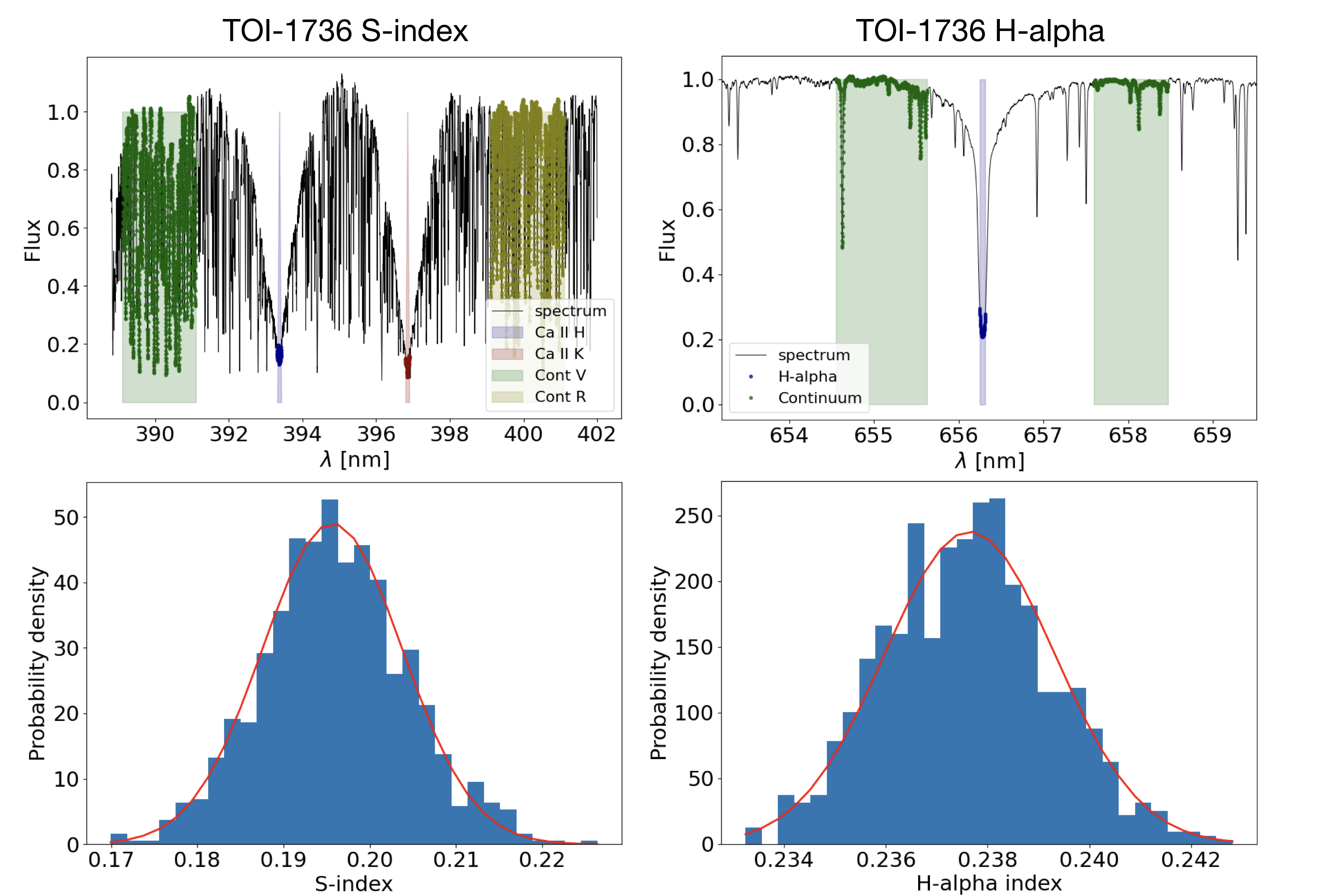}
\caption{TOI-1736 measurements of S-index and H$_\alpha$ from SOPHIE spectra. The top panels show the SOPHIE template spectrum around the S-index and H$_\alpha$ regions, showing the integration windows for the core of lines and continuum regions. The bottom panels show the posterior distribution obtained from the Monte Carlo samples calculated in the measurement of each quantity. }
\label{fig:toi1736_sindex_halpha}
\end{figure*}

\begin{figure*}
\centering
\includegraphics[width=1.0\hsize]{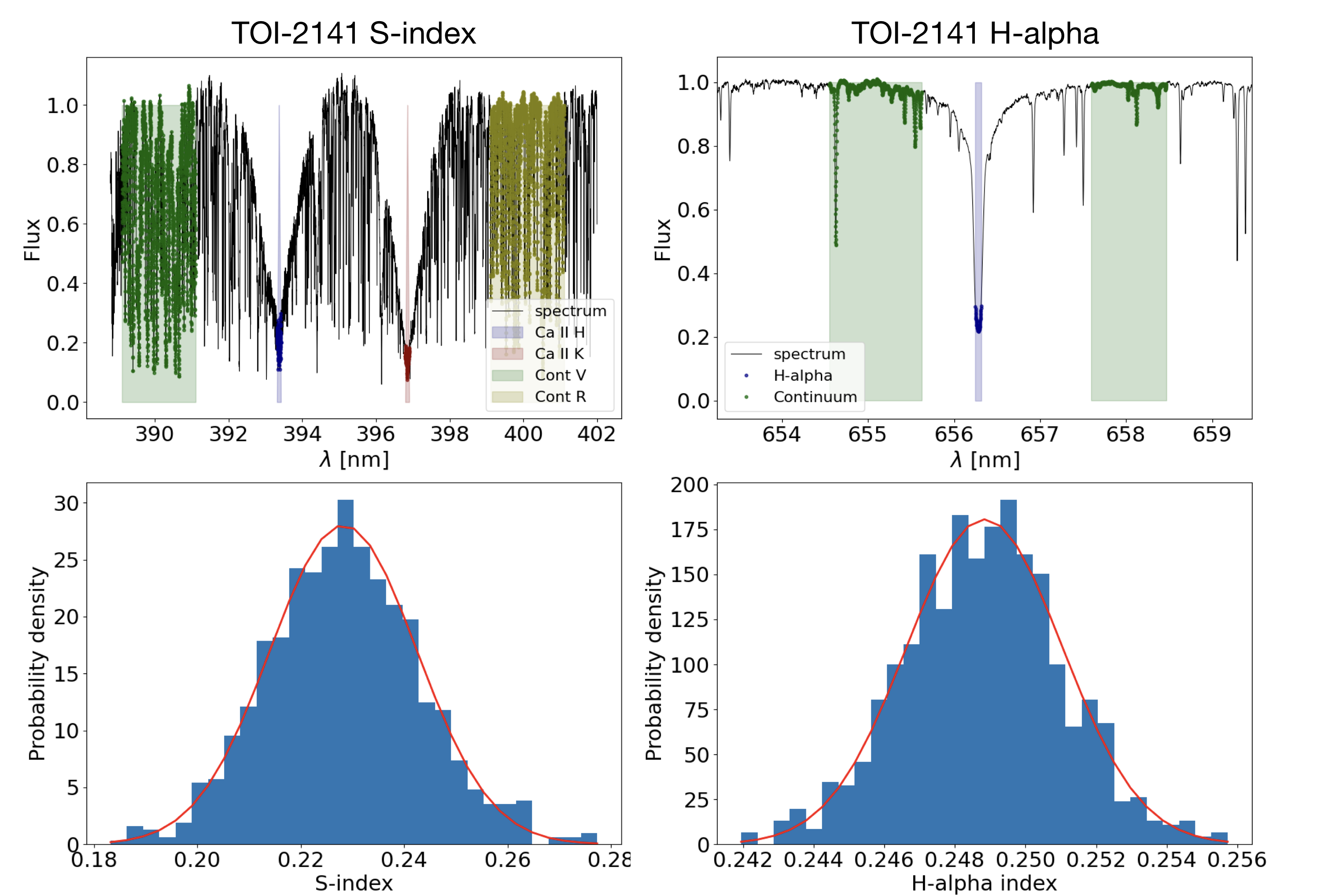}
\caption{TOI-2141 measurements of S-index and H$_\alpha$ from SOPHIE spectra. The top panels show the SOPHIE template spectrum around the S-index and H$_\alpha$ regions, showing the integration windows for the core of lines and continuum regions. The bottom panels show the posterior distribution obtained from the Monte Carlo samples calculated in the measurement of each quantity. }
\label{fig:toi2141_sindex_halpha}
\end{figure*}

The H$\alpha$ index is calculated from the flux at the center of the H$\alpha$ line as described by \cite{Boisse2009}, that is,

\begin{equation}
H_\alpha = \frac{F_{H_\alpha}}{F_1 + F_2},
\end{equation}

where $F_{H_\alpha}$ is the integrated flux measured within a 0.068~nm window centered at 656.2808~nm, $F_1$ and $F_2$ are the integrated fluxes measured near the edge of the H$\alpha$ wings centered at 655.087~nm and 658.031~nm with rectangular band-passes in 1.075~nm and 0.875~nm wide windows, as illustrated in Figures \ref{fig:toi1736_sindex_halpha} and  \ref{fig:toi2141_sindex_halpha}. In a similar way to the S-index calculation, the integration of spectral regions for H$\alpha$ calculation employs a Monte Carlo sampling approach. The H$\alpha$ indices measured in the template spectra are $H_\alpha = 0.2376\pm0.0016$ for TOI-1736 and $H_\alpha = 0.2488\pm0.0023$ for TOI-2141, and the individual measurements per exposure are listed in Tables \ref{tab:sophiervstoi1736} and \ref{tab:sophiervstoi2141}.

\subsection{Time series of activity indices}

\begin{figure*}
\centering
\includegraphics[width=1.0\hsize]{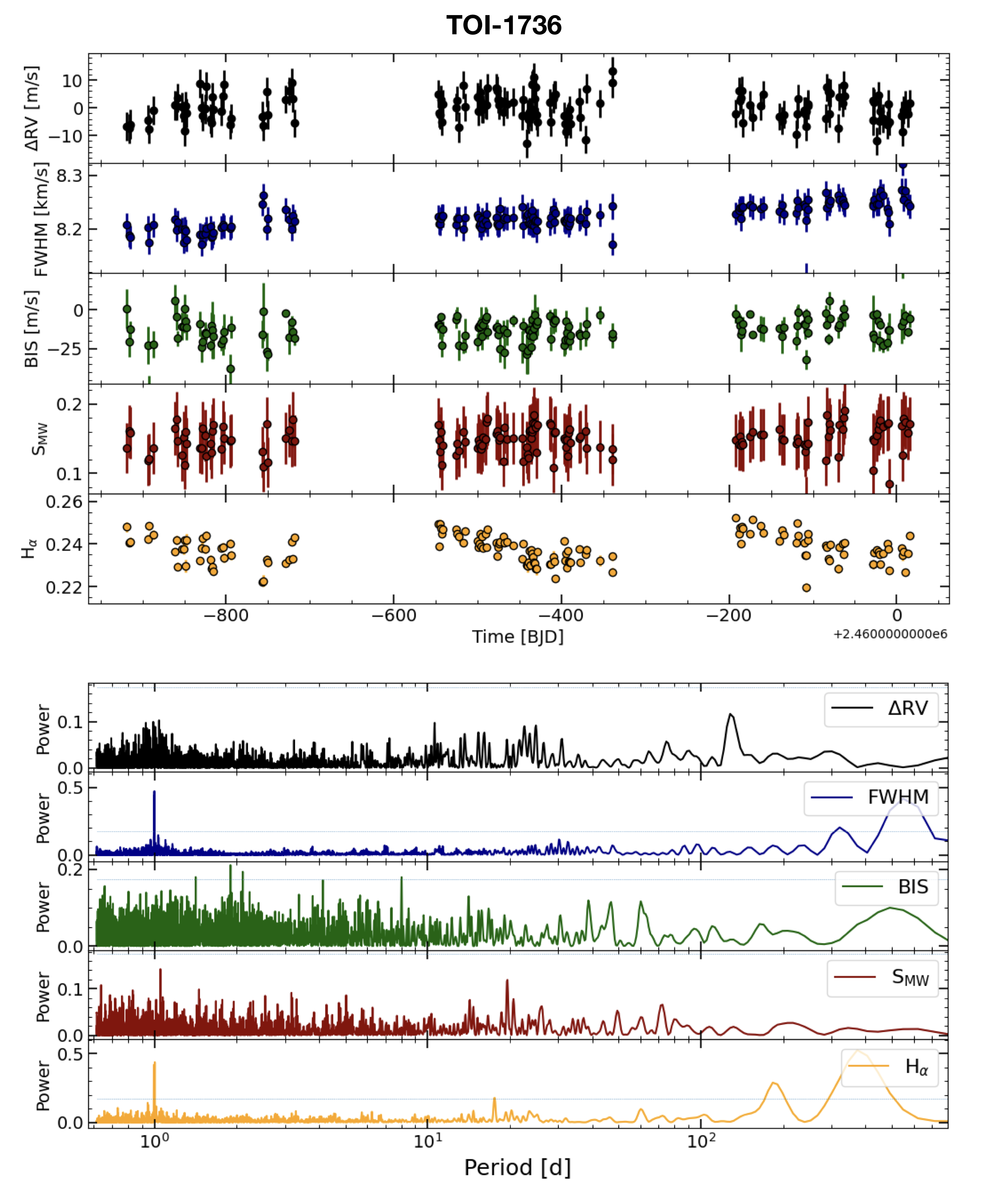}
\caption{Time series and respective GLS periodograms for the TOI-1736 SOPHIE RVs subtracted from the best-fit model and for the following activity indices: CCF FWHM, bisector span, H$_\alpha$, and S-index.}
\label{fig:toi1736_activity_timeseries}
\end{figure*}

\begin{figure*}
\centering
\includegraphics[width=1.0\hsize]{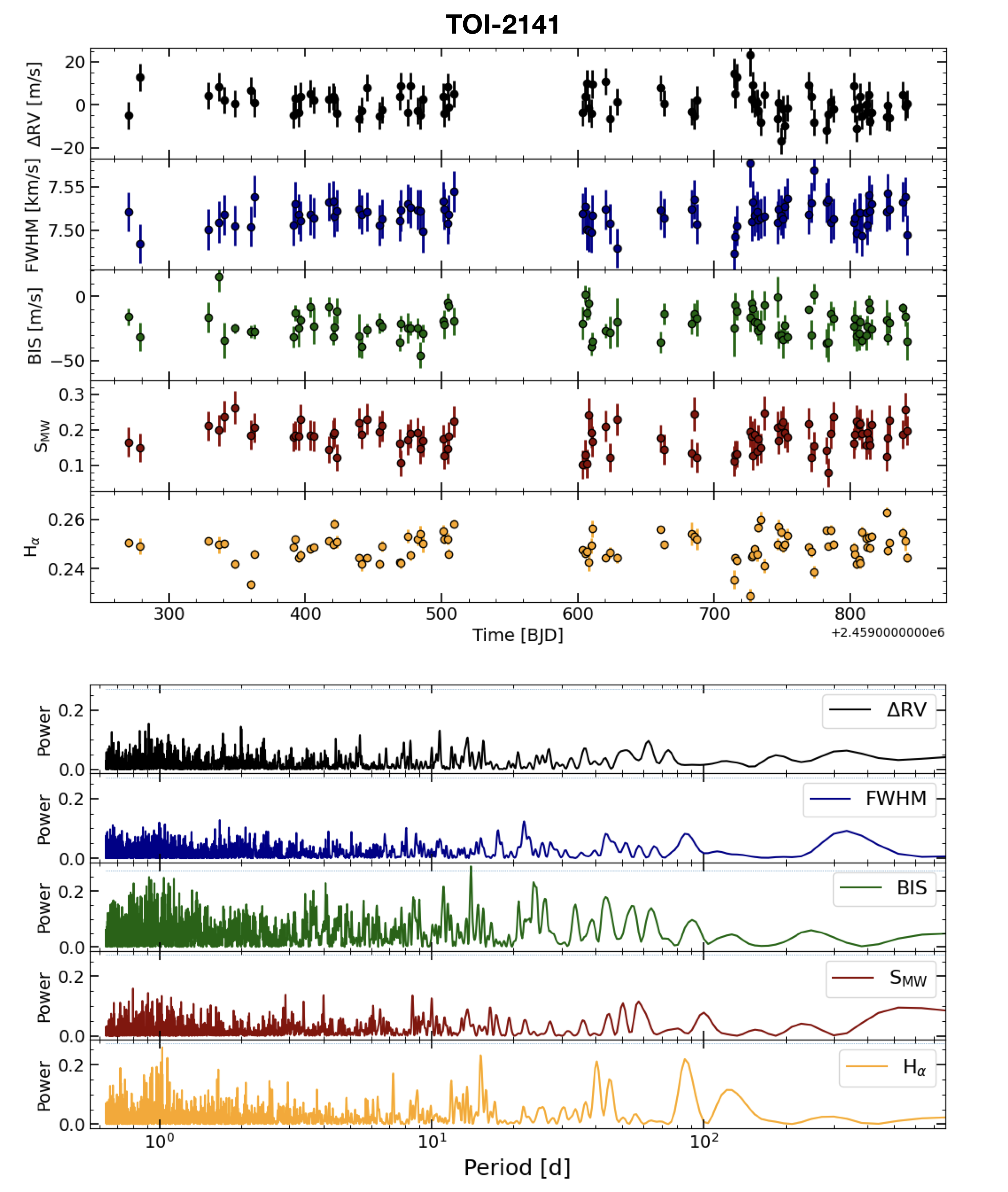}
\caption{Time series and respective GLS periodograms for the TOI-2141 SOPHIE RVs subtracted from the best-fit model and for the following activity indices: CCF FWHM, bisector span, H$_\alpha$, and S-index.}
\label{fig:toi2141_activity_timeseries}
\end{figure*}

As star rotation can modulate RVs and activity indices, we can analyze these quantities together and compare their time series and respective GLS periodograms in an attempt to find possible signals that can be used to infer the star rotation period. Figures \ref{fig:toi1736_activity_timeseries} and \ref{fig:toi2141_activity_timeseries} show the time series data and the GLS periodograms for both stars, where we cannot identify any significant periodicity. The residual RVs of TOI-1736 show a peak with a false alarm probability greater than 0.001 at 25~d, which may be related to star rotation, but this peak does not appear in any other indices.  As such, we could not find any strong evidence for the star's rotation period from these data.

\subsection{Correlations between activity indices and RVs}

\begin{figure*}
\centering
\includegraphics[width=1.0\hsize]{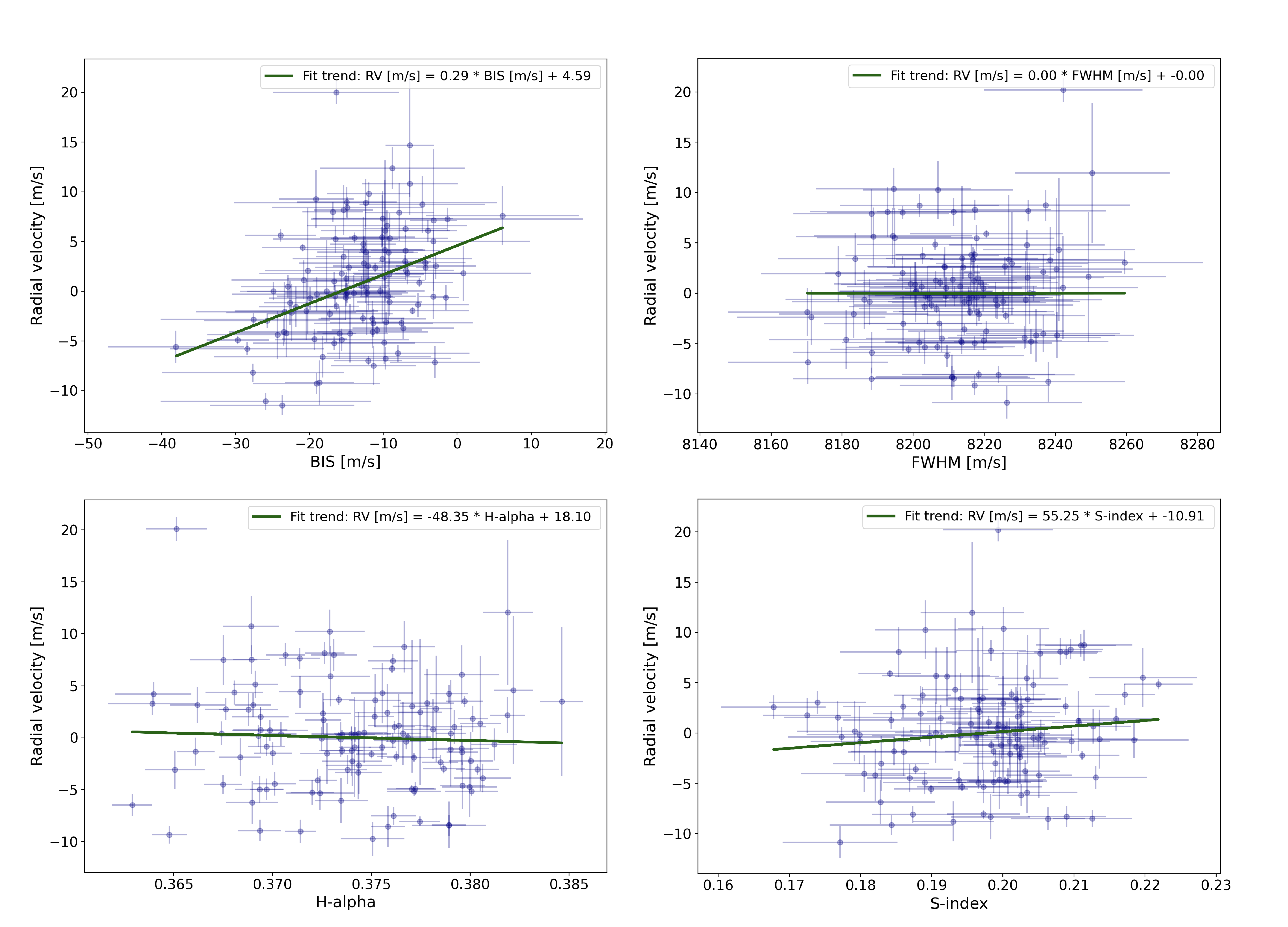}
\caption{Correlations between the TOI-1736 SOPHIE RVs subtracted from the best-fit model and the activity indices CCF FWHM, bisector span, H$_\alpha$, and S-index.}
\label{fig:toi1736_rv_activity_correlations}
\end{figure*}

\begin{figure*}
\centering
\includegraphics[width=1.0\hsize]{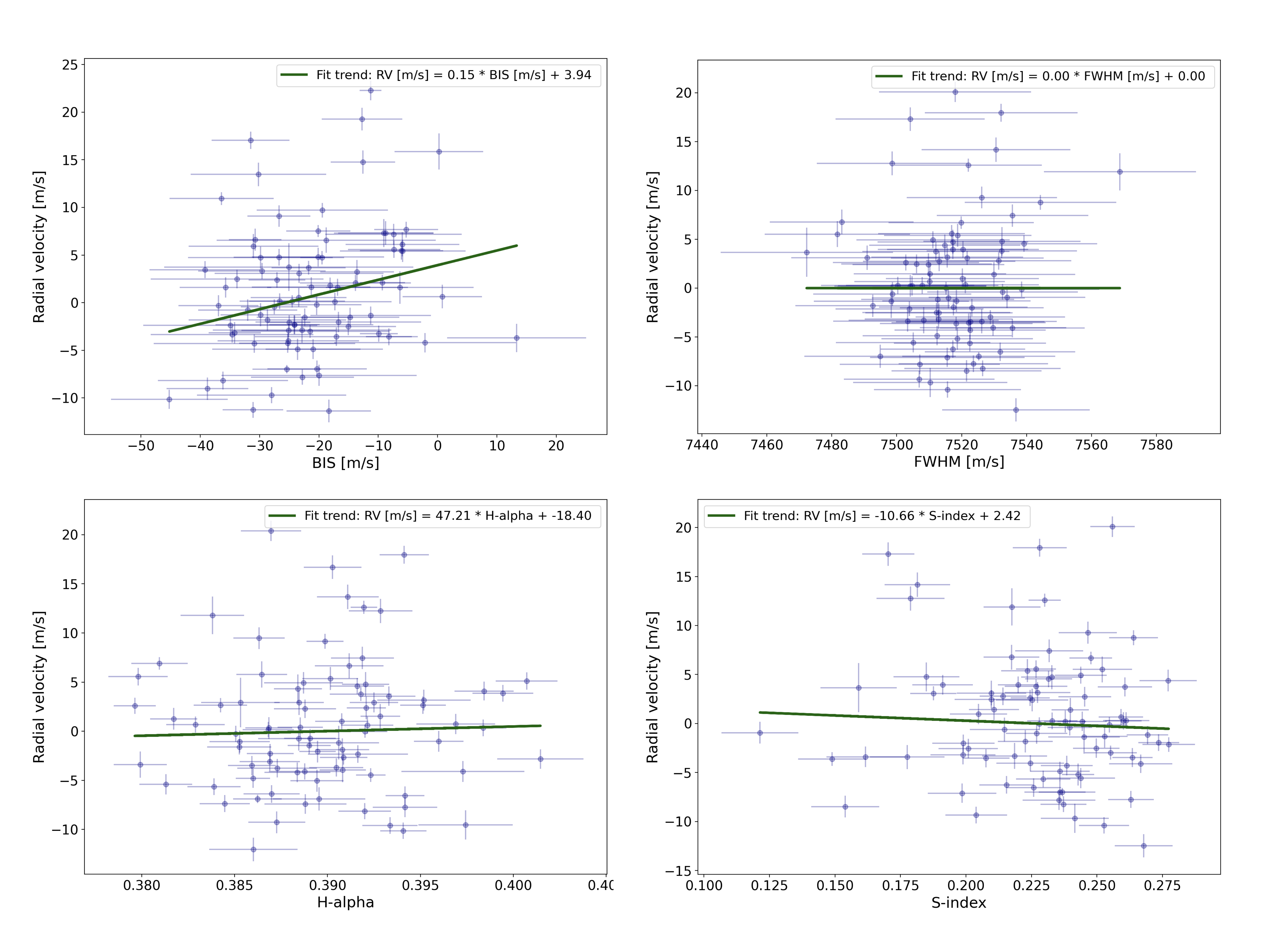}
\caption{Correlations between the TOI-2141 SOPHIE RVs subtracted from the best-fit model and the activity indices CCF FWHM, bisector span, H$_\alpha$, and S-index.}
\label{fig:toi2141_rv_activity_correlations}
\end{figure*}

Stellar activity can change the shape of line profiles and lead to deterioration in the detection of planetary signals, or even to spurious detections. One way to inspect the impact of certain types of activity on RV measurements is to look at the correlation between the RVs subtracted from the best-fit RV orbit model and the activity indices. Here, we present these correlations for the CCF FWHM, bisector span, H$_\alpha$, and S-index, as illustrated in Figures \ref{fig:toi1736_rv_activity_correlations} and \ref{fig:toi2141_rv_activity_correlations}. As pointed out in the main text, there is no strong correlation between these quantities, which indicates that our RVs are not strongly affected by stellar activity. 

\section{Priors and posterior distributions of model parameters}
\label{app:priors-and-posteriors}

\begin{figure*}
\centering
\includegraphics[width=0.95\hsize]{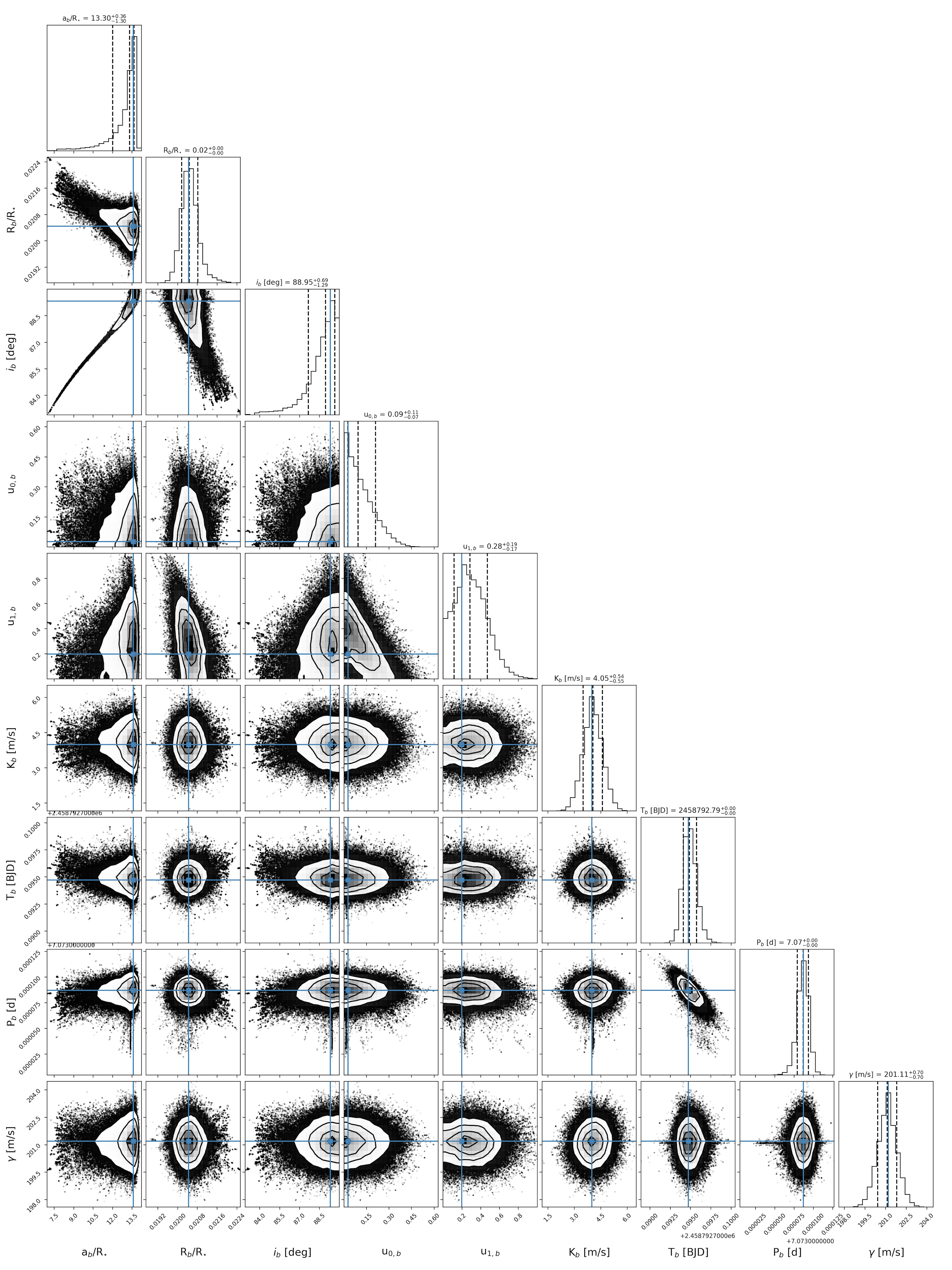}
\caption{Pairs plot showing the MCMC samples and posterior distributions of the free parameters in our joint analysis of the TESS photometry and the SOPHIE RV data of TOI-1736.  The contours mark the 1$\sigma$, 2$\sigma$, and 3$\sigma$ regions of the distribution. The blue crosses indicate the best-fit values for each parameter and the dashed vertical lines in the projected distributions show the median values and the 1$\sigma$ uncertainty (34\% on each side of the median).}
\label{fig:toi-1736_pairsplot}
\end{figure*}

\begin{figure*}
\centering
\includegraphics[width=0.95\hsize]{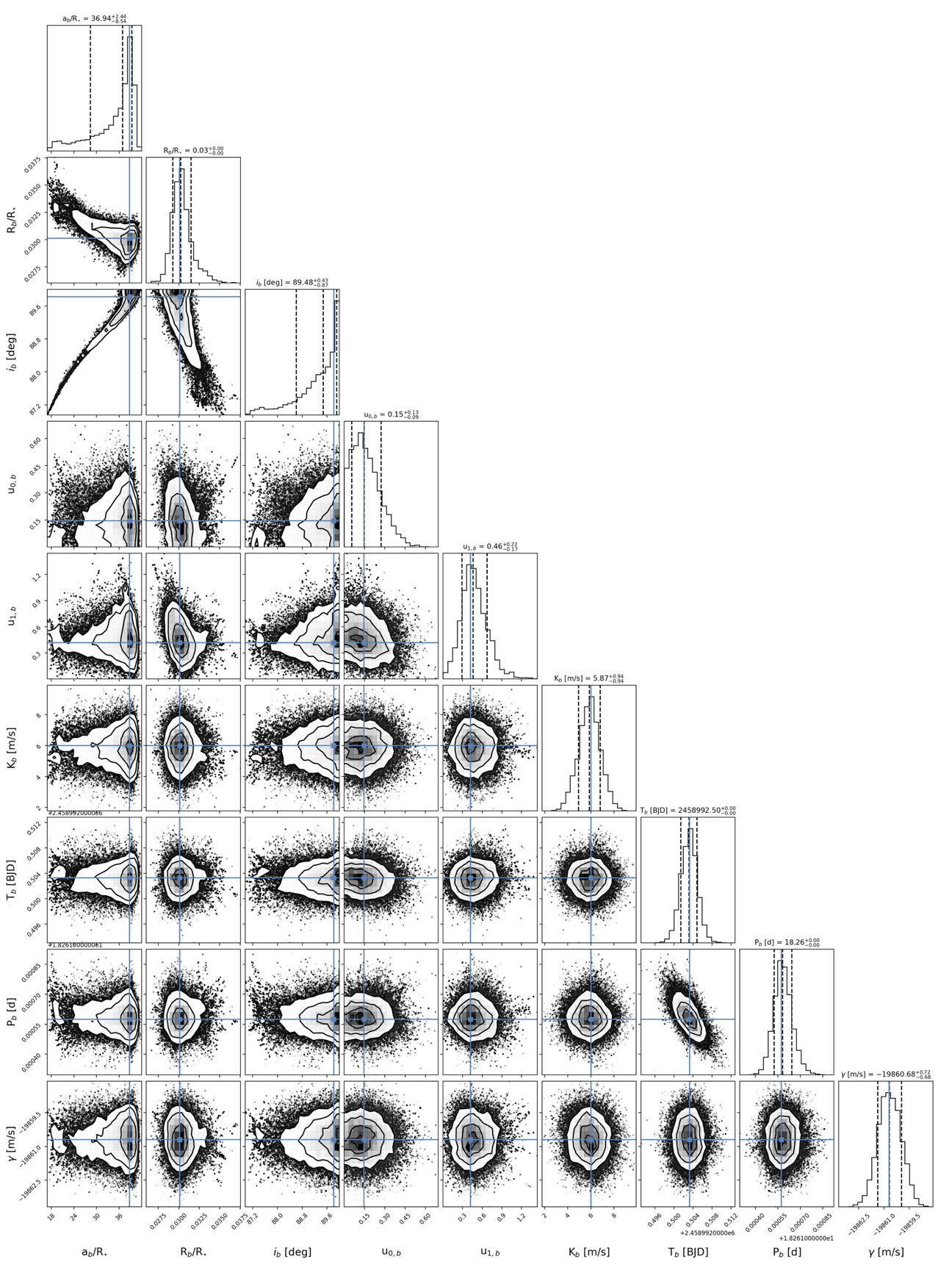}
\caption{Same as Figure \ref{fig:toi-1736_pairsplot} but for the analysis of TOI-2141 data.}
\label{fig:toi-2141_pairsplot}
\end{figure*}

This appendix presents in Table \ref{tab:planetsparamspriors} the prior distributions adopted for each model parameter that we considered in our analysis. Figures \ref{fig:toi-1736_pairsplot} and \ref{fig:toi-2141_pairsplot} illustrate the MCMC samples and the final posterior distributions of free parameters used  in our analysis. 

\begin{table*}
\centering
\caption{Prior distributions.}
\label{tab:planetsparamspriors}
\begin{tabular}{lccc}
\hline
Parameter & TOI-1736~b & TOI-1736~c & TOI-2141~b\\
\hline
 % BTJD = BJD - 2457000
time of conjunction, $T_{c}$ (BJD) & $\mathcal{U}(2458791,2458795)$ & $\mathcal{U}(2455300,2455330)$ & $\mathcal{U}(2458990,2458996)$ \\
orbital period, $P$ (d) & $\mathcal{U}(7.07,7.08)$ & $\mathcal{U}(200,1500)$ & $\mathcal{U}(18.0,18.5)$ \\
eccentricity, $e$  & FIXED (0)  & $\mathcal{U}(0,1)$ & FIXED (0) \\
argument of periastron, $\omega$ (deg) & FIXED (90) & $\mathcal{U}(0,360)$ & FIXED (90) \\
%-------------------------------------------------------
normalized semimajor axis, $a/R_{\star}$ & $\mathcal{U}(1,100)$ &  & $\mathcal{U}(1,100)$ \\
orbital inclination, $i_{p}$  (deg) &  $\mathcal{U}(80,90)$ &  & $\mathcal{U}(0,90)$ \\
%-------------------------------------------------------
planet-to-star radius ratio, $R_{p}/R_{\star}$  & $\mathcal{U}(0,1)$ &  & $\mathcal{U}(0,1)$ \\
%-------------------------------------------------------
velocity semi-amplitude, $K_{p}$ (m\,s$^{-1}$) & $\mathcal{U}(0,100)$ & $\mathcal{U}(0,1000)$ & $\mathcal{U}(0,100)$ \\
%-------------------------------------------------------
linear limb dark. coef., $u_{0}$  & $\mathcal{U}(0,1)$ &  & $\mathcal{U}(0,1)$ \\
quadratic limb dark. coef., $u_{1}$ & $\mathcal{U}(0,1)$ &  & $\mathcal{U}(0,1)$ \\
%-------------------------------------------------------
systemic radial velocity, $\gamma$ (km\,s$^{-1}$) & \multicolumn{2}{c}{$\mathcal{U}(-\infty,+\infty)$}  & $\mathcal{U}(-\infty,+\infty)$ \\
slope of linear trend, $\alpha_{\rm trend}$ (m\,s$^{-1}\,d^{-1}$) & \multicolumn{2}{c}{$\mathcal{U}(-\infty,+\infty)$}  &  FIXED (0) \\
%-------------------------------------------------------
GP phot. mean, $\mu$ (ppm) & $\mathcal{U}(-\infty,+\infty)$ &  & $\mathcal{U}(-\infty,+\infty)$ \\
GP phot. white noise, $\sigma_{\rm phot}$ (ppm) & $\mathcal{U}(0,+\infty)$ &  & $\mathcal{U}(0,+\infty)$ \\
GP phot. amplitude, $\alpha$ (ppm) & $\mathcal{U}(0,+\infty)$ &   &  $\mathcal{U}(0,+\infty)$ \\
GP phot. decay time, $l$ (d) & FIXED (10)  &  & FIXED (10)  \\
GP phot. smoothing factor, $\beta$ & FIXED (0.1)  &  & FIXED (0.1) \\
GP phot. period, $P$ (d) & $\mathcal{U}(2,1000)$ &   &  $\mathcal{U}(2,1000)$ \\
%-------------------------------------------------------
\hline
\end{tabular}
\end{table*}

\begin{table*}
\centering
\caption{GP QP kernel parameter posteriors for TESS flux baseline fitting.}
\label{tab:photgpposteriors}
\begin{tabular}{lcc}
\hline
Parameter & TOI-1736 & TOI-2141\\
\hline
%-------------------------------------------------------
GP phot. mean, $\mu$ (ppm) & $1.000015(13)$ &  $1.000014(9)$ \\
GP phot. white noise, $\sigma_{\rm phot}$ (ppm) & $0.000035(2)$ & $0.000064(7)$  \\
GP phot. amplitude, $\alpha$ (ppm) & $0.000133(9)$ &  $0.000060(9)$ \\
GP phot. decay time, $l$ (d) & 10  & 10 \\
GP phot. smoothing factor, $\beta$ & 0.1 & 0.1  \\
GP phot. period, $P$ (d) & $23^{+1.5}_{-1.7}$ &  $10.0^{+0.6}_{-3.3}$ \\
%-------------------------------------------------------
\hline
\end{tabular}
\end{table*}

\end{appendix}

\end{document}